%% file: document.tex
\documentclass[a4paper]{extarticle}
\usepackage{graphicx} 
\usepackage{subfig}
\usepackage[english]{babel}
\usepackage[hcentering, bindingoffset = 10mm, right = 20 mm, left = 10 mm, top=20 mm, bottom = 20 mm]{geometry}
\usepackage{amssymb, amsmath, amsfonts, mathtools}
\usepackage{bbm}
\usepackage{makecell}
\usepackage{tcolorbox}
\usepackage{xcolor}
\usepackage{wrapfig}
\usepackage{tikz}
\usepackage{hyperref}
\usepackage{tikz}
\usepackage{adjustbox}
\usepackage[table]{xcolor}

\definecolor{linkcolor}{rgb}{0,0,1}
\definecolor{citecolor}{rgb}{0.1,0.5,0.1}
\definecolor{urlcolor}{rgb}{0.7,0.1,0.1}

\usetikzlibrary{positioning,fit,shapes.geometric}
\usetikzlibrary {decorations.pathreplacing}
\usetikzlibrary {arrows.meta}
\usetikzlibrary {calc}
\usetikzlibrary{knots}
\usepackage{ytableau}
\usepackage{myfunc}

\usepackage{amsthm}
\theoremstyle{plain}
\newtheorem{lemma}{Lemma}[section]
\newtheorem{corollary}[lemma]{Corollary}

\hypersetup{pdfstartview=FitH,  linkcolor=linkcolor,urlcolor=urlcolor,citecolor=citecolor, colorlinks=true}

\numberwithin{equation}{section}

\newcommand{\jonesplus}{%
	\vcenter{\hbox{%
			\begin{tikzpicture}[
				x=0.18cm,
				y=0.18cm,
				line cap=round,
				wire/.style={
					draw=black,
					line width=0.5pt,
					-{Latex[length=1mm,width=0.8mm]}
				}
				]
				\draw[wire]
				(1,-1) -- (-1,1);
				
				\draw[white,line width=2.2pt]
				(-1,-1) -- (1,1);
				
				\draw[wire]
				(-1,-1) -- (1,1);
			\end{tikzpicture}%
	}}%
}

\newcommand{\jonesminus}{%
	\vcenter{\hbox{%
			\begin{tikzpicture}[
				x=0.18cm,
				y=0.18cm,
				line cap=round,
				wire/.style={
					draw=black,
					line width=0.5pt,
					-{Latex[length=1mm,width=0.8mm]}
				}
				]
				\draw[wire]
				(-1,-1) -- (1,1);
				
				\draw[white,line width=2.2pt]
				(1,-1) -- (-1,1);
				
				\draw[wire]
				(1,-1) -- (-1,1);
			\end{tikzpicture}%
	}}%
}

\newcommand{\joneszero}{%
	\vcenter{\hbox{%
			\begin{tikzpicture}[
				x=0.18cm,
				y=0.18cm,
				line cap=round,
				wire/.style={
					draw=black,
					line width=0.5pt,
					-{Latex[length=1mm,width=0.8mm]}
				}
				]
				\draw[wire]
				(-1,-1)
				.. controls
				(-0.55,-0.35)
				and
				(-0.55,0.35)
				..
				(-1,1);
				
				\draw[wire]
				(1,-1)
				.. controls
				(0.55,-0.35)
				and
				(0.55,0.35)
				..
				(1,1);
			\end{tikzpicture}%
	}}%
}

\title{Ising anyons in the $SU(2)_2$ Chern--Simons theory}
\author{Artem Belov$^{a,b}$, Andrey Morozov$^{a,b,c}$}
\date{}

\begin{document}
\ytableausetup{mathmode, boxframe=0.06em, boxsize=0.55em}
\maketitle
\vspace{-5.5cm}

\begin{center}
	\hfill ITEP/TH - 25/26
	
	\hfill IITP/TH - 23/26
	
	\hfill MIPT/TH 21/26
\end{center}

\vspace{3.0cm}
\begin{center}
	$^a${\small {\it Moscow Institute of Physics and Technology, Dolgoprudny 141701, Russia }}\\
	$^b${\small {\it Institute for Information Transmission Problems, Moscow 127994, Russia}}\\
	$^c${\small {\it ITEP, Moscow, 123182, Russia}}
\end{center}

\vspace{1cm}
\begin{abstract}
	Ising anyons and $SU(2)_2$ Chern--Simons theory share the fusion rules and
	$F$-symbols, and their $R$-symbols agree up to braiding orientation and an
	overall phase. The quantum-group description, $U_q(\mathfrak{sl}_2)$ at
	$q = e^{i\pi/4}$, is nevertheless not semisimple: tensor powers of the
	fundamental representation contain an additional spin-$3/2$ representation
	and reducible indecomposable representations, all of zero quantum dimension.
	We construct explicit highest-weight bases for up to four strands, compute
	the Racah and braiding matrices, and show that for four strands changes of
	fusion basis and braiding mix the physical sector with an indecomposable
	summand. We prove that this mixing affects neither the braid-closure
	invariant nor the plat-closure amplitude for any number of strands, so that
	each braiding matrix can be replaced by its physical block. The full
	representation theory thus reproduces the Ising-anyon data for all
	observables relevant to topological quantum computation.
\end{abstract}

\section{Introduction}

Quantum computing emerged from attempts to exploit the laws of quantum mechanics for information processing and the simulation of quantum systems~\cite{Feynman1982,Deutsch1985}. Interest in this field increased substantially following the development of quantum algorithms for integer factorization~\cite{Shor1997} and for searching an unstructured database~\cite{Grover1996}. Unlike a classical bit, a quantum bit can exist in a superposition of states, which in certain problems can provide a computational advantage over classical algorithms.

The practical implementation of quantum computing faces a number of fundamental challenges. One of them is that a quantum system inevitably interacts with its environment. Such interactions, together with imperfections in quantum operations and measurements, give rise to quantum noise and decoherence~\cite{Zurek2003,Preskill1998}. Various methods of quantum error correction have been developed to mitigate these effects. A distinct approach is provided by topological quantum computation~\cite{Kitaev2003,Nayak2008}. In this framework, computational information is encoded in global properties of the state of several spatially separated quasiparticles. Therefore, a local perturbation in the vicinity of one quasiparticle should not change the total fusion channel of the system. It is precisely this topological protection that makes such systems attractive for the robust implementation of quantum algorithms.

One of the best-known models arising in this context is that of Ising anyons~\cite{Nayak2008}. In the conformal field theory description, they correspond to the three primary chiral fields of the minimal model $\mathcal{M}(4,3)$, also known as the Ising minimal model: $\mathbbm{1}, \sigma, \psi$.

These fields possess properties that are essential for topological quantum computation: they can fuse through several channels, braid with one another, and admit different orders of successive fusion. In the language of topological quantum computation, these properties are described by fusion rules, braiding matrices, and transition matrices between different fusion bases. Their origin and physical meaning will be discussed in detail below.

Another description of systems related to anyons and topological quantum
computation is provided by Chern--Simons theory~\cite{Witten1989,KolganovMorozov2020}.
It is a three-dimensional topological gauge theory whose action does not
depend on the choice of metric on $(2+1)$-dimensional spacetime. In the
present work, we consider Chern--Simons theory with gauge group $SU(2)$. For
an $\mathfrak{su}(2)$-valued gauge field $A$ on an oriented three-manifold
$M$, the action reads
\begin{equation}\label{eq:CS_action_intro}
	S_{\mathrm{CS}}[A]
	=
	\frac{k}{4\pi}\int_M
	\operatorname{Tr}\left(A\wedge dA+\frac23\,A\wedge A\wedge A\right),
\end{equation}
where $\operatorname{Tr}$ is the trace in the adjoint representation.
The parameter $k$ is called the Chern--Simons level. Under gauge
transformations that are not homotopic to the identity, the action changes
by an integer multiple of $2\pi k$, so that the quantum theory is well
defined only for integer $k$. Such a theory is therefore commonly denoted
as $SU(2)_k$ Chern--Simons theory.

The main topological observables in this theory are Wilson loops. Let $\mathcal{K}$ be a closed contour (a knot or link) in $(2+1)$-dimensional spacetime, and let $\mathfrak{R}$ be a representation of the group $SU(2)$. Then the Wilson loop operator is given by

\begin{equation}
W_\mathfrak{R}(\mathcal{K})=
\operatorname{Tr}_\mathfrak{R}\,
\mathcal P\exp\left(\oint_\mathcal{K} A\right),
\end{equation}

where $A$ is the gauge field and $\mathcal P$ denotes path ordering along the contour. The corresponding expectation value $\langle W_\mathfrak{R}(\mathcal{K})\rangle$ yields a topological invariant of the corresponding labeled knot or link. Thus, Wilson loops provide a natural connection between Chern--Simons theory, knot theory, and topological quantum computation.

Direct evaluation of a Wilson loop expectation value using the functional integral is, in general, a difficult problem. One way to obtain a more convenient description is to choose a spatial slice of the three-dimensional spacetime. If Wilson lines intersect the chosen surface at several points, these points become punctures of the surface labeled by representations of $SU(2)$. The state space of $SU(2)_k$ Chern--Simons theory on such a punctured surface is identified with the space of chiral conformal blocks of the $SU(2)_k$ WZW model~\cite{Witten1989,LabastidaRamallo1989}. In other words, the two-dimensional conformal field theory describes not the planar section of the Wilson loop itself, but rather the state space associated with this section. Braiding of Wilson lines acts on this state space. In the conformal field theory description it corresponds to transformations of the conformal blocks.

There is also another approach to computing expectation values of Wilson loops in $SU(2)_k$ Chern--Simons theory. Witten showed that such expectation values are related to the Jones polynomial of the knot or link traced by the integration contour~\cite{Witten1989}. The Jones polynomial itself can be computed in several different ways. For example, it may be evaluated using skein relations. However, for knots represented as closures of braids, the Reshetikhin--Turaev construction is particularly convenient~\cite{RT1990}. In this approach, the geometric action of braiding the strands is replaced by the action of $R$-matrices on tensor products of representations of the quantum algebra $U_q(\mathfrak{sl}_2)$, while the braid closure is expressed in terms of the quantum trace.

The parameter $q$ of the quantum algebra $U_q(\mathfrak{sl}_2)$ is called the deformation parameter. In the context of Chern--Simons theory, its value is determined by the level $k$~\cite{Witten1989,Nayak2008}. In the conventions adopted below,

\begin{equation}
q=\exp\left(\frac{\pi i}{k+2}\right).
\end{equation}

Therefore, for a fixed level $k$, the parameter $q$ takes a fixed complex value. In particular, for $k=2$,

\begin{equation}
q=e^{\pi i/4},
\qquad q^8=1.
\end{equation}

Thus, the parameter of the quantum algebra is a root of unity. It should be noted that the precise relation between $q$ and the variable of the Jones polynomial depends on normalization conventions and is sometimes written with the replacement $q\mapsto q^2$. In what follows, we consistently use the convention specified above.

The representation theory of $U_q(\mathfrak{sl}_2)$ for $q$ not equal to a root of unity closely resembles the ordinary representation theory of $U(\mathfrak{sl}_2)$~\cite{Jimbo1985,Arnaudon1994}. When $q$ is a root of unity, however, the situation changes. In the class of representations considered in this paper, generated by the fundamental representation, tensor products cease to be completely reducible~\cite{Arnaudon1994}. Alongside irreducible representations of $\operatorname{spin}$ type, reducible but indecomposable representations of $\operatorname{ind}$ type appear.

Other classes of representations also exist at roots of unity, such as cyclic and semicyclic representations~\cite{Arnaudon1994,BishlerMironovMorozov2023}; however, they are not of the scope the present work.

It is also important for the present work that the minimal model $\mathcal{M}(4,3)$ and the $SU(2)_2$ WZW model are closely related at the level of the data relevant to topological quantum computation. In both theories, the sector under consideration contains three types of topological charges: $\mathbbm{1}, \sigma, \psi$ in the minimal model and $j=0, j=1/2, j=1$ in the $SU(2)_2$ WZW model. They can be identified as follows:

\begin{equation}
\mathbbm{1}\longleftrightarrow j=0,\qquad
\sigma\longleftrightarrow j=\frac12,\qquad
\psi\longleftrightarrow j=1.
\end{equation}

The fusion rules and the transition matrices between different fusion orders coincide in these two descriptions. The braiding matrices agree up to the choice of braiding orientation and an overall phase. At the same time, $\mathcal{M}(4,3)$ and the $SU(2)_2$ WZW model are not literally the same conformal field theory: their central charges are, respectively,

\begin{equation}
c_{\mathcal{M}(4,3)}=\frac12,
\qquad
c_{SU(2)_2}=\frac32.
\end{equation}

Therefore, in what follows we will not speak of a complete equivalence between these theories, but rather of a correspondence between the data relevant to the observables considered here.

Thus, one obtains the chain of interconnected structures shown in Fig.~\ref{fig:ising_wzw_cs_uq}.
\begin{figure}[!ht]
	\centering
	\begin{tikzpicture}[
		relation/.style={
			<->,
			>=Stealth,
			draw=black!75!black,
			line width=0.85pt
		},
		object/.style={
			align=center,
			font=\normalsize
		}
		]
		
		\node[object, text width=4.7cm] (ising) at (0,2.65) {
			Ising anyons\\
			in the minimal model $\mathcal{M}(4,3)$
		};
		
		\node[object, text width=4.3cm] (wzw) at (8.4,2.65) {
			$SU(2)_2$ WZW\\
			conformal field theory
		};
		
		\node[object, text width=5.0cm] (uq) at (0,0) {
			Highest-weight representations\\
			of $U_q(\mathfrak{sl}_2)$ at
			$q=\exp(i\pi/4)$
		};
		
		\node[object, text width=4.8cm] (wilson) at (8.4,0) {
			Wilson loops\\
			in $SU(2)_2$ Chern--Simons\\
			theory
		};
		
		\draw[relation] (ising.east) -- (wzw.west);
		\draw[relation] (wzw.south) -- (wilson.north);
		\draw[relation] (uq.east) -- (wilson.west);
		
	\end{tikzpicture}
	
	\caption{\centering
		Diagram illustrating the relations between Ising anyons, the WZW model, Wilson loops, and representations of the quantum algebra used in the present work.
	}
	\label{fig:ising_wzw_cs_uq}
\end{figure}
The two ends of this chain are particularly important for the present work. On the one hand, the theory of Ising anyons provides the expected fusion rules, braiding operations, and transformations between fusion bases. On the other hand, the quantum algebra $U_q(\mathfrak{sl}_2)$ allows the corresponding objects to be computed within representation theory and applied to observables in Chern--Simons theory. The main focus of this work is the study of representations of $U_q(\mathfrak{sl}_2)$ at $q=e^{\pi i/4}$.

First, a table of tensor-product decompositions for representations of
$\operatorname{spin}$ type will be presented (Sec.~\ref{sec:spin_type},
Table~\ref{tab:spin_tensor_products}; products involving representations of
$\operatorname{ind}$ type are collected in Table~\ref{tab:mult_table_spin_ind}).
It reproduces the fusion rules of Ising anyons~\eqref{eq:ising_fusion_rules} up
to representations of $\operatorname{ind}$ type (Sec.~\ref{sec:ind_type}) and
one additional representation of $\operatorname{spin}$ type,
$\operatorname{spin}(3/2,+)$, all of which have zero quantum dimension,
Eqs.~\eqref{eq:qdim_spin_three_half} and~\eqref{eq:qdim_ind_zero}; the
identification of the remaining three representations with the charges
$\mathbbm{1},\sigma,\psi$ is given in~\eqref{eq:rep_ising_notation}. We will then
show why the contributions of these representations cancel in the observables
considered here (Secs.~\ref{sec:three_strands}, \ref{sec:four_strands}
and~\ref{sec:arbitrary-strands}).

Next, the Racah matrix will be computed for the tensor product of three
fundamental representations, Eqs.~\eqref{eq:three_strands_racah}
and~\eqref{eq:three_strands_ising_F} in Sec.~\ref{sec:three_strands}. After
fixing the normalization conventions, it coincides with the fusion matrix of
Ising anyons~\eqref{eq:ising_fusion_matrix}. Using the universal $R$-matrix of
$U_q(\mathfrak{sl}_2)$~\eqref{eq:universal_R_matrix}, the braiding matrix will
then be constructed, Eqs.~\eqref{eq:two_strands_Uq_R}
and~\eqref{eq:three_strands_R}; it is related to the braiding matrix of Ising
anyons~\eqref{eq:ising_R_matrix} by the choice of braiding orientation and an
overall phase, Eqs.~\eqref{eq:two_strands_R_comparison}
and~\eqref{eq:three_strands_R_comparison}.

The most substantial part of the present work concerns the tensor product of
four fundamental representations, that is, a system of four strands
(Sec.~\ref{sec:four_strands}). In this case the decomposition acquires an
indecomposable summand~\eqref{eq:4strand_real_decomp}, and apparent
contradictions arise between the full representation-theoretic structure of the
quantum algebra and the data expected from the Ising theory: the transition
matrix between fusion bases~\eqref{eq:LP_transition} and the braiding
operator~\eqref{eq:R2_pairwise} mix the physical sector with this summand. We
will show how these apparent contradictions are resolved at the level of the
observables considered here, separately for the plat closure
(Sec.~\ref{sec:four_strands_plat},
Eqs.~\eqref{eq:R2_plat_blocks}--\eqref{eq:plat_amplitude_four}) and for the
braid closure (Sec.~\ref{sec:four_strands_braid},
Eqs.~\eqref{eq:ind_vanishes_quantum_trace} and~\eqref{eq:trace_vs_ising_four}).
In the final part (Sec.~\ref{sec:arbitrary-strands}) this mechanism will be
generalized to an arbitrary number of strands,
Eqs.~\eqref{eq:s34:trace-physical-reduction}
and~\eqref{eq:s34:plat-physical-reduction}.

The general statement that representations of zero quantum dimension can be
consistently discarded, so that the remaining representations of
$U_q(\mathfrak{sl}_2)$ at a root of unity reproduce the fusion rules,
transition matrices, and braiding matrices of $SU(2)_k$ Chern--Simons theory,
is known~\cite{Andersen1992,AndersenParadowski1995,KirbyMelvin1991,EtingofOstrik2022};
the precise relation between the $SU(2)_2$ theory and the Ising anyon model
is also known~\cite{Kitaev2006}. The contribution of the present work is an
explicit, matrix-level realization of this statement in bases of
highest-weight vectors: we construct the transition and braiding matrices
including the representations of $\operatorname{ind}$ type, show that these
representations mix with the physical ones under changes of fusion basis and
under braiding, and give an elementary proof that this mixing does not
affect the observables obtained by the braid closure and by the plat closure
for any number of strands.

\subsection{Minimal model $\mathcal{M}(4,3)$}
Minimal models are two-dimensional conformal field theories whose spectra
consist of a finite number of irreducible representations of the Virasoro
algebra~\cite{BPZ1984}. Recall that in two dimensions the infinitesimal local
conformal transformations $z\mapsto z+\epsilon(z)$,
$\bar z\mapsto\bar z+\bar\epsilon(\bar z)$ split into a holomorphic and an
antiholomorphic part, each generated by the Witt algebra
$[\ell_m,\ell_n]=(m-n)\ell_{m+n}$ with $\ell_n=-z^{n+1}\partial_z$. In the
quantum theory this symmetry acts projectively, and the holomorphic part is
represented by the modes $L_n$ of the energy--momentum tensor
$T(z)=\sum_n L_n z^{-n-2}$, which satisfy the central extension of the Witt
algebra, the Virasoro algebra
\begin{equation}
	[L_m,L_n]=(m-n)L_{m+n}+\frac{c}{12}\,m(m^2-1)\,\delta_{m+n,0},
\end{equation}
where $c$ is the central charge; the antiholomorphic modes $\bar L_n$ form a
second, commuting copy. In what follows only the holomorphic (chiral) copy is
needed. Irreducible highest-weight representations of the Virasoro algebra are
characterized by the central charge $c$ and the highest weight $h$. For a fixed
value of $c$, an infinite set of values of $h$ is generally
possible. However, Belavin, Polyakov, and Zamolodchikov showed~\cite{BPZ1984}
that for special discrete values of the central charge $c<1$ there exist
conformal field theories with only a finite number of primary fields. These
theories are known as minimal models.

In the present work, we consider the standard series of Virasoro minimal models, denoted by
\begin{equation}
\mathcal M(p',p),\qquad
p,p'\geq 2,\qquad
\gcd(p,p')=1.
\end{equation}
Their central charge is given by
\begin{equation}
c_{p',p}
=
1-\frac{6(p'-p)^2}{pp'}.
\end{equation}
The primary fields are denoted by $\phi_{r,s}$, where $1\leq r\leq p-1$ and $1\leq s\leq p'-1$.
Their conformal weights are given by the Kac formula
\begin{equation}
h_{r,s}
=
\frac{(p'r-ps)^2-(p'-p)^2}{4pp'}.
\end{equation}

At the same time, the fields $\phi_{r,s}\sim\phi_{p-r,p'-s}$ describe the same irreducible representation~\cite{BPZ1984}. Therefore, the number of distinct primary fields is
\begin{equation}
\frac{(p-1)(p'-1)}{2}.
\end{equation}

Among minimal models, the unitary ones are of particular physical importance. This is essential for interpreting the states as physical quantum states. For minimal models, the unitarity condition is satisfied if and only if $p'=p+1$~\cite{FQS1984,GKO1986}. Therefore, in what follows we restrict our attention to the unitary series

\begin{equation}
\mathcal M(p+1,p),
\qquad
c=1-\frac{6}{p(p+1)}.
\end{equation}

For $p=2$, one obtains the model $\mathcal M(3,2)$ with $c=0$. It contains only the vacuum sector and therefore does not give rise to a nontrivial anyon model. 

The next model in the unitary series is $\mathcal \mathcal{M}(4,3)$, or the Ising minimal model. Its central charge is $c=1/2$ and contains three distinct primary chiral fields. This model is the first unitary minimal model in which the fusion of two identical nontrivial fields can proceed through two different channels. The corresponding anyon theory is therefore non-Abelian. Substituting $p'=4$ and $p=3$ into the Kac formula, we obtain three primary fields $\mathbbm{1}$, $\sigma$ and $\psi$ (see table \ref{tab:ising_primary_fields}).
\begin{table}[h!]
	\centering
	\begin{tabular}{c|c|c}
		\textbf{Notation} & \textbf{Representative in the Kac table} & $\boldsymbol{h}$ \\ \hline
		$\mathbbm{1}$ & $\phi_{1,1}\sim\phi_{2,3}$ & $0$ \\
		$\sigma$ & $\phi_{1,2}\sim\phi_{2,2}$ & $1/16$ \\
		$\psi$ & $\phi_{2,1}\sim\phi_{1,3}$ & $1/2$
	\end{tabular}
	\caption{\centering Primary fields of the Ising minimal model $\mathcal{M}(4,3)$, their representatives in the Kac table, and the corresponding conformal weights.}
	\label{tab:ising_primary_fields}
\end{table}
Here, $\mathbbm{1}$ denotes the vacuum field. The field $\sigma$ is the chiral component of the spin field of the critical Ising model. The field with $h=1/2$ is commonly denoted by $\psi$, since it represents a chiral Majorana fermion. In the full, nonchiral Ising model, it corresponds to the energy-density field $\varepsilon(z,\bar z)$ with conformal weights
\begin{equation}
(h,\bar h)=\left(\frac12,\frac12\right).
\end{equation}
In what follows, we will be interested specifically in the chiral sector, and therefore we use the notation $\psi$ for the corresponding topological charge. The basic fusion rules follow from the general formula for minimal models:
\begin{equation}
\phi_{r_1,s_1}\times\phi_{r_2,s_2}
=
\sum_{\substack{
		r=|r_1-r_2|+1\\
		\text{step }2
}}^{\min(r_1+r_2-1,\;2p-r_1-r_2-1)}
\quad
\sum_{\substack{
		s=|s_1-s_2|+1\\
		\text{step }2
}}^{\min(s_1+s_2-1,\;2p'-s_1-s_2-1)}
\phi_{r,s}.
\end{equation}
In particular,
\begin{equation}
\phi_{1,2}\times\phi_{1,2}
=
\phi_{1,1}+\phi_{1,3}
\sim
\mathbbm{1}+\psi.
\end{equation}
Similarly, the remaining fusion rules are obtained:
\begin{equation}\label{eq:ising_fusion_rules}
\sigma\times\sigma=\mathbbm{1}+\psi,
\qquad
\sigma\times\psi=\sigma,
\qquad
\psi\times\psi=\mathbbm{1}.
\end{equation}
The first of these relations is essential: the outcome of fusing two $\sigma$-anyons is not uniquely determined. They may fuse either into the vacuum channel $\mathbbm{1}$ or into the fermionic channel $\psi$. Therefore, $\sigma$ is a non-Abelian anyon. Graphically, the fusion process is represented by a trivalent vertex: the two upper lines correspond to the initial anyons, while the lower line corresponds to a possible outcome of their fusion. All rules~\eqref{eq:ising_fusion_rules} are represented in this notation in Fig.~\ref{fig:ising_fusion_rules}.
\begin{figure}[!ht]
	\centering
	\begin{tikzpicture}[
		anyon/.style={
			draw=blue!75!black,
			line width=1.05pt,
			line cap=round
		},
		every node/.style={
			font=\small,
			inner sep=1.2pt
		}
		]
		
		\newcommand{\isingFusionTree}[4]{%
			\begin{scope}[xshift=#1]
				\draw[anyon] (-0.65,1.10) -- (0,0.38) -- (0,-0.28);
				\draw[anyon] ( 0.65,1.10) -- (0,0.38);
				
				\node[above] at (-0.65,1.10) {$#2$};
				\node[above] at ( 0.65,1.10) {$#3$};
				\node[below] at (0,-0.35) {$#4$};
			\end{scope}
		}
		
		\isingFusionTree{-4.8cm}{\sigma}{\sigma}{\mathbbm{1}}
		\isingFusionTree{-1.6cm}{\sigma}{\sigma}{\psi}
		\isingFusionTree{ 1.6cm}{\sigma}{\psi}{\sigma}
		\isingFusionTree{ 4.8cm}{\psi}{\psi}{\mathbbm{1}}
		
	\end{tikzpicture}
	
	\caption{\centering
		Graphical representation of the fusion rules~\eqref{eq:ising_fusion_rules}: \\
		$\sigma\times\sigma\to\mathbbm{1}$,
		$\sigma\times\sigma\to\psi$,
		$\sigma\times\psi\to\sigma$, and
		$\psi\times\psi\to\mathbbm{1}$.
	}
	\label{fig:ising_fusion_rules}
\end{figure}

The quantum dimensions follow from the fusion rules. If $N_{ab}^{\;c}$ denotes the number of fusion channels $a\times b\to c$, then the quantum dimensions $\mathfrak{d}_a$ are defined as the positive solution of the system~\cite{Verlinde1988,Nayak2008}
\begin{equation}\label{eq:ising_Q_dim_global_rule}
	\mathfrak{d}_a \mathfrak{d}_b=\sum_c N_{ab}^{\;c}\mathfrak{d}_c.
\end{equation}
The normalization of the vacuum sector is chosen as $\mathfrak{d}_{\mathbbm{1}}=1$. Then, from the fusion rules~\eqref{eq:ising_fusion_rules}, it follows that
\begin{equation}
	\mathfrak{d}_\psi^2=\mathfrak{d}_{\mathbbm{1}},
	\qquad
	\mathfrak{d}_\sigma^2=\mathfrak{d}_{\mathbbm{1}}+\mathfrak{d}_\psi,
\end{equation}
and, choosing the positive solution, we obtain
\begin{equation}\label{eq:ising_Q_dim}
	\mathfrak{d}_\psi=1,
	\qquad
	\mathfrak{d}_\sigma=\sqrt{2}.
\end{equation}

The quantum dimension does not represent the number of internal states of a single anyon. Rather, it characterizes the asymptotic growth of the dimension of the fusion space as the number of anyons increases. Let $\mathcal H_{\sigma^{\otimes N}\to c}$ denote the state space of $N$ $\sigma$-anyons with fixed total topological charge $c$. Repeated application of the fusion rules~\eqref{eq:ising_fusion_rules} gives
\begin{equation}
	\dim\mathcal H_{\sigma^{\otimes 2n}\to\mathbbm{1}}
	=
	\dim\mathcal H_{\sigma^{\otimes 2n}\to\psi}
	=
	2^{n-1},
	\qquad
	n\geq 1,
\end{equation}
\begin{equation}
	\dim\mathcal H_{\sigma^{\otimes(2n+1)}\to\sigma}
	=
	2^n,
	\qquad
	n\geq 0.
\end{equation}
In particular, four $\sigma$-anyons with total topological charge $\mathbbm{1}$ form a two-dimensional state space. Each additional pair of $\sigma$-anyons doubles its dimension. Therefore,
\begin{equation}
	\lim_{n\to\infty}
	\left(
	\dim\mathcal H_{\sigma^{\otimes 2n}\to\mathbbm{1}}
	\right)^{\frac{1}{2n}}
	=
	\sqrt{2}
	=
	\mathfrak{d}_\sigma.
\end{equation}
Thus, the value $\mathfrak{d}_\sigma=\sqrt{2}>1$ quantitatively reflects the non-Abelian nature of the $\sigma$-anyon: the number of independent fusion channels of a system containing many such anyons grows exponentially. By contrast, the quantum dimensions $\mathfrak{d}_{\mathbbm{1}}=\mathfrak{d}_\psi=1$ correspond to Abelian sectors, whose fusion does not generate additional channels.

The same fusion rules~\eqref{eq:ising_fusion_rules} can also be read off from the operator product expansions~\cite{BPZ1984,Nayak2008}. For the three chiral fields of the Ising model, they take the form
\begin{equation}
\begin{aligned}
	\sigma(z)\sigma(0)
	&\sim
	z^{-1/8}\bigl[\mathbbm{1}+\ldots\bigr]
	+
	z^{3/8}\bigl[\psi(0)+\ldots\bigr],\\
	\psi(z)\sigma(0)
	&\sim
	z^{-1/2}\bigl[\sigma(0)+\ldots\bigr],\\
	\psi(z)\psi(0)
	&\sim
	z^{-1}\bigl[\mathbbm{1}+\ldots\bigr].
\end{aligned}
\end{equation}
The powers of $z$ in these expressions are determined by the conformal weights of the fields. For example, the two leading channels in the expansion of $\sigma(z)\sigma(0)$ have exponents
\begin{equation}
h_{\mathbbm{1}}-2h_\sigma=-\frac18,
\qquad
h_\psi-2h_\sigma=\frac38.
\end{equation}

The operator product expansion also allows one to determine the phases arising from the braiding of fields~\cite{MooreSeiberg1989,Nayak2008}. For a positive exchange of two $\sigma$-fields, which locally corresponds to the analytic continuation $z\longmapsto e^{i\pi}z$, the two fusion channels acquire different phases:
\begin{equation}\label{eq:ising_R_symbols}
R_{\sigma\sigma}^{\mathbbm{1}}
=
e^{-i\pi/8},
\qquad
R_{\sigma\sigma}^{\psi}
=
e^{3i\pi/8}.
\end{equation}
Therefore, in the basis where the two $\sigma$-anyons are fused first, the braiding matrix takes the form
\begin{equation}\label{eq:ising_R_matrix}
R_{\sigma\sigma}
=
\begin{pmatrix}
	e^{-i\pi/8} & 0\\
	0 & e^{3i\pi/8}
\end{pmatrix}.
\end{equation}
A graphical representation of the action of the braiding operator is shown in Fig.~\ref{fig:ising_R_action}. The operator $R_{\sigma\sigma}$, acting on two $\sigma$-strands, preserves their fusion channel while multiplying the corresponding state by its eigenphase. In the channel $\sigma\times\sigma\to\mathbbm{1}$, this phase is $e^{-i\pi/8}$, whereas in the channel $\sigma\times\sigma\to\psi$, it is $e^{3i\pi/8}$.
\begin{figure}[h!]
	\centering
	\begin{tikzpicture}[
		anyon/.style={
			draw=blue!75!black,
			line width=1.05pt,
			line cap=round
		},
		transition/.style={
			-{Stealth[length=2.2mm]},
			draw=blue!75!black,
			line width=0.9pt
		},
		every node/.style={font=\small, inner sep=1.2pt}
		]
		
		\newcommand{\drawIsingRChannel}[2]{%
			
			\draw[anyon] (-0.65,1.12) -- (0,0.38) -- (0,-0.28);
			\draw[anyon] ( 0.65,1.12) -- (0,0.38);
			
			\node[above] at (-0.65,1.12) {$\sigma$};
			\node[above] at ( 0.65,1.12) {$\sigma$};
			\node[below] at (0,-0.35) {$#1$};
			
			\draw[transition] (0.98,0.43) -- (2.10,0.43)
			node[midway,above,text=black] {$R_{\sigma\sigma}$};
			
			\draw[anyon]
			(3.95,1.12)
			.. controls (4.05,0.82) and (2.80,0.94) ..
			(2.90,0.58)
			.. controls (2.96,0.46) and (3.12,0.39) ..
			(3.30,0.38);
			
			\draw[
			anyon,
			preaction={draw=white,line width=3.8pt}
			]
			(2.65,1.12)
			.. controls (2.55,0.82) and (3.80,0.94) ..
			(3.70,0.58)
			.. controls (3.64,0.46) and (3.48,0.39) ..
			(3.30,0.38);
			
			\draw[anyon] (3.30,0.38) -- (3.30,-0.28);
			
			\node[above] at (2.65,1.12) {$\sigma$};
			\node[above] at (3.95,1.12) {$\sigma$};
			\node[below] at (3.30,-0.35) {$#1$};
			
			\node at (4.55,0.38) {$=$};
			\node at (5.25,0.38) {$#2$};
			
			\draw[anyon] (6.05,1.12) -- (6.70,0.38) -- (6.70,-0.28);
			\draw[anyon] (7.35,1.12) -- (6.70,0.38);
			
			\node[above] at (6.05,1.12) {$\sigma$};
			\node[above] at (7.35,1.12) {$\sigma$};
			\node[below] at (6.70,-0.35) {$#1$};
		}
		
		\drawIsingRChannel{\mathbbm{1}}{e^{-i\pi/8}}
		
		\begin{scope}[yshift=-3.15cm]
			\drawIsingRChannel{\psi}{e^{3i\pi/8}}
		\end{scope}
		
	\end{tikzpicture}
	
	\caption{\centering
		Action of the positive braiding operator $R_{\sigma\sigma}$ on the two allowed fusion channels $\sigma\times\sigma=\mathbbm{1}+\psi$.
	}
	\label{fig:ising_R_action}
\end{figure}

In the same conventions,
\begin{equation}
R_{\psi\psi}^{\mathbbm{1}}=-1,
\qquad
R_{\psi\sigma}^{\sigma}
=
R_{\sigma\psi}^{\sigma}
=
e^{-i\pi/2}.
\end{equation}

Finally, let us consider three $\sigma$-anyons with total topological charge $\sigma$. Their state space is two-dimensional: the intermediate result of fusing the first two anyons can be either $\mathbbm{1}$ or $\psi$. Two natural bases can be chosen: in the first basis, anyons $1$ and $2$ are fused first, while in the second basis, anyons $2$ and $3$ are fused first.

To determine the transition matrix between these bases, it is convenient to consider the four-point correlation function of the field $\sigma$. The fourth field represents the total topological charge $\sigma$ of the three anyons: it is the out-state $\langle\sigma|$, realized by the insertion $\sigma(\infty)=\lim_{w\to\infty}w^{2h_\sigma}\sigma(w)$~\cite{BPZ1984,Nayak2008}. By means of a global conformal transformation, three of the four insertion points can be placed at $\infty$, $1$ and $0$, with only one coordinate $z$ remaining:
\begin{equation}\label{eq:four_point_function}
	\left\langle
	\sigma(\infty)\sigma(1)\sigma(z)\sigma(0)
	\right\rangle.
\end{equation}

\begin{figure}[h!]
	\centering
	\begin{tikzpicture}[
		plane/.style={draw=black, line width=0.6pt, fill=blue!8},
		axis/.style={draw=black!60, line width=0.5pt},
		fixed/.style={circle, fill=blue!65!black, inner sep=2pt},
		free/.style={circle, draw=blue!75!black, fill=white, line width=1.0pt, inner sep=1.8pt},
		fusion/.style={draw=blue!75!black, line width=1.05pt, dashed, line cap=round},
		motion/.style={-{Stealth[length=2mm]}, draw=blue!75!black, line width=0.9pt},
		every node/.style={font=\small, inner sep=1.2pt}
		]
		
		\newcommand{\fourPointPlane}[5]{%
			\begin{scope}[xshift=#1]
				\draw[plane] (-1.2,-1.0) rectangle (3.4,2.0);
				\node[anchor=north west] at (-1.15,1.95) {$\mathbb C$};
				\draw[axis] (-1.2,0) -- (3.4,0);
				
				\node[fixed, label={[label distance=1pt]below:$\sigma(0)$}] at (0,0) {};
				\node[fixed, label={[label distance=1pt]below:$\sigma(1)$}] at (2.2,0) {};
				\node[fixed, label={[label distance=1pt]above right:$\sigma(\infty)$}] at (3.4,2.0) {};
				
				\draw[fusion] (#3) ellipse (1.0 and 0.95);
				\draw[motion] (#2) to[bend right=20] (#4);
				
				\node[free, label={[label distance=2pt]above:$\sigma(z)$}] at (#2) {};
				
				\node at (1.1,-1.35) {#5};
			\end{scope}
		}
		
		\fourPointPlane{0cm}{0.75,0.85}{0.35,0.25}{0.2,0.25}{$s$-channel: $\sigma(z)\sigma(0)$, $z\to0$}
		\fourPointPlane{6.4cm}{1.45,0.85}{1.85,0.25}{2.0,0.25}{$t$-channel: $\sigma(1)\sigma(z)$, $z\to1$}
		
	\end{tikzpicture}
	
	\caption{\centering
		The four-point function~\eqref{eq:four_point_function}. The insertion points $0$, $1$ and $\infty$ are fixed by a global conformal transformation, while $z$ remains free. The dashed region encloses the pair of fields fused first: $\sigma(z)\sigma(0)$ in the $s$-channel ($z\to0$) and $\sigma(1)\sigma(z)$ in the $t$-channel ($z\to1$).
	}
	\label{fig:four_point_function_channels}
\end{figure}

The fusion of the fields $\sigma(z)$ and $\sigma(0)$ is referred to as the $s$-channel, while the fusion of $\sigma(1)$ and $\sigma(z)$ defines the $t$-channel~\cite{MooreSeiberg1989}(Fig.~\ref{fig:four_point_function_channels}). The field $\sigma$ is degenerate at the second level, and therefore the Ward identities reduce the chiral conformal blocks of this correlation function to solutions of a second-order differential equation~\cite{BPZ1984}. Its two solutions in the $s$-channel correspond to the intermediate fields $\mathbbm{1}$ and $\psi$:
\begin{equation}
	\begin{aligned}
		\mathcal{F}^{(s)}_{\mathbbm{1}}(z)
		&=
		\frac{1}{\sqrt{2}},[z(1-z)]^{-1/8}
		\sqrt{1+\sqrt{1-z}}, \\
		\mathcal{F}^{(s)}_{\psi}(z)
		&=
		\frac{1}{\sqrt{2}},[z(1-z)]^{-1/8}
		\sqrt{1-\sqrt{1-z}}.
	\end{aligned}
\end{equation}
In the $t$-channel, the fields $\sigma(1)$ and $\sigma(z)$ are fused first. The corresponding conformal blocks take the form
\begin{equation}
	\begin{aligned}
		\mathcal{F}^{(t)}_{\mathbbm{1}}(z)
		&=
		\frac{1}{\sqrt{2}},[z(1-z)]^{-1/8}
		\sqrt{1+\sqrt{z}}, \\
		\mathcal{F}^{(t)}_{\psi}(z)
		&=
		\frac{1}{\sqrt{2}},[z(1-z)]^{-1/8}
		\sqrt{1-\sqrt{z}}.
	\end{aligned}
\end{equation}

\begin{figure}[h!]
	\centering
	\begin{tikzpicture}[baseline={([yshift=-.4ex]current bounding box.center)}, scale=0.6, font=\large]
		\node[inner sep=4pt] (A) at (-1,-2.5) {$\mathcal{F}^{(s)}_i$};
		\draw[thick] (A.north west) -- (A.north east) -- (A.south east);
		
		\node[inner sep=4pt] (A) at (13,-2.5) {$\mathcal{F}^{(t)}_j$};
		\draw[thick] (A.north east) -- (A.north west) -- (A.south west);
		
		\draw[blue, thick] (2,-2)--(0,0);
		\draw[blue, thick] (3,-1)--(2,0);
		\draw[blue, thick] (3,-1)--(4,0);
		\draw[blue, thick] (2,-2)--(3,-1);
		\draw[blue, thick] (2,-2)--(3,-3);

		\node[black] at (4, 0.5) {$\sigma$};
		\node[black] at (2, 0.5) {$\sigma$};
		\node[black] at (0, 0.5) {$\sigma$};
		
		\node[black] at (1.8, -2.6) {$\sigma$};
		\node[black] at (3, -1.6) {$i$};
		
		\draw[->, ultra thick] (4,-1.5)--(8,-1.5)node[midway, above]{$\hat{\mathcal{F}}^{\tiny\sigma\sigma\sigma}_{\tiny\sigma}$};
		
		\draw[blue, thick] (10,-2)--(8,0);
		\draw[blue, thick] (9,-1)--(10,0);
		\draw[blue, thick] (11,-1)--(12,0);
		\draw[blue, thick] (10,-2)--(11,-1);
		\draw[blue, thick] (10,-2)--(11,-3);
		
		\node[black] at (12, 0.5) {$\sigma$};
		\node[black] at (10, 0.5) {$\sigma$};
		\node[black] at (8, 0.5) {$\sigma$};
		
		\node[black] at (9.8, -2.6) {$\sigma$};
		
		\node[black] at (9, -1.6) {$j$};
	\end{tikzpicture}
	\caption{\centering
		The fusion matrix $\hat{\mathcal{F}}_{\sigma}^{\sigma\sigma\sigma}$ implements the transformation from the $s$-channel to the $t$-channel, that is, between the two fusion orders of three $\sigma$-anyons. The indices $i,j\in\{\mathbbm{1},\psi\}$ label the intermediate fusion channels.
	}
	\label{fig:fusion_matrix_def}
\end{figure}
For $0<z<1$, choosing the positive branches of the square roots, the $s$- and $t$-channel conformal blocks can be expressed in terms of one another. The transition matrix is
\begin{equation}\label{eq:ising_fusion_matrix}
	\hat{\mathcal{F}}_{\sigma}^{\sigma\sigma\sigma}
	=
	\frac{1}{\sqrt{2}}
	\begin{pmatrix}
		1 & 1 \\
		1 & -1
	\end{pmatrix}.
\end{equation}
In the convention adopted here, the conformal blocks are related by
\begin{equation}
	\begin{pmatrix}
		\mathcal{F}^{(s)}_{\mathbbm{1}} \\
		\mathcal{F}^{(s)}_{\psi}
	\end{pmatrix}
	=
	\hat{\mathcal{F}}_{\sigma}^{\sigma\sigma\sigma}
	\begin{pmatrix}
		\mathcal{F}^{(t)}_{\mathbbm{1}} \\
		\mathcal{F}^{(t)}_{\psi}
	\end{pmatrix}.
\end{equation}
Thus, the matrix $\hat{\mathcal{F}}_{\sigma}^{\sigma\sigma\sigma}$ is the transition matrix between the bases defined by the two fusion orders of three $\sigma$-anyons. The rows and columns of this matrix are labeled by the intermediate channels $\mathbbm{1}$ and $\psi$. This change of basis is shown graphically in Fig.~\ref{fig:fusion_matrix_def}.

It is precisely the fusion rules, the matrix $\hat{\mathcal{F}}_{\sigma}^{\sigma\sigma\sigma}$, and the $R$-matrices that will be compared below with tensor-product decompositions, Racah matrices, and $R$-matrices of the quantum algebra $U_q(\mathfrak{sl}_2)$ at $q=e^{i\pi/4}$.

\subsection{$SU(2)_2$ WZW-conformal field theory}
In the previous section, we considered the Ising minimal model, a conformal
field theory whose symmetry is governed by the Virasoro algebra. However,
two-dimensional conformal field theories may possess richer symmetry
structures. An important example is provided by the Wess--Zumino--Witten (WZW)
models. These are two-dimensional nonlinear sigma models whose target space is
a compact Lie group $G$ (principal chiral models), supplemented by the
Wess--Zumino term, which first arose in the study of anomalous Ward
identities~\cite{WessZumino1971}. Witten showed~\cite{Witten1984} that for a
specific value of the coupling constant $\Lambda$ of the sigma model, fixed by
the integer coefficient $k$ of the Wess--Zumino term, the model becomes
conformally invariant. At this conformal value of the coupling constant,
holomorphic currents $J^a(z)$ exist in addition to the energy--momentum tensor,
and these currents generate an affine Lie algebra $\widehat{\mathfrak g}_k$; the
corresponding conformal field theory was constructed by Knizhnik and
Zamolodchikov~\cite{KZ1984}.

In the present work, we will be interested in the case $G=SU(2)$, where the
field
\begin{equation}
	g:\Sigma\longrightarrow SU(2)
\end{equation}
takes values in the group $SU(2)$, while $\Sigma$ is a two-dimensional surface.

The action of the $SU(2)$ WZW model consists of two parts. The first part is
the action of the principal chiral model,
\begin{equation}
	S_{\text{CM}}[g]=
	\frac{1}{4\Lambda^{2}}\int_\Sigma\operatorname{Tr}\bigl(\partial_\mu g^{-1}\,
\partial^\mu g\bigr)\,d^2x ,
\end{equation}
with coupling constant $\Lambda$. The second part,
the Wess--Zumino term, requires extending the field $g$ from the surface
$\Sigma$ to a three-dimensional manifold $B$ whose boundary is
$\Sigma=\partial B$. At the conformal value of the coupling constant, which in
the normalization adopted below corresponds to
\begin{equation}
	\frac{1}{4\Lambda^{2}}=\frac{k}{8\pi},
\end{equation}
the action takes the form
\begin{equation}
S_{\mathrm{WZW}}[g]
=
\frac{k}{8\pi}
\int_{\Sigma}
\operatorname{Tr}\bigl(\partial_\mu g^{-1}\,\partial^\mu g\bigr)\,d^2x
+
\frac{k}{12\pi}
\int_B
\operatorname{Tr}\bigl(g^{-1}dg\bigr)^3.
\end{equation}

Under a change of the extension of the field into $B$, the action changes by an integer multiple of $2\pi k$, where $k\in\mathbb{Z}$. Therefore, for $k\in\mathbb{Z}$, the quantum theory is independent of the choice of extension. This integer parameter is called the level of the WZW model~\cite{Witten1984}. In what follows, we denote the corresponding conformal field theory by $SU(2)_k$ WZW.

As already mentioned, the WZW model at the conformal point possesses holomorphic currents $J^a(z)$, which generate the affine algebra $\widehat{\mathfrak{su}}(2)_k$~\cite{KZ1984}. Therefore, the primary fields of the WZW model are classified not by ordinary representations of $\mathfrak{su}(2)$, but by integrable representations of the affine algebra~\cite{GepnerWitten1986}. For a positive integer level $k$, these representations are labeled by spins
\begin{equation}
j=0,\frac12,1,\ldots,\frac{k}{2}.
\end{equation}
Thus, for a fixed value of $k$, the theory contains a finite number of primary chiral sectors. The central charge of the WZW model is determined by the Sugawara construction~\cite{KZ1984}:
\begin{equation}
c
=
\frac{3k}{k+2}.
\end{equation}
The conformal weight of a primary field with spin $j$ is
\begin{equation}
h_j
=
\frac{j(j+1)}{k+2}.
\end{equation}
The fusion rules are a truncated version of the ordinary angular-momentum addition rules in $SU(2)$~\cite{GepnerWitten1986,Verlinde1988}:
\begin{equation}
j_1\times j_2
=
\bigoplus_{j=|j_1-j_2|}^{\min(j_1+j_2,\;k-j_1-j_2)}
j,
\end{equation}
where the summation proceeds in steps of $1$, and the allowed values of $j$ satisfy the condition
\begin{equation}
j-j_1-j_2\in\mathbb Z.
\end{equation}
The upper bound
\begin{equation}
j\leq k-j_1-j_2
\end{equation}
is absent in the ordinary representation theory of $SU(2)$. It is a consequence of the finite level $k$ and makes the set of allowed fusion channels finite. The quantum dimension of the sector with spin $j$ \cite{Witten1989} is
\begin{equation}
d_j
=
\frac{
	\sin\left(\dfrac{(2j+1)\pi}{k+2}\right)
}{
	\sin\left(\dfrac{\pi}{k+2}\right)
}.
\end{equation}
This quantity will be particularly important below when comparing the WZW model with representations of the quantum algebra $U_q(\mathfrak{sl}_2)$. 

In the present work we need the case $k=2$. Then exactly three spin values are allowed:
\begin{equation}
j=0,\qquad
j=\frac12,\qquad
j=1.
\end{equation}
Their main characteristics are
\begin{equation}
c_{SU(2)_2}=\frac32,
\end{equation}
\begin{equation}
h_0=0,\qquad
h_{1/2}=\frac{3}{16},\qquad
h_1=\frac12,
\end{equation}
\begin{equation}
d_0=1,\qquad
d_{1/2}=\sqrt2,\qquad
d_1=1.
\end{equation}
The fusion rules take the form
\begin{equation}
\frac12\times\frac12=0+1,
\qquad
\frac12\times1=\frac12,
\qquad
1\times1=0.
\end{equation}
Therefore, at the level of the fusion rules and quantum dimensions, one obtains the correspondence
\begin{equation}
\mathbbm{1}\longleftrightarrow j=0,
\qquad
\sigma\longleftrightarrow j=\frac12,
\qquad
\psi\longleftrightarrow j=1.
\end{equation}
In particular,
\begin{equation}
\frac12\times\frac12=0+1
\end{equation}
corresponds exactly to the Ising fusion rule
\begin{equation}
\sigma\times\sigma=\mathbbm{1}+\psi.
\end{equation}
The only nontrivial transition matrix between the two fusion orders of three fields with spin $j=1/2$ can also be chosen in the form \eqref{eq:ising_fusion_matrix}.

In the WZW model, braiding matrices arise as monodromies of chiral conformal blocks~\cite{KZ1984,MooreSeiberg1989}. Under analytic continuation of the conformal blocks along a path that exchanges two insertion points, the blocks transform linearly. This transformation gives the action of the generators of the braid group. In the basis where the two fields with spin $j=1/2$ are fused first, the corresponding braiding operator is diagonal, since it preserves the intermediate channel $j=0$ or $j=1$. Its eigenvalues are determined by the exponents of the operator product expansion, as in~\eqref{eq:ising_R_symbols} for the Ising model, but with one additional ingredient. Unlike the Ising field $\sigma$, the WZW primary field of spin $1/2$ carries an $\mathfrak{su}(2)$ index, and the exchange of two such fields permutes their indices as well as their positions. In the formalism of Moore and Seiberg~\cite{MooreSeiberg1989}, the braiding eigenvalue in the channel $j$ is therefore the product of the half-monodromy phase $e^{\pm i\pi(h_j-h_{j_1}-h_{j_2})}$ and the eigenvalue $\pm1$ of the permutation of the two legs of the three-point coupling (Eqs.~(2.13), (3.8),
(4.2) and footnote~7 therein). For the coupling $\tfrac12\otimes\tfrac12\to j$
this eigenvalue equals $(-1)^{1-j}$: $+1$ for the symmetric triplet and $-1$
for the antisymmetric singlet. The same sign appears in the quantum-group
$R$-matrix, see Theorem~3.1 of~\cite{KirillovReshetikhin1989}. Hence
\begin{equation}\label{eq:WZW_R_symbols}
	R^{\,j}_{\frac12\times\frac12}
	=(-1)^{\,1-j}\,e^{\,i\pi\left(h_j-2h_{1/2}\right)},
	\qquad j=0,1,
\end{equation}
and for $k=2$, with $h_{1/2}=3/16$, $h_0=0$, $h_1=1/2$, the braiding matrix
in the basis $(j=0,\,j=1)$ takes the form
\begin{equation}
	R_{\frac{1}{2}\times\frac{1}{2}}
	=
	\begin{pmatrix}
		e^{5\pi i/8} & 0 \\
		0 & e^{\pi i/8}
	\end{pmatrix}.
\end{equation}
The braiding operator in the opposite direction is given by the inverse matrix
\begin{equation}
	\left(R_{\frac{1}{2}\times\frac{1}{2}}\right)^{-1}
	=
	\begin{pmatrix}
		e^{-5\pi i/8} & 0 \\
		0 & e^{-\pi i/8}
	\end{pmatrix}.
\end{equation}
The braiding matrix of Ising anyons introduced in the previous section is related to the WZW braiding matrix by
\begin{equation}\label{eq:R_ising_R_WZW}
	[R_{\sigma\sigma}]_\text{ising}
	=
	e^{i\pi/2}\left[R_{\frac{1}{2}\times\frac{1}{2}}\right]^{-1}_{SU(2)_2}.
\end{equation}
This relation can be read off from~\eqref{eq:WZW_R_symbols}: since $h_1=h_\psi$ and $h_0=h_{\mathbbm 1}$, the half-monodromy phases of the two theories differ only by the overall factor $e^{-2\pi i(h_{1/2}-h_\sigma)}=e^{-i\pi/4}$, and it is the additional sign in the singlet channel that reverses the relative phase of the two channels and thereby turns this overall factor into the inverse in~\eqref{eq:R_ising_R_WZW}. Therefore, after choosing the braiding orientation, the matrices in the two theories differ only by an overall phase.

Nevertheless, the $SU(2)_2$ WZW model and the Ising minimal model $\mathcal{M}(4,3)$ are not the same conformal field theory. Their central charges are different:
\begin{equation}
c_{SU(2)_2}=\frac32,
\qquad
c_{\mathcal \mathcal{M}(4,3)}=\frac12.
\end{equation}
In addition, the conformal weight of the $j=1/2$ field does not coincide with the conformal weight of the $\sigma$-anyon
\begin{equation}
[h_{1/2}]_{SU(2)_2}=\frac{3}{16},
\qquad
[h_\sigma]_{\mathcal{M}(4,3)}=\frac1{16}.
\end{equation}
Therefore, the absolute braiding phases in the two theories are different. After choosing the braiding orientation and multiplying each braid-group generator by a common phase, their actions on the fusion spaces agree up to an overall phase factor.

\subsection{The coset theory $SU(2)_2/U(1)$ and the Ising model}\label{subsec:coset}
A more precise relation between these theories is formulated in terms of the coset theory$SU(2)_2/U(1)$ \cite{GKO1985,Nayak2008}. The $U(1)$ sector has central charge $1$, and therefore the central charge of the coset theory is
\begin{equation}
	c_{\mathrm{coset}}
	=
	c_{SU(2)_2}-c_{U(1)}
	=
	\frac{3}{2}-1
	=
	\frac{1}{2}.
\end{equation}
Upon passing to the coset theory, the contribution of the corresponding $U(1)$ sector is subtracted from the conformal weight of the WZW field:
\begin{equation}
	[h_{j,m}]_{\mathrm{coset}}
	=
	[h_j]_{SU(2)_2}
	-
	[h_m]_{U(1)},
\end{equation}
where
\begin{equation}
	[h_j]_{SU(2)_2}
	=
	\frac{j(j+1)}{4},
	\qquad
	[h_m]_{U(1)}
	=
	\frac{m^2}{2}.
\end{equation}
At $k=2$, the three primary fields of the Ising model correspond to the following components of the WZW fields:
\begin{equation}
	\begin{aligned}
		1
		&\longleftrightarrow
		(j,l)=(0,0),
		\\
		\sigma
		&\longleftrightarrow
		\left(j,l\right)
		=
		\left(\frac{1}{2},\pm\frac{1}{2}\right),
		\\
		\psi
		&\longleftrightarrow
		(j,l)=(1,0).
	\end{aligned}
\end{equation}
Both signs $l=\pm 1/2$ correspond to the same Ising sector $\sigma$. Therefore, the conformal weights in the coset theory are
\begin{equation}
	\begin{aligned}
		h_1
		&=
		0,
		\\
		h_\sigma
		&=
		\frac{3}{16}-\frac{1}{8}
		=
		\frac{1}{16},
		\\
		h_\psi
		&=
		\frac{1}{2}-0
		=
		\frac{1}{2}.
	\end{aligned}
\end{equation}
Thus, among the chosen representatives, only the conformal weight of the $\sigma$ field changes, since it is the only one carrying a nonzero $U(1)$ charge. The fields $1$ and $\psi$ are represented by components with $l=0$, and therefore their conformal weights remain unchanged upon passing to the coset theory.

It is in this sense that the Ising minimal model is identified with the $SU(2)_2/U(1)$ coset theory, rather than with the full $SU(2)_2$ WZW model.

\subsection{$SU(2)_k$ Chern-Simons theory}\label{subsec:ChSbasics}
Chern--Simons theory occupies a special place among three-dimensional gauge theories. Its original geometric form was introduced by S. S. Chern and J. Simons in the study of secondary characteristic classes of fiber bundles~\cite{ChernSimons1974}. This construction was later interpreted as a quantum field theory in $2+1$ dimensions~\cite{DJT1982,Witten1989}. Unlike conventional gauge theories, the Chern--Simons action does not require a choice of metric on spacetime. As a result, after quantization the theory is naturally related to topological invariants of three-dimensional manifolds, knots, and links~\cite{Witten1989,RT1991}.

This feature makes Chern--Simons theory particularly important for topological quantum computation~\cite{KolganovMorozov2020,KolganovMironovMorozov2023}. In this interpretation, Wilson lines describe the worldlines of quasiparticles, while their braiding in spacetime corresponds to operations on the states of an anyonic system~\cite{Nayak2008,KolganovMorozov2020}. Since the result depends on the topological class of the braiding rather than on its detailed geometric shape, small local deformations of the trajectories should not affect the computational outcome. The braiding matrices and state spaces associated with Wilson lines thus serve as the basic building blocks of topological quantum algorithms.

To define Chern--Simons theory, let us consider an oriented three-dimensional manifold $M$ and a gauge field
\begin{equation}
A\in\Omega^1(M,\mathfrak{su}(2)).
\end{equation}
In other words, $A$ is an $\mathfrak{su}(2)$-valued differential $1$-form, that is, a connection with gauge group $SU(2)$. In the present work, we will focus specifically on this gauge group. The action of $SU(2)$ Chern--Simons theory is given by
\begin{equation}
S_{\mathrm{CS}}[A]
=
\frac{k}{4\pi}
\int_M
\operatorname{Tr}
\left(
A\wedge dA
+
\frac23 A\wedge A\wedge A
\right).
\end{equation}
Here, $d$ denotes the exterior derivative, $\wedge$ is the exterior product of differential forms, and $\operatorname{Tr}$ is an invariant bilinear form on $\mathfrak{su}(2)$, which in what follows will be written as the trace in the adjoint representation.

The action does not involve a metric on $M$. Therefore, the classical theory depends on the topological structure of the manifold rather than on distances or angles defined on it. The parameter $k$ is called the level of the Chern--Simons theory. Under gauge transformations that are not homotopic to the identity, the action may change by an integer multiple of $2\pi$. Therefore, the quantum theory is well defined when $k \in \mathbb{Z}$. In what follows, we will denote this theory by $SU(2)_k$ Chern--Simons theory.

\begin{figure}[h!]
	\centering
	\begin{tikzpicture}[
		line cap=round,
		line join=round,
		wilson/.style={
			draw=blue,
			line width=1.15pt
		},
		overpass/.style={
			draw=blue,
			line width=1.15pt,
			preaction={
				draw=white,
				line width=4pt,
				line cap=butt
			}
		},
		puncture/.style={
			circle,
			fill=blue!65!black,
			inner sep=2.2pt
		}
		]
		
		\coordinate (R1) at (-2.40,0);
		\coordinate (R2) at (-0.80,0);
		\coordinate (R3) at ( 0.80,0);
		\coordinate (R4) at ( 2.40,0);
		
		
		\draw[wilson]
		(R1)
		.. controls (-1.95,-0.70) and (-2.85,-1.00) ..
		(-2.85,-1.55)
		.. controls (-2.85,-2.10) and (-2.30,-2.20) ..
		(-2.30,-2.75);
		
		\draw[wilson]
		(R4)
		.. controls (1.95,-0.70) and (2.85,-1.00) ..
		(2.85,-1.55)
		.. controls (2.85,-2.10) and (2.30,-2.20) ..
		(2.30,-2.75);
		
		\draw[wilson]
		(R2)
		.. controls (-0.70,-0.20) and (-0.62,-0.50) ..
		(-0.48,-0.65)
		.. controls (-0.32,-0.78) and (-0.16,-0.93) ..
		(0,-1.05)
		.. controls (0.16,-1.16) and (0.32,-1.28) ..
		(0.48,-1.38)
		.. controls (0.70,-1.75) and (0.80,-2.45) ..
		(0.80,-2.85);
		
		\draw[wilson]
		(R3)
		.. controls (0.70,-0.20) and (0.62,-0.50) ..
		(0.48,-0.65)
		.. controls (0.32,-0.78) and (0.16,-0.93) ..
		(0,-1.05)
		.. controls (-0.16,-1.16) and (-0.32,-1.28) ..
		(-0.48,-1.38)
		.. controls (-0.70,-1.75) and (-0.80,-2.45) ..
		(-0.80,-2.85);
		
		\draw[overpass]
		(-0.48,-0.65)
		.. controls (-0.32,-0.78) and (-0.16,-0.93) ..
		(0,-1.05)
		.. controls (0.16,-1.16) and (0.32,-1.28) ..
		(0.48,-1.38);
		
		\draw[wilson]
		(-2.30,-2.75)
		.. controls (-2.30,-3.70) and (-0.80,-3.75) ..
		(-0.80,-2.85);
		
		\draw[wilson]
		(0.80,-2.85)
		.. controls (0.80,-3.75) and (2.30,-3.70) ..
		(2.30,-2.75);
		
		
		\fill[blue!8]
		(-4.30,-0.48)
		--
		(3.70,-0.48)
		--
		(4.30,0.48)
		--
		(-3.70,0.48)
		--
		cycle;
		
		\draw[black,line width=0.7pt]
		(-4.30,-0.48)
		--
		(3.70,-0.48)
		--
		(4.30,0.48)
		--
		(-3.70,0.48)
		--
		cycle;
		
		
		\draw[wilson]
		(R1)
		.. controls (-2.85,0.70) and (-2.85,1.00) ..
		(-2.85,1.55)
		.. controls (-2.85,2.10) and (-2.30,2.20) ..
		(-2.30,2.75);
		
		\draw[wilson]
		(R4)
		.. controls (2.85,0.70) and (2.85,1.00) ..
		(2.85,1.55)
		.. controls (2.85,2.10) and (2.30,2.20) ..
		(2.30,2.75);
		
		\draw[wilson]
		(R2)
		.. controls (-0.90,0.20) and (-0.65,0.42) ..
		(-0.48,0.55)
		.. controls (-0.32,0.65) and (-0.16,0.75) ..
		(0,0.85)
		.. controls (0.16,0.95) and (0.32,1.04) ..
		(0.48,1.12)
		.. controls (0.65,1.22) and (0.85,1.25) ..
		(0.85,1.42)
		.. controls (0.85,1.58) and (0.65,1.65) ..
		(0.48,1.72)
		.. controls (0.32,1.82) and (0.16,1.92) ..
		(0,2.00)
		.. controls (-0.16,2.10) and (-0.32,2.20) ..
		(-0.48,2.28)
		.. controls (-0.65,2.38) and (-0.80,2.60) ..
		(-0.80,2.85);
		
		\draw[wilson]
		(R3)
		.. controls (0.90,0.20) and (0.65,0.42) ..
		(0.48,0.55)
		.. controls (0.32,0.65) and (0.16,0.75) ..
		(0,0.85)
		.. controls (-0.16,0.95) and (-0.32,1.04) ..
		(-0.48,1.12)
		.. controls (-0.65,1.22) and (-0.85,1.25) ..
		(-0.85,1.42)
		.. controls (-0.85,1.58) and (-0.65,1.65) ..
		(-0.48,1.72)
		.. controls (-0.32,1.82) and (-0.16,1.92) ..
		(0,2.00)
		.. controls (0.16,2.10) and (0.32,2.20) ..
		(0.48,2.28)
		.. controls (0.65,2.38) and (0.80,2.60) ..
		(0.80,2.85);
		
		\draw[overpass]
		(0.48,0.55)
		.. controls (0.32,0.65) and (0.16,0.75) ..
		(0,0.85)
		.. controls (-0.16,0.95) and (-0.32,1.04) ..
		(-0.48,1.12);
		
		\draw[overpass]
		(0.48,1.72)
		.. controls (0.32,1.82) and (0.16,1.92) ..
		(0,2.00)
		.. controls (-0.16,2.10) and (-0.32,2.20) ..
		(-0.48,2.28);
		
		\draw[wilson]
		(-2.30,2.75)
		.. controls (-2.30,3.70) and (-0.80,3.75) ..
		(-0.80,2.85);
		
		\draw[wilson]
		(0.80,2.85)
		.. controls (0.80,3.75) and (2.30,3.70) ..
		(2.30,2.75);
		
		
		\node[puncture] at (R1) {};
		\node[puncture] at (R2) {};
		\node[puncture] at (R3) {};
		\node[puncture] at (R4) {};
		
		\node[below left=1pt]  at (R1) {$\mathfrak{R}_1$};
		\node[below left=1pt]  at (R2) {$\mathfrak{R}_2$};
		\node[below right=1pt] at (R3) {$\mathfrak{R}_3$};
		\node[below right=1pt] at (R4) {$\mathfrak{R}_4$};
		
		\node at (-3.7,0.15) {$\Sigma$};
		
		\draw[
		-{Stealth[length=3mm]},
		line width=0.8pt
		]
		(4.65,-3.30)
		--
		(4.65,3.40)
		node[above] {$\mathbb{R}$};
		
	\end{tikzpicture}
	\caption{\centering A closed Wilson line and its intersection with the spatial surface $\Sigma$. The plat closure shown forms a trefoil knot. The intersection points are punctures of the surface labeled by the representations $\mathfrak{R}_1,\ldots,\mathfrak{R}_4$.}
	\label{fig:wilson_trefoil_slice}
	
\end{figure}

The main topological observables in this theory are Wilson loops~\cite{Witten1989}. Let $\mathcal{K}\subset M$ be a closed contour, and let $\mathfrak{R}$ be a representation of the group $SU(2)$ labeling this contour. The Wilson loop operator is defined by
\begin{equation}
W_\mathfrak{R}(\mathcal{K})
=
\operatorname{Tr}_\mathfrak{R}
\mathcal P
\exp\left(\oint_\mathcal{K} A\right),
\end{equation}
where $\mathcal P$ denotes path ordering along the contour. Its expectation value is given by the functional integral

\begin{equation}\label{eq:Wilson_loop_expect}
\langle W_\mathfrak{R}(\mathcal{K})\rangle
=
\frac{1}{\mathcal{Z}_{\text{CS}}}
\int\mathcal DA\,
e^{iS_{\mathrm{CS}}[A]}\,
W_\mathfrak{R}(\mathcal{K}).
\end{equation}

Direct computation of the functional integral for a Wilson loop expectation value is, in general, a difficult problem. One way to obtain a more convenient description is to choose a spatial slice of the three-dimensional spacetime~\cite{Witten1989}. Let
\begin{equation}
M=\Sigma\times\mathbb R,
\end{equation}
where $\Sigma$ is a spatial surface, while $\mathbb{R}$ is interpreted as time. If Wilson lines intersect $\Sigma$ (Fig.~\ref{fig:wilson_trefoil_slice}), the intersection points become punctures of the surface labeled by representations
\begin{equation}
\mathfrak{R}_1,\ldots,\mathfrak{R}_n.
\end{equation}
Under canonical quantization, the state space of Chern--Simons theory on such a punctured surface is identified with the space of chiral conformal blocks of the $SU(2)_k$ WZW model~\cite{Witten1989,LabastidaRamallo1989}:
\begin{equation}
\mathcal H_{\mathrm{CS}}
\bigl(\Sigma;\mathfrak{R}_1,\ldots,\mathfrak{R}_n\bigr)
\simeq
\operatorname{CB}_{SU(2)_k}
\bigl(\Sigma;\mathfrak{R}_1,\ldots,\mathfrak{R}_n\bigr).
\end{equation}
Thus, the spatial slice does not compute the Wilson loop expectation value directly, but instead reformulates the problem in the language of two-dimensional conformal field theory. Braiding of Wilson lines acts on the space of conformal blocks, and in the WZW description this action is represented by braiding matrices and transition matrices between different fusion orders, as discussed in the previous sections.

On the other hand, Wilson loop expectation values are related to invariants of knot theory. Witten's result~\cite{Witten1989}, which was given a rigorous mathematical formulation in the Reshetikhin--Turaev construction~\cite{RT1990,RT1991}, can be stated as follows.

For $SU(2)_k$ Chern--Simons theory on $S^3$, the expectation value of a Wilson loop labeled by the fundamental representation $j=1/2$ coincides, up to normalization and an overall phase, with the Jones polynomial of the corresponding knot evaluated at a root of unity:

\begin{equation}
\frac{
	\bigl\langle W_{1/2}[\mathcal{K}]\bigr\rangle
}{
	\bigl\langle W_{1/2}[\text{unknot}]\bigr\rangle
}
=
\mathcal{J}[\mathcal{K}](t),
\qquad
t=
\exp\left(
\frac{2\pi i}{k+2}
\right).
\end{equation}
Here, 'unknot' denotes the trivial knot. For links and for arbitrary representations labeling their components, the corresponding colored Jones polynomial is obtained.

The Jones polynomial can be computed using skein relations~\cite{Jones1985}. With the normalization
\begin{equation}
\mathcal{J}[{\text{unknot}}](t)=1
\end{equation}
the skein relation takes the form
\begin{equation}
	t^{-1}
	\mathcal{J}\!\left[\jonesplus\right](t)
	-
	t\,
	\mathcal{J}\!\left[\jonesminus\right](t)
	=
	\left(
	t^{1/2}-t^{-1/2}
	\right)
	\mathcal{J}\!\left[\;\joneszero\;\right](t).
	\label{eq:jones_skein_relation}
\end{equation}
Skein relations allow the computation of the Jones polynomial of a complicated knot to be reduced recursively to the computation of polynomials for simpler links. However, for knots represented by long braids, this approach quickly becomes cumbersome. For knots given as braid closures, a matrix-based approach is also convenient, in which operators of braiding are assigned to strand crossings~\cite{MorozovSmirnov2010}.

A more systematic method for computing knot invariants is based on the quantum algebra $U_q(\mathfrak{sl}_2)$~\cite{RT1990,BishlerEtAl2018}. It is a $q$-deformation of the universal enveloping algebra $U(\mathfrak{sl}_2)$~\cite{Jimbo1985}, and in the limit $q\to1$ it reduces to the ordinary universal enveloping algebra. In the standard $\hbar$-notation, the algebra $U_q(\mathfrak{sl}_2)$ is generated by the elements $E$, $F$, $K$, and $K^{-1}$.

In addition to multiplication, the quantum algebra is equipped with a coproduct
\begin{equation}
\Delta:
U_q(\mathfrak{sl}_2)
\longrightarrow
U_q(\mathfrak{sl}_2)\otimes U_q(\mathfrak{sl}_2).
\end{equation}
It defines the action of the generators on a tensor product of representations, that is, on a system of several strands. The coproduct is chosen as a $q$-deformation of the usual rule for the action of a Lie algebra on a tensor product and must be compatible with multiplication in $U_q(\mathfrak{sl}_2)$. In the conventions used here, the deformation parameter is related to the Chern--Simons level and to the parameter of the Jones polynomial by
\begin{equation}
	q
	=
	\exp\left(
	\frac{\pi i}{k+2}
	\right),
	\qquad
	t=q^2.
	\label{eq:q_t_cs_relation}
\end{equation}
In particular, for $k=2$,
\begin{equation}
	q=e^{\pi i/4},
	\qquad
	q^8=1.
	\label{eq:q_eighth_root}
\end{equation}
For this reason, in what follows we will have to consider representations of the quantum algebra $U_q(\mathfrak{sl}_2)$ at roots of unity~\cite{Arnaudon1994,BishlerMironovMorozov2023}.

For a generic value of $q$, the quantum algebra $U_q(\mathfrak{sl}_2)$ possesses a universal $R$-matrix $\mathcal R$ satisfying the relation
\begin{equation}
	\mathcal R\,\Delta(x)
	=
	\Delta^{\mathrm{op}}(x)\,\mathcal R,
	\qquad
	x\in U_q(\mathfrak{sl}_2),
	\label{eq:universal_R_intertwining}
\end{equation}
where $\Delta^{\mathrm{op}}$ denotes the opposite coproduct, in which the tensor factors are interchanged.
\begin{figure}[h!]
	\centering
	\subfloat[$5_2$ knot]{
		\makebox[0.30\textwidth][c]{
			\input{pictures/5_1_knot.tex}
		}\label{fig:4_1_3D_representation}}
	\hfill
	\subfloat[$5_2$ knot as a braid closure]{
		\makebox[0.30\textwidth][c]{
			\input{pictures/4_1_braid.tex}
		}\label{fig:4_1_closure_1}}
	\hfill
	\subfloat[$5_2$ knot as a plat closure ]{
		\makebox[0.30\textwidth][c]{
			\input{pictures/4_1_morse_link.tex}
		}\label{fig:4_1_closure_2}}
	\caption{The $5_2$ knot from the Rolfsen table in different representations.}
	\label{fig:4_1_representations}
\end{figure}
A more detailed definition of $U_q(\mathfrak{sl}_2)$, its coproduct, and highest-weight representations at roots of unity will be given in section~\ref{sec:HWrepUq}. At this stage, we only need the fact that the $R$-matrix defines a braiding operator. Let $\mathfrak{R}_1$ and $\mathfrak{R}_2$ be two representations of $U_q(\mathfrak{sl}_2)$. Then the operator corresponding to the geometric braiding of two strands is defined by
\begin{equation}
	\check{R}
	=
	P\circ\mathcal{R}
	\qquad:\qquad
	\mathfrak{R}_1\otimes \mathfrak{R}_2
	\longrightarrow
	\mathfrak{R}_2\otimes \mathfrak{R}_1.
	\label{eq:braiding_from_universal_R}
\end{equation}
The permutation operator $P$ is necessary because, under geometric braiding, the strands not only interact but also exchange their positions. If all strands are labeled by the same representation $\mathfrak{R}$, then the generators of the braid group $B_n$ are represented by the operator
\begin{equation}
	\check R_i
	=
	\operatorname{id}^{\otimes(i-1)}
	\otimes
	\check R
	\otimes
	\operatorname{id}^{\otimes(n-i-1)}.
	\label{eq:braid_generator_representation}
\end{equation}
The negative crossing corresponds to the inverse operator $\check R_i^{-1}$.

To illustrate how braiding operators are used in the computation of Wilson loop expectation values, let us consider a specific example. Suppose that the Wilson loop contour forms the knot $5_2$ from the Rolfsen table (Fig.~\ref{fig:4_1_representations}). This knot can be represented both as the ordinary closure of a three-strand braid and as the plat closure of a four-strand braid.

Let each strand carry the fundamental representation of the algebra $U_q(\mathfrak{sl}_2)$. Then the two braid words correspond to the operators
\begin{equation}
	\check{B}_{\mathrm{br}}
	=
	\check R_1^{3}
	\check R_2
	\check R_1^{-1}
	\check R_2,
\end{equation}
and
\begin{equation}
	\check{B}_{\mathrm{pl}}
	=
	\check R_2^{-2}
	\check R_3^{2}
	\check R_2^{-1}.
\end{equation}
For the ordinary braid closure, the Wilson loop expectation value is expressed in terms of the quantum trace~\cite{RT1990,MorozovSmirnov2010}:
\begin{equation}
	\left\langle W_\mathfrak{R}(5_2)\right\rangle
	\propto
	\operatorname{Tr}_q\!\left(\check{B}_{\mathrm{br}}\right).
\end{equation}
In the plat representation, the same expectation value is written as a matrix element between vacuum states~\cite{KolganovMorozov2020,KolganovMironovMorozov2023}:
\begin{equation}
	\left\langle W_\mathfrak{R}(5_2)\right\rangle
	\propto
	\left\langle \mathrm{\varnothing\otimes\varnothing}\right|
	\check{B}_{\mathrm{pl}}
	\left|\mathrm{\varnothing\otimes\varnothing}\right\rangle.
\end{equation}
Here, $\left|\mathrm{\varnothing\otimes\varnothing}\right\rangle$ denotes the state in which the strand pairs $(1,2)$ and $(3,4)$ fuse into the trivial representation. After matching the normalization conventions, both constructions yield the same invariant of the knot $5_2$.

\section{Highest-weight representations of the quantum group $U_q(\mathfrak{sl}_2)$ at roots of unity}\label{sec:HWrepUq}

In the previous section, we showed that the computation of Wilson loop expectation values in $SU(2)_k$ Chern--Simons theory can be reduced to calculations in the quantum algebra $U_q(\mathfrak{sl}_2)$. In the conventions used here, the deformation parameter is related to the level $k$ by
\begin{equation}
	q=\exp\left(\frac{\pi i}{k+2}\right).
\end{equation}
Therefore, for the $SU(2)_2$ theory, which is related to Ising anyons, one must consider the value
\begin{equation}
	q=e^{\pi i/4},
	\qquad
	q^8=1.
\end{equation}
Thus, in what follows we will deal with representations of the quantum algebra at a root of unity.

The quantum algebra $U_q(\mathfrak{sl}_2)$ is a $q$-deformation of the universal enveloping algebra $U(\mathfrak{sl}_2)$. In the limit $q\to1$, its defining relations reduce to the ordinary commutation relations of the algebra $\mathfrak{sl}_2$. In the standard $K$-notation, it is generated by the elements $E$, $F$, $K$, and $K^{-1}$, where $E$ and $F$ play the role of raising and lowering operators, while $K$ plays the role of the exponential of the weight operator. The defining relations are
\begin{equation}
	KK^{-1}=K^{-1}K=1,
\end{equation}
\begin{equation}
	\begin{aligned}
		KEK^{-1}&=q^2E, \qquad
		KFK^{-1}&=q^{-2}F,
	\end{aligned}
\end{equation}
\begin{equation}
	[E,F]
	=
	\frac{K-K^{-1}}{q-q^{-1}}.
\end{equation}
In what follows, we will also use the $q$-numbers
\begin{equation}
	[n]_q
	=
	\frac{q^n-q^{-n}}{q-q^{-1}}.
\end{equation}
In the limit $q\to1$, they reduce to the ordinary integers:
\begin{equation}
	[n]_q\longrightarrow n.
\end{equation}

It is sometimes convenient to use the operator $H$, which is related to $K$ by
\begin{equation}
	K=q^H=e^{\hbar H},
	\qquad
	q=e^\hbar.
\end{equation}
Then, formally, the defining relations can be written as

\begin{equation}
	\begin{aligned}
		[H,E]&=2E, \qquad
		[H,F]&=-2F, \qquad
		[E,F]
		&=
		\frac{e^{\hbar H}-e^{-\hbar H}}{e^\hbar-e^{-\hbar}}.
	\end{aligned}
\end{equation}

The algebra written in this formal $\hbar$-notation is often denoted by $U_\hbar(\mathfrak{sl}_2)$. However, when $q$ is specialized to a root of unity, the element $H$ is not globally determined by $K$ alone. Therefore, throughout this work the $\hbar$-notation will be used as the primary definition of the algebra.

To define the action of the quantum algebra on a tensor product of representations, one must specify a coproduct. In the conventions used here, it takes the form
\begin{equation}\label{eq:Uq_coproduct}
	\begin{aligned}
		\Delta(E)
		&=
		E\otimes K+\mathbf{1}\otimes E, \\
		\Delta(F)
		&=
		F\otimes\mathbf{1}+K^{-1}\otimes F, \\
		\Delta(K)
		&=
		K\otimes K, \\
		\Delta(K^{-1})
		&=
		K^{-1}\otimes K^{-1}.
	\end{aligned}
\end{equation}
If the $\hbar$-notation is used, then

\begin{equation}
	\Delta(H)
	=	
	H\otimes\mathbf{1}
	+
	\mathbf{1}\otimes H.
\end{equation}

The coproduct defines the action of the generators of the quantum algebra on the tensor product of two representations, that is, on a system of two strands. It must be compatible with the multiplication of elements of the algebra:
\begin{equation}
	\Delta(xy)
	=
	\Delta(x)\Delta(y),
	\qquad
	\Delta(1)
	=
	1\otimes 1.
\end{equation}
In particular, it follows that
\begin{equation}
	\Delta([x,y])
	=
	[\Delta(x),\Delta(y)].
\end{equation}
Therefore, the operators $\Delta(E)$, $\Delta(F)$, and $\Delta(K)$ satisfy the same defining relations as the original generators $E$, $F$, and $K$. In other words, the coproduct preserves the commutation relations and thereby defines a consistent action of $U_q(\mathfrak{sl}_2)$ on the tensor product of representations. The factors of $K$ appearing in the expressions for $\Delta(E)$ and $\Delta(F)$ are precisely what ensure this compatibility. In the limit $q\to1$, these factors disappear, and the coproduct reduces to the usual rule for the action of a generator on a tensor product.

The action on three or more strands is constructed by repeated application of the coproduct. The result must not depend on which two strands are combined first. This is guaranteed by the condition

\begin{equation}
	(\Delta\otimes\operatorname{id})\Delta(x)
	=
	(\operatorname{id}\otimes\Delta)\Delta(x).
\end{equation}

In the ordinary universal enveloping algebra, interchanging the two tensor factors in the coproduct does not change the total action of the generators. In the quantum algebra, however, this is generally no longer the case. To describe the action of the coproduct after exchanging the strands, we introduce the opposite coproduct $\Delta^{\mathrm{op}}(x)$, which differs from the coproduct \eqref{eq:Uq_coproduct} by the reversed order of the tensor factors. The difference between $\Delta$ and $\Delta^{\mathrm{op}}$ is compensated by the universal $R$-matrix $\mathcal{R}$:
\begin{equation}
	\mathcal{R}\Delta(x)
	=
	\Delta^{\mathrm{op}}(x)\mathcal{R},
	\qquad
	x\in U_q(\mathfrak{sl}_2),
	\label{eq:universal_R_intertwining_sec_2}
\end{equation}
which can be written in the following form:
\begin{equation}\label{eq:universal_R_matrix}
	\mathcal{R}
	=
	e^{\frac{\hbar}{2}(H\otimes H)}
	\sum_{n=0}^{\infty}
	\frac{
		q^{\frac{n(n+1)}{2}}
		\left(1-q^{-2}\right)^n
	}{
		[n]_q!
	}
	E^n\otimes F^n,
	\qquad
	q=e^\hbar.
\end{equation}
To obtain an operator that not only changes the state but also exchanges the strands themselves, one must additionally apply the permutation operator. Thus, the elementary braiding of two strands is represented by the operator
\begin{equation}
	\check{R}
	=
	P \circ \mathcal{R}.
	\label{eq:braiding_operator}
\end{equation}
If both strands carry the same representation $\mathfrak{R}$, then $\check{R}$ acts on $\mathfrak{R}\otimes\mathfrak{R}$ and commutes with the total action of the quantum algebra:
\begin{equation}
	\check{R}\Delta(x)
	=
	\Delta(x)\check{R}.
	\label{eq:R_commutes_with_coproduct}
\end{equation}
As before, the action of the $R$-matrix on the space $\mathfrak{R}^{\otimes n}$ is represented by the operator
\begin{equation}
	\check R_i
	=
	\operatorname{id}^{\otimes(i-1)}
	\otimes
	\check R
	\otimes
	\operatorname{id}^{\otimes(n-i-1)},
	\label{eq:braid_generator_representation_2}
\end{equation}
which exchanges the $i$-th and $(i+1)$-th strands. The universal $R$-matrix also ensures that the Yang--Baxter equation is satisfied. This equation coincides with the fundamental relation among the generators of the braid group. Therefore, the operators $\check{R}_i$, acting on neighboring tensor factors, make it possible to construct the matrix corresponding to an arbitrary braid and then use it to compute the invariant of the associated closed knot.

For values of $q$ that are not roots of unity, finite-dimensional highest-weight representations behave in much the same way as ordinary representations of $\mathfrak{sl}_2$, and their tensor products decompose into direct sums of irreducible representations. However, in the case of interest here,
\begin{equation}
	q=e^{i\pi/4},
	\qquad
	[4]_q=0,
\end{equation}
 in tensor products, invariant subspaces may arise that cannot be separated as independent direct summands. The corresponding representations are reducible but indecomposable.
 
 In what follows, we will need two types of representations that arise in
 tensor powers of the fundamental representation of $U_q(\mathfrak{sl}_2)$.
 Irreducible highest-weight representations will be referred to as
 representations of $\operatorname{spin}$ type; for generic $q$, they are
 quantum analogues of the ordinary spin representations of $\mathfrak{sl}_2$.
 Such reducible but indecomposable representations, which arise in tensor
 products of $\operatorname{spin}$-type representations, will be referred to
 as representations of $\operatorname{ind}$ type; their appearance is the main
 distinction between the case considered here and the ordinary semisimple
 representation theory. The explicit structure of both types, the action of
 the generators, and the tensor-product decomposition rules will be discussed
 in detail in the following subsections.

\subsection{Representations of spin type}\label{sec:spin_type}

\begin{wrapfigure}[28]{r}{0.4\textwidth}
	\vspace{-0.5cm}
	\centering
	\begin{tikzpicture}[
		-Stealth, 
		thick, 
		]
		
		\node[draw, circle, blue] (v0) at (0, 6) {$v_0$};
		\node[draw, circle, blue] (v1) at (0, 4) {$v_1$};
		\node[draw, circle, blue] (v2) at (0, 2) {$v_2$};
		\node[draw, circle, blue] (v3) at (0, 0) {$v_3$};
		
		\draw (v1) to[bend left=0] node[right, draw=none]{$Ev_1=[\gamma]_qv_0$} (v0);
		\draw (v0) to[bend right=30] node[left]{$Fv_0=v_1$} (v1);
		\draw (v2) to[bend left=0] node[right, draw=none]{$Ev_2=[2]_q[\gamma-1]_qv_1$} (v1);
		\draw (v1) to[bend right=30] node[left]{$Fv_1=v_2$} (v2);
		\draw (v3) to[bend left=0] node[right] {$Ev_3=[3]_q[\gamma-2]_qv_2$} (v2);
		\draw (v2) to[bend right=30] node[left]{$Fv_2=v_3$} (v3);
		\draw (v0) to[loop left] node[left] {$Hv_0 = \gamma v_0$} (v0);
		\draw (v1) to[loop right] node[right] {$Hv_1 = (\gamma-2) v_1$} (v1);
		\draw (v2) to[loop right] node[right] {$Hv_2 = (\gamma-4) v_2$} (v2);
		\draw (v3) to[loop right] node[right] {$Hv_3 = (\gamma-6) v_3$} (v3);

		\draw[Stealth-] (v3) -- +(0, -1);
		\draw[-](v3)..controls+(-0.4, -0.8)..+(-0.4, -1);
		\node[draw=none] at (-0.2, -1.2){$\vdots$};
	\end{tikzpicture}
	
	\caption{\centering A highest-weight representation of the quantum algebra $U_\hbar(\mathfrak{sl}_2)$. The relations to the left of the diagram are postulated, while those to the right are derived from the postulates.}
	\label{fig:HW_rep_qsl2_pic}
\end{wrapfigure}

Let us begin with finite-dimensional highest-weight representations of the quantum algebra $U_q(\mathfrak{sl}_2)$. They are quantum analogues of the ordinary irreducible representations of $\mathfrak{sl}_2$ and will henceforth be referred to as representations of $\operatorname{spin}$ type~\cite{KlimykSchmudgen1997}. For their construction, it is convenient to use the generators $E$, $F$, and $H$, with $H$ related to the generator $K$ by $K=q^H$.

Let $v_0$ be a highest-weight vector. The action of the raising operator $E$, the Cartan operator $H$, and the generator $K$ on this vector is specified by
\begin{equation}
	Ev_0=0,\qquad
	Hv_0=\gamma v_0,\qquad
	Kv_0=q^\gamma v_0,
\end{equation}
where $\gamma$ is called the highest weight. The remaining vectors of the representation are obtained by successive action of the lowering operator $F$:
\begin{equation}
	v_n=F^n v_0,
	\qquad
	Fv_n=v_{n+1}.
\end{equation}
It follows from the commutation relations of the quantum algebra that
\begin{equation}
	Hv_n=(\gamma-2n)v_n,
	\qquad
	Kv_n=q^{\gamma-2n}v_n.
\end{equation}
In addition, for $n\geq 1$, the action of the operator $E$ is given by
\begin{equation}
Ev_n= \sum\limits_{k=0}^{n-1}[\gamma-2k]_qv_{n-1}=[n]_q[\gamma-(n-1)]_qv_{n-1}
\end{equation}
Thus, the highest-weight representation has the form of a chain of vectors $v_0,v_1,\ldots$: the operator $F$ maps each vector to the next one, while the operator $E$ acts in the opposite direction with coefficients determined by the quantum numbers. To obtain a finite-dimensional representation of dimension $d$, one must impose
\begin{equation}\label{eq:F_zero_rule}
	Fv_{d-1}=0,
\end{equation}
and choose the highest weight to be
\begin{equation}
	\gamma=d-1.
\end{equation}
Then, in addition to \eqref{eq:F_zero_rule}, one also has
\begin{equation}
	Ev_{d}=0.
\end{equation}
The corresponding representation will be denoted by $\mathrm{spin}\left(\frac{\gamma}{2},+\right)$. When $q$ is not a root of unity, this construction reproduces all finite-dimensional irreducible highest-weight representations, just as in the ordinary representation theory of $\mathfrak{sl}_2$.

The situation is substantially different when $q$ is a root of unity. Let $m=\operatorname{ord}(q)$ be the smallest positive integer such that
\begin{equation}
	q^m=1.
\end{equation}
We also define the number
\begin{equation}
	m'=
	\begin{cases}
		m, & \text{if } m \text{ is odd},\\
		m/2, & \text{if } m \text{ is even}.
	\end{cases}
\end{equation}
For $q\neq\pm1$, the number $m'$ is the smallest positive integer for which the corresponding quantum number vanishes:
\begin{equation}\label{eq:D_constraint}
	[m']_q=0.
\end{equation}
This implies that only finitely many irreducible highest-weight representations exist. From the viewpoint of figure~\ref{fig:HW_rep_qsl2_pic}, equation~\eqref{eq:D_constraint} means that it is impossible to construct an irreducible indecomposable representation of dimension $d$ greater than $m'$.

In the case of interest below,
\begin{equation}
	q=\exp\left(\frac{i\pi}{4}\right)
\end{equation}
we have $m=8$ and $m'=4$. Therefore, there exist four irreducible $\operatorname{spin}$-type representations with positive sign, corresponding to the spins
\begin{equation}
	j=0,\quad \frac12,\quad 1,\quad \frac32.
\end{equation}
The last of these has zero quantum dimension; its role will be discussed in more detail in Sec.~\ref{sec:three_strands}.

In addition to the representations $\mathrm{spin}(\gamma/2,+)$, there also exist representations with a negative sign, $\mathrm{spin}(\gamma/2,-)$. Their distinction can conveniently be defined through the action of $K$ on the highest-weight vector:
\begin{equation}
	Kv_0=q^\gamma v_0
	\qquad\text{for}\qquad
	\mathrm{spin}\left(\frac{\gamma}{2},+\right),
\end{equation}
\begin{equation}
	Kv_0=-q^\gamma v_0
	\qquad\text{for}\qquad
	\mathrm{spin}\left(\frac{\gamma}{2},-\right).
\end{equation}
Accordingly, for an arbitrary weight vector we have
\begin{equation}
	Kv_n=\pm q^{\gamma-2n}v_n.
\end{equation}
The sign is part of the structure of the representation and, in general, cannot be removed by a simple change of basis. Its role is particularly important when $m$ is even. In this case, $q^{m'}=-1$, and therefore
\begin{equation}
	-q^\gamma=q^{\gamma+m'}.
\end{equation}
By contrast, when $m$ is odd, $m'=m$, and one has
\begin{equation}
	q^{\gamma+m'}=q^\gamma.
\end{equation}
This distinction leads to the appearance of representations with both signs in tensor-product decompositions at roots of unity. In addition, a representation with a negative sign can be obtained by taking tensor product with the one-dimensional representation $\mathrm{spin}(0,-)$:

\begin{equation}
	\mathrm{spin}\left(\frac{\gamma}{2},-\right)
	\simeq
	\mathrm{spin}\left(\frac{\gamma}{2},+\right)
	\otimes
	\mathrm{spin}(0,-).
\end{equation}

Finally, the quantum dimension is an important characteristic of $\operatorname{spin}$-type representations. In the convention where the quantum trace is defined using $K$, for a representation of dimension $d$ we obtain
\begin{equation}
	\operatorname{qdim}
	\left(
	\mathrm{spin}\left(\frac{\gamma}{2},+\right)
	\right)
	=
	\operatorname{Tr}_{\mathfrak{R}}(K).
\end{equation}
In particular, for the representations with positive sign at $q=\exp(i\pi/4)$, we have

\begin{equation}
	\operatorname{qdim}\bigl(\mathrm{spin}(0,+)\bigr)=1,
	\qquad
	\operatorname{qdim}\left(\mathrm{spin}\left(\frac12,+\right)\right)=\sqrt{2},
\end{equation}

\begin{equation}\label{eq:qdim_spin_three_half}
	\operatorname{qdim}\bigl(\mathrm{spin}(1,+)\bigr)=1,
	\qquad
	\operatorname{qdim}\left(\mathrm{spin}\left(\frac32,+\right)\right)=0.
\end{equation}

The first three values coincide with the quantum dimensions of the vacuum, $\sigma$, and $\psi$ sectors of the Ising model. The vanishing quantum dimension of the last representation indicates that its contribution requires separate consideration. In the next subsection, we introduce reducible but indecomposable representations of $\operatorname{ind}$ type, which arise naturally in tensor products of $\operatorname{spin}$-type representations~\cite{Arnaudon1992,Bishler2022,BishlerMironovMorozov2023}.

\subsection{Representations of ind type}\label{sec:ind_type}

At a root of unity, tensor products of $\operatorname{spin}$-type representations do not always decompose into direct sums of irreducible representations. In particular, reducible but indecomposable representations may arise. In the present work, such representations will be denoted by
\begin{equation}
	\operatorname{ind}\left(\frac{\gamma}{2},\pm\right).
\end{equation}
The symbol $\operatorname{ind}$ emphasizes that the representation is indecomposable. It contains an irreducible subrepresentation isomorphic to
\begin{equation}
	\mathrm{spin}\left(\frac{\gamma}{2},\pm\right),
\end{equation}
but it does not decompose into a direct sum of this subrepresentation and another representation. For the family of $\operatorname{ind}$-type representations considered here, the ordinary dimension is
\begin{equation}
	\dim \operatorname{ind}\left(\frac{\gamma}{2},\pm\right)=2m'.
\end{equation}
For an $\operatorname{ind}$ representation, the quantum dimension is given by
\begin{equation}
	\operatorname{qdim}
	\left(
	\operatorname{ind}\left(\frac{\gamma}{2},+\right)
	\right)
	=
	\left(
	[\gamma+1]_q+[2m'-\gamma-1]_q
	\right).
\end{equation}

Since $q^{2m'}=1$, the following identity holds:
\begin{equation}
	[2m'-\gamma-1]_q=-[\gamma+1]_q.
\end{equation}
Therefore,
\begin{equation}\label{eq:qdim_ind_zero}
	\operatorname{qdim}
	\left(
	\operatorname{ind}\left(\frac{\gamma}{2},+\right)
	\right)
	=0.
\end{equation}

Thus, $\operatorname{ind}$-type representations arise because, at a root of unity, some tensor-product channels that would normally be independent merge into a single reducible but indecomposable representation. In the problem considered below, such representations first appear in tensor products involving a sufficiently large number of fundamental representations.

A vanishing quantum dimension does not by itself imply that the quantum trace of an arbitrary operator on an $\operatorname{ind}$ representation is zero. Therefore, in what follows, $\operatorname{ind}$-type representations will not be discarded automatically. Their contribution to the braiding operators of interest and to the observables of Chern--Simons theory will be examined separately~\cite{Arnaudon1992,Bishler2022,BishlerMironovMorozov2023}.

\section{Highest-weight representations of $U_q(\mathfrak{sl}_2)$ at $q=\exp(\pi i/4)$}

In the previous section, we introduced the minimal information required about $\operatorname{spin}$- and $\operatorname{ind}$-type representations at a general root of unity. We now fix the value of the deformation parameter corresponding to $SU(2)_2$ Chern--Simons theory:
\begin{equation}
q=e^{\pi i/4}.
\end{equation}
The first quantum numbers are then
\begin{equation}
[1]_q=1,\qquad [2]_q=\sqrt{2},\qquad [3]_q=1,\qquad [4]_q=0.
\end{equation}

The construction on figure~\ref{fig:HW_rep_qsl2_pic} allows one to build a sequence of irreducible highest-weight representations $\operatorname{spin}(j,+)$. The relation $[4]_q=0$ is responsible for the main differences from the ordinary representation theory of $\mathfrak{sl}_2$. The representation with spin $j=3/2$ still exists, but its quantum dimension is zero. The sequence of irreducible highest-weight representations $\operatorname{spin}(j,+)$ terminates at this point.

For the subsequent comparison with Ising anyons, it is necessary to determine not only the representations themselves, but also the decomposition rules for their tensor products. In the quantum-algebraic description, these decompositions play the role of fusion rules: they specify which representations can arise when two lines labeled by given representations are combined.

Let us consider tensor products of two $\operatorname{spin}$-type representations. Their decompositions at $q=e^{\pi i/4}$ are collected in table~\ref{tab:spin_tensor_products}.
\begin{table}[!ht]
	\centering
	\renewcommand{\arraystretch}{1.15}
	
	\begin{tabular}{
			|>{\columncolor{blue!15}}m{2cm}|
			m{2.4cm}|
			m{2.4cm}|
			m{2.4cm}|
			m{2.4cm}|
		}
		\hline
		\rowcolor{blue!15}
		&
		$\operatorname{spin}(0,+)$ &
		$\operatorname{spin}(1/2,+)$ &
		$\operatorname{spin}(1,+)$ &
		$\operatorname{spin}(3/2,+)$
		\\
		\hline
		
		$\operatorname{spin}(0,+)$ &
		$\operatorname{spin}(0,+)$ &
		$\operatorname{spin}(1/2,+)$ &
		$\operatorname{spin}(1,+)$ &
		$\operatorname{spin}(3/2,+)$
		\\
		\hline
		
		$\operatorname{spin}(1/2,+)$ &
		$\operatorname{spin}(1/2,+)$ &
		$\operatorname{spin}(1,+)\oplus\operatorname{spin}(0,+)$ &
		$\operatorname{spin}(3/2,+)\oplus\operatorname{spin}(1/2,+)$ &
		$\operatorname{ind}(1,+)$
		\\
		\hline
		
		$\operatorname{spin}(1,+)$ &
		$\operatorname{spin}(1,+)$ &
		$\operatorname{spin}(3/2,+)\oplus\operatorname{spin}(1/2,+)$ &
		$\operatorname{ind}(1,+)\oplus\operatorname{spin}(0,+)$ &
		$\operatorname{ind}(1/2,+)\oplus\operatorname{spin}(3/2,+)$
		\\
		\hline
		
		$\operatorname{spin}(3/2,+)$ &
		$\operatorname{spin}(3/2,+)$ &
		$\operatorname{ind}(1,+)$ &
		$\operatorname{ind}(1/2,+)\oplus\operatorname{spin}(3/2,+)$ &
		$\operatorname{ind}(0,+)\oplus\operatorname{ind}(1,+)$
		\\
		\hline
	\end{tabular}
	
	\caption{\centering Decomposition rules for tensor products of $\operatorname{spin}$-type representations at $q=e^{\pi i/4}$.}
	\label{tab:spin_tensor_products}
\end{table}
Let us consider the representations $\operatorname{spin}\left(1/2,+\right)$, $\operatorname{spin}\left(1,+\right)$, and $\operatorname{spin}\left(3/2,+\right)$. Their diagrams are shown in Fig.~\ref{fig:all_spin_rep}.
\begin{figure}[h!]
		\centering
		\begin{tikzpicture}
			\begin{scope}[shift={(0,-1.5)}]
				\input{pictures/spin_fund.tex}
			\end{scope}
			
			\begin{scope}[shift={(4,-0.75)}]
				\input{pictures/spin_sym.tex}
			\end{scope}
			
			\begin{scope}[shift={(8,0)}]
				\input{pictures/spin_zero.tex}
			\end{scope}
		\end{tikzpicture}
		\caption{\centering $\operatorname{Spin}$ representations. Each arrow corresponds to a nonzero action of one of the generators $E$ or $F$. The subscripts are introduced solely to emphasize that the vectors
			$\lvert \uparrow\rangle_{1/2}$ , $\lvert \uparrow\rangle_{1}$  and so on,
			belong to different vector spaces.}
		\label{fig:all_spin_rep}
	\end{figure}
	
These representations themselves do not differ substantially from representations of the classical algebra $\mathfrak{sl}_2$. However, as can be seen from table~\ref{tab:spin_tensor_products}, tensor-product decompositions of these representations give rise to representations of $\operatorname{ind}$ type. In particular, the tensor product $\operatorname{spin}(1,+)\otimes\operatorname{spin}(3/2,+)$ contains the representation $\operatorname{ind}(1/2,+)$ in its direct-sum decomposition. Figure~\ref{fig:rep_ind_1/2} illustrates the structure of this representation. The representation $\operatorname{ind}\left(1/2,+\right)$ contains a subrepresentation isomorphic to $\operatorname{spin}\left(1/2,+\right)$. At the same time, the full representation $\operatorname{ind}\left(1/2,+\right)$ does not decompose into a direct sum of this subrepresentation and another invariant subspace.

\begin{figure}[h!]
	\centering
	\begin{adjustbox}{angle=0}
		\begin{tikzpicture}
			\input{pictures/ind_fund.tex}
		\end{tikzpicture}
	\end{adjustbox}
	\caption{\centering The $\operatorname{ind}(1/2,+)$ representation. Each arrow corresponds to a nonzero action of one of the generators $E$ or $F$. A bifurcating arrow indicates that the action of the corresponding generator produces a linear combination of the vectors shown at the ends of its branches.}
	\label{fig:rep_ind_1/2}
\end{figure}
 
 Let us also note a property that will be important in the construction of the $R$-matrices. From figure~\ref{fig:rep_ind_1/2}, one can see that
 \begin{equation}
 	E^4=0,
 	\qquad
 	F^4=0.
 \end{equation}
 This property is not unique to the representation $\operatorname{ind}(1/2,+)$. It holds for all highest-weight representations of $\operatorname{spin}$ and $\operatorname{ind}$ type at $q=\exp(i\pi/4)$. This guarantees that in the expression for the universal $R$-matrix
 \begin{equation}
 	\mathcal{R}=
 	e^{\frac{\hbar}{2}(H\otimes H)}
 	\sum
 	q^{\frac{n(n+1)}{2}}
 	\frac{(1-q^{-2})^n}{[n]_q!}\,
 	E^n\otimes F^n,
 	\qquad q=e^\hbar
 \end{equation}
 no division by zero occurs.
 
 \begin{table}[h!]
 	\centering
 	\renewcommand{\arraystretch}{1.15}
 	\setlength{\tabcolsep}{4pt}
 	
 	\begin{tabular}{|c|c|c|c|}
 		\hline
 		\cellcolor{blue!15}
 		&
 		\cellcolor{blue!15}$\operatorname{Ind}(0,+)$
 		&
 		\cellcolor{blue!15}$\operatorname{Ind}(1/2,+)$
 		&
 		\cellcolor{blue!15}$\operatorname{Ind}(1,+)$
 		\\
 		\hline
 		
 		\cellcolor{blue!15}$\operatorname{spin}(0,+)$
 		&
 		$\operatorname{Ind}(0,+)$
 		&
 		$\operatorname{Ind}(1/2,+)$
 		&
 		$\operatorname{Ind}(1,+)$
 		\\
 		\hline
 		
 		\cellcolor{blue!15}$\operatorname{spin}(1/2,+)$
 		&
 		$\begin{array}{c}
 			2\,\operatorname{spin}(3/2,-)\\
 			\oplus\,\operatorname{Ind}(1/2,+)
 		\end{array}$
 		&
 		$\begin{array}{c}
 			\operatorname{Ind}(0,+)\\
 			\oplus\,\operatorname{Ind}(1,+)
 		\end{array}$
 		&
 		$\begin{array}{c}
 			\operatorname{Ind}(1/2,+)\\
 			\oplus\,2\,\operatorname{spin}(3/2,+)
 		\end{array}$
 		\\
 		\hline
 		
 		\cellcolor{blue!15}$\operatorname{spin}(1,+)$
 		&
 		$\begin{array}{c}
 			2\,\operatorname{Ind}(1,-)\\
 			\oplus\,\operatorname{Ind}(1,+)
 		\end{array}$
 		&
 		$\begin{array}{c}
 			2\,\operatorname{spin}(3/2,-)\\
 			\oplus\,\operatorname{Ind}(1/2,+)\\
 			\oplus\,2\,\operatorname{spin}(3/2,+)
 		\end{array}$
 		&
 		$\begin{array}{c}
 			\operatorname{Ind}(0,+)\\
 			\oplus\,2\,\operatorname{Ind}(1,+)
 		\end{array}$
 		\\
 		\hline
 		
 		\cellcolor{blue!15}$\operatorname{spin}(3/2,+)$
 		&
 		$\begin{array}{c}
 			2\,\operatorname{Ind}(1/2,-)\\
 			\oplus\,2\,\operatorname{spin}(3/2,-)\\
 			\oplus\,2\,\operatorname{spin}(3/2,+)
 		\end{array}$
 		&
 		$\begin{array}{c}
 			2\,\operatorname{Ind}(1,-)\\
 			\oplus\,2\,\operatorname{Ind}(1,+)
 		\end{array}$
 		&
 		$\begin{array}{c}
 			2\,\operatorname{spin}(3/2,-)\\
 			\oplus\,2\,\operatorname{Ind}(1/2,+)\\
 			\oplus\,2\,\operatorname{spin}(3/2,+)
 		\end{array}$
 		\\
 		\hline\hline
 		
 		\cellcolor{blue!15}$\operatorname{Ind}(0,+)$
 		&
 		$\begin{array}{c}
 			4\,\operatorname{Ind}(1,+)\\
 			\oplus\,2\,\operatorname{Ind}(1,-)\\
 			\oplus\,2\,\operatorname{Ind}(0,+)
 		\end{array}$
 		&
 		$\begin{array}{c}
 			4\,\operatorname{spin}(3/2,+)\\
 			\oplus\,2\,\operatorname{Ind}(1/2,-)\\
 			\oplus\,4\,\operatorname{spin}(3/2,-)\\
 			\oplus\,2\,\operatorname{Ind}(1/2,+)
 		\end{array}$
 		&
 		$\begin{array}{c}
 			2\,\operatorname{Ind}(0,-)\\
 			\oplus\,4\,\operatorname{Ind}(1,-)\\
 			\oplus\,2\,\operatorname{Ind}(1,+)
 		\end{array}$
 		\\
 		\hline
 		
 		\cellcolor{blue!15}$\operatorname{Ind}(1/2,+)$
 		&
 		$\begin{array}{c}
 			4\,\operatorname{spin}(3/2,+)\\
 			\oplus\,2\,\operatorname{Ind}(1/2,-)\\
 			\oplus\,4\,\operatorname{spin}(3/2,-)\\
 			\oplus\,2\,\operatorname{Ind}(1/2,+)
 		\end{array}$
 		&
 		$\begin{array}{c}
 			2\,\operatorname{Ind}(0,-)\\
 			\oplus\,2\,\operatorname{Ind}(1,-)\\
 			\oplus\,2\,\operatorname{Ind}(0,+)\\
 			\oplus\,2\,\operatorname{Ind}(1,+)
 		\end{array}$
 		&
 		$\begin{array}{c}
 			2\,\operatorname{Ind}(1/2,-)\\
 			\oplus\,4\,\operatorname{spin}(3/2,-)\\
 			\oplus\,2\,\operatorname{Ind}(1/2,+)\\
 			\oplus\,4\,\operatorname{spin}(3/2,+)
 		\end{array}$
 		\\
 		\hline
 		
 		\cellcolor{blue!15}$\operatorname{Ind}(1,+)$
 		&
 		$\begin{array}{c}
 			2\,\operatorname{Ind}(0,-)\\
 			\oplus\,4\,\operatorname{Ind}(1,-)\\
 			\oplus\,2\,\operatorname{Ind}(1,+)
 		\end{array}$
 		&
 		$\begin{array}{c}
 			2\,\operatorname{Ind}(1/2,-)\\
 			\oplus\,4\,\operatorname{spin}(3/2,-)\\
 			\oplus\,2\,\operatorname{Ind}(1/2,+)\\
 			\oplus\,4\,\operatorname{spin}(3/2,+)
 		\end{array}$
 		&
 		$\begin{array}{c}
 			2\,\operatorname{Ind}(1,-)\\
 			\oplus\,2\,\operatorname{Ind}(0,+)\\
 			\oplus\,4\,\operatorname{Ind}(1,+)
 		\end{array}$
 		\\
 		\hline
 	\end{tabular}
 	
 	\caption{\centering Tensor-product decompositions of representations of $\operatorname{spin}$ and $\operatorname{Ind}$ type.}
 	\label{tab:mult_table_spin_ind}
 \end{table}
 
Taking into account the direct-sum decomposition table~\ref{tab:spin_tensor_products} and the equality of the quantum dimensions
\begin{equation}
	\mathfrak{d}_\mathbbm{1} = \mathfrak{d}_{\operatorname{spin}(0,+)}=1, \qquad
	\mathfrak{d}_\mathbbm{\sigma} = \mathfrak{d}_{\operatorname{spin}(1/2,+)}=\sqrt{2}, 
	\qquad
	\mathfrak{d}_\mathbbm{\psi} = \mathfrak{d}_{\operatorname{spin}(1,+)}=1,
\end{equation} 
the following correspondence holds:
\begin{equation}\label{eq:rep_ising_notation}
		\mathbbm{1}  \longleftrightarrow  \text{spin}(0,+), \qquad
		\sigma  \longleftrightarrow   \text{spin}(1/2,+), \qquad
		\psi  \longleftrightarrow    \text{spin}(1,+) 
\end{equation}
between the Ising anyons $\mathbbm{1}$, $\sigma$, and $\psi$ and the representations $\operatorname{spin}(0,+)$, $\operatorname{spin}(1/2,+)$, and $\operatorname{spin}(1,+)$ of the quantum algebra $U_q(\mathfrak{sl}_2)$. The fusion rule $\sigma\times\sigma$ is then reproduced exactly, while the fusion rules $\sigma\times\psi$ and $\psi\times\psi$ are reproduced up to the omission of direct-summand representations with zero quantum dimension.

The quantum dimensions of representations of the quantum algebra satisfy the same basic properties as the quantum dimensions of anyons~\eqref{eq:ising_Q_dim_global_rule}. For any two representations $\mathfrak{R}_1$ and $\mathfrak{R}_2$, one has
\begin{equation}
	\operatorname{qdim}(\mathfrak{R}_1\otimes \mathfrak{R}_2)
	=
	\operatorname{qdim}(\mathfrak{R}_1)\operatorname{qdim}(\mathfrak{R}_2).
\end{equation}
If the representation $\mathfrak{R}$ decomposes into a direct sum
\begin{equation}
\mathfrak{R}\simeq\bigoplus_a n_a \mathfrak{R}_a,
\end{equation}
then
\begin{equation}
	\operatorname{qdim}(\mathfrak{R})
	=
	\sum_a n_a\operatorname{qdim}(\mathfrak{R}_a).
\end{equation}
The first property follows from the relation
$\Delta(K)=K\otimes K$, while the second follows from the additivity of the quantum trace with respect to direct sums. Thus, for the quantum dimensions of representations of $U_q(\mathfrak{sl}_2)$, one obtains a direct analogue of formula~\eqref{eq:ising_Q_dim_global_rule} for Ising anyons. In particular, if at least one of the tensor factors has zero quantum dimension, then the quantum dimension of their tensor product is also zero. In the tables considered here, all representations with nonzero quantum dimension have quantum dimension $1$ or $\sqrt{2}$.

The representations $\operatorname{spin}(0,-)$,
$\operatorname{spin}(1/2,-)$, and $\operatorname{spin}(1,-)$, which could contribute with the opposite sign, do not appear in these decompositions. Therefore, cancellation between nonzero quantum dimensions is impossible: if a tensor product has zero quantum dimension, then all representations appearing in its decomposition must also have zero quantum dimension.

A physical interpretation of this phenomenon is the following. If one assumes otherwise, a contradiction would arise with formula~\eqref{eq:Wilson_loop_expect} for the Wilson loop expectation value. This expression involves the quantum trace, which weights the state associated with each representation by its quantum dimension. If the decomposition were to contain a representation with nonzero quantum dimension, one would obtain an additional state contributing to the observable without a corresponding contribution in the original tensor product.

It should also be noted that none of the decompositions listed in tables~\ref{tab:spin_tensor_products} and~\ref{tab:mult_table_spin_ind} contains the representations $\operatorname{spin}(0,-)$, $\operatorname{spin}(1/2,-)$, or $\operatorname{spin}(1,-)$. In the context of the present work, this means that an arbitrarily large tensor product built from the representations $\operatorname{spin}(0,+)$, $\operatorname{spin}(1/2,+)$, and $\operatorname{spin}(1,+)$ decomposes only into the representations $\operatorname{spin}(0,+)$, $\operatorname{spin}(1/2,+)$, $\operatorname{spin}(1,+)$, together with representations of zero quantum dimension.

In what follows, we will be primarily interested in tensor powers of the fundamental representation $\operatorname{spin}\left(1/2,+\right)$, since it is this representation that labels the Wilson lines.

\subsection{Two strands}

For two strands, we need to verify the correspondence between the braiding matrix for Ising anyons $[R_{\sigma\sigma}]_\text{Ising}$ and the braiding matrix for the tensor product
\begin{equation}\label{eq:two_strands_decomp}
\operatorname{spin}(1/2,+)\otimes\operatorname{spin}(1/2,+) 
=
\operatorname{spin}(0,+)
\oplus
\operatorname{spin}(1,+).
\end{equation}
Figure~\ref{fig:spin_half_tensor_spin_half} shows the decomposition \eqref{eq:two_strands_decomp} graphically.
\begin{figure}[h!]
	\centering
	\begin{tikzpicture}[scale=1.1, every node/.style={font=\scriptsize}]
		\draw[blue, thick] (0,0) -- (-0.6,0.9);
		\draw[blue, thick] (0,0) -- (0.6,0.9);
		\draw[blue, thick] (0,0) -- (0,-1.0);
		\node at (-0.95,1.05) {$\operatorname{spin}(1/2,+)$};
		\node at (0.95,1.05) {$\operatorname{spin}(1/2,+)$};
		\node at (0,-1.2) {$\operatorname{spin}(0,+)$};
		
		\draw[blue, thick] (5,0) -- (4.4,0.9);
		\draw[blue, thick] (5,0) -- (5.6,0.9);
		\draw[blue, thick] (5,0) -- (5,-1.0);
		\node at (4.05,1.05) {$\operatorname{spin}(1/2,+)$};
		\node at (5.95,1.05) {$\operatorname{spin}(1/2,+)$};
		\node at (5,-1.2) {$\operatorname{spin}(1,+)$};
	\end{tikzpicture}
	\caption{\centering Graphical representation of the decomposition
		$\mathrm{spin}(1/2,+)\otimes \mathrm{spin}(1/2,+)
		= \mathrm{spin}(0,+)\oplus \mathrm{spin}(1,+)$.}
	\label{fig:spin_half_tensor_spin_half}
\end{figure}
It exactly reproduces the fusion rule for two $\sigma$-anyons:
\begin{equation}\label{eq:ising_two_decomp}
	\sigma\times\sigma
	=
	\mathbbm{1}+\psi.
\end{equation}

Recall that in the quantum algebra the braiding operator $\check{R}$ is constructed as the composition of the permutation operator $P$ and the universal $R$-matrix
\begin{equation}
	\mathcal{R}
	=
	e^{\frac{\hbar}{2}(H\otimes H)}
	\sum_{n=0}^{3}
	\frac{
		q^{\frac{n(n+1)}{2}}
		\left(1-q^{-2}\right)^n
	}{
		[n]_q!
	}
	E^n\otimes F^n,
	\qquad
	q=e^\hbar.
	\label{eq:two_strands_universal_R}
\end{equation}

We now construct the operator $\check{R}^{\bullet\bullet}$ acting on the space \eqref{eq:two_strands_decomp}, that is, on two strands. The circles in the upper-right corner indicate the number of strands. They will also be useful below to denote the fusion order. The operator
$\check R^{\bullet\bullet}$ commutes with the total action of the quantum algebra:
\begin{equation}
	\check R^{\bullet\bullet}\,\Delta(x)
	=
	\Delta(x)\,\check R^{\bullet\bullet},
	\qquad
	x\in U_q(\mathfrak{sl}_2).
	\label{eq:two_strands_R_commutes}
\end{equation}
Therefore, if a highest-weight vector of a given representation is an eigenvector of $\check R^{\bullet\bullet}$, then $\check R^{\bullet\bullet}$ acts with the same eigenvalue on the entire representation generated by this vector. Hence, in this section and in the following ones, we will write the braiding matrix acting on the highest-weight vectors.

In the representation $\operatorname{spin}(1/2,+)$, the basis vectors $\mid\uparrow\rangle_{1/2}$ and $\mid\downarrow\rangle_{1/2}$ satisfy the following relations:
\begin{equation}
	\begin{aligned}
		K\mid\uparrow\rangle_{1/2} &= q\mid\uparrow\rangle_{1/2},
		&
		K\mid\downarrow\rangle_{1/2} &= q^{-1}\mid\downarrow\rangle_{1/2},
		\\
		E\mid\uparrow\rangle_{1/2} &= 0,
		&
		E\mid\downarrow\rangle_{1/2} &= \mid\uparrow\rangle_{1/2},
		\\
		F\mid\uparrow\rangle_{1/2} &= \mid\downarrow\rangle_{1/2},
		&
		F\mid\downarrow\rangle_{1/2} &= 0.
	\end{aligned}
	\label{eq:fundamental_representation_action}
\end{equation}
This basis is consistent with Fig.~\ref{fig:all_spin_rep}. Since in what follows we will only consider strands carrying the fundamental representation, that is, tensor powers of the fundamental representation, we introduce the notation
\begin{equation}
\mid a_1\rangle_{1/2}\otimes
\mid a_2\rangle_{1/2}\otimes
\dots\otimes
\mid a_p\rangle_{1/2} 
\to 
\mid a_1 a_2 \dots a_p\rangle.
\end{equation}
In accordance with the notation introduced above, the direct-sum decomposition of the tensor product \eqref{eq:two_strands_decomp} contains two highest-weight vectors:
\begin{equation}
	v^{\bullet\bullet}_{0}
	=
	\mid\downarrow\uparrow\rangle
	-
	q\mid\uparrow\downarrow\rangle,
	\qquad
	v^{\bullet\bullet}_{1}
	=
	\mid\uparrow\uparrow\rangle,
	\label{eq:two_strands_highest_vectors}
\end{equation}
Substitution into Eq.~\eqref{eq:two_strands_universal_R} gives
\begin{equation}
	\check R^{\bullet\bullet}\,v^{\bullet\bullet}_{0}
	=
	-q^{-3/2}v^{\bullet\bullet}_{0},
	\qquad
	\check R^{\bullet\bullet}\,v^{\bullet\bullet}_{1}
	=
	q^{1/2}v^{\bullet\bullet}_{1}.
	\label{eq:two_strands_R_eigenvalues}
\end{equation}
Therefore, in the basis of highest-weight vectors $(\operatorname{spin}(0,+), \operatorname{spin}(1,+))$, the braiding matrix is
\begin{equation}
	\check R^{\bullet\bullet}
	=
	\begin{pmatrix}
		-q^{-3/2} & 0\\
		0 & q^{1/2}
	\end{pmatrix}
	=
	\begin{pmatrix}
		e^{5i\pi/8} & 0\\
		0 & e^{i\pi/8}
	\end{pmatrix}.
	\label{eq:two_strands_Uq_R}
\end{equation}

Finally, one can verify directly that
\begin{equation}
	[R_{\sigma\sigma}]_\text{ising}
	=
	e^{i\pi/2}
	\left(
	\check R^{\bullet\bullet}
	\right)^{-1}.
	\label{eq:two_strands_R_comparison}
\end{equation}
Indeed, the right-hand side is equal to
\begin{equation}
e^{i\pi/2}
\begin{pmatrix}
	e^{-5i\pi/8} & 0\\
	0 & e^{-i\pi/8}
\end{pmatrix}
=
\begin{pmatrix}
	e^{-i\pi/8} & 0\\
	0 & e^{3i\pi/8}
\end{pmatrix}.
\end{equation}
Thus, the braiding matrix obtained from the universal $R$-matrix of $U_q(\mathfrak{sl}_2)$ reproduces Eq.~\eqref{eq:R_ising_R_WZW}. The inverse matrix corresponds to the opposite choice of braiding orientation, while the factor $e^{i\pi/2}$ is a common phase for both channels. As before, no essential discrepancy arises between the observables constructed in $SU(2)_k$ Chern--Simons theory and those of the Ising minimal model, since an overall phase does not affect the corresponding measurement probabilities.

\subsection{Three strands}\label{sec:three_strands}

For three strands, one has to consider the tensor product of three fundamental representations $\operatorname{spin}(1/2,+)$. This tensor product decomposes into a direct sum
\begin{equation}\label{eq:three_strands_decomposition}	
	\operatorname{spin}(1/2,+)\otimes
	\operatorname{spin}(1/2,+)\otimes
	\operatorname{spin}(1/2,+)
	=
	2
	\operatorname{spin}(1/2,+)
	\oplus
	\operatorname{spin}(3/2,+).
\end{equation}
Thus, the fundamental representation appears twice, while no representations of ind type arise for three strands. This leads to an apparent discrepancy between the decomposition \eqref{eq:three_strands_decomposition} and the corresponding decomposition for Ising anyons
\begin{equation}
	\sigma\times\sigma\times\sigma
	=
	\sigma+\sigma.
\end{equation}
This apparent discrepancy is easily explained by the fact that the quantum dimension of $\operatorname{spin}(3/2,+)$ is zero and therefore does not contribute to the total quantum dimension. As will be shown below, both braiding matrices in \eqref{eq:three_strands_R} have a block-diagonal structure and hence do not mix vectors belonging to $2\operatorname{spin}(1/2,+)$ with those belonging to $\operatorname{spin}(3/2,+)$. Since the observable involves the quantum trace of a product of braiding matrices, the contribution of the representation $\operatorname{spin}(3/2,+)$ is ultimately multiplied by its quantum dimension and therefore vanishes. Thus, no contradiction arises, and for three strands the representation $\operatorname{spin}(3/2,+)$ can be omitted.

The decomposition \eqref{eq:three_strands_decomposition} can be performed in two different orders. In particular, one may first decompose either the left or the right pair in the tensor product
\begin{equation}
	\Big(\operatorname{spin}(1/2,+)\otimes\operatorname{spin}(1/2,+)\Big)\otimes\operatorname{spin}(1/2,+)
	\cong
	\operatorname{spin}(1/2,+)\otimes\Big(\operatorname{spin}(1/2,+)\otimes\operatorname{spin}(1/2,+)\Big),
\end{equation}
Here, the symbol $\cong$ is used instead of the equality sign $=$ in order to emphasize that the spaces obtained in the final decomposition are isomorphic, while the corresponding bases of vectors are different. The transition matrix between the two decomposition orders in the space $2\operatorname{spin}(1/2,+)$ should coincide with the fusion matrix for Ising anyons \eqref{eq:ising_fusion_matrix}.

As before, it is convenient to represent the fusion order graphically by means of diagrams. Examples of such diagrams for three strands are shown in Fig.~\ref{fig:three_strands_fusion}, where the notations $(\bullet\bullet)\bullet$ and $\bullet(\bullet\bullet)$ are introduced for the highest-weight bases corresponding to the left and right decomposition orders, respectively.

\begin{figure}[htbp]
	\centering
	\begin{tikzpicture}[x=1.3cm, line width=0.8pt]
		
		\begin{scope}
			\node[above] at (0,2.3) {$\operatorname{spin}(1/2,+)$};
			\node[above] at (2,2.3) {$\operatorname{spin}(1/2,+)$};
			\node[above] at (4,2.3) {$\operatorname{spin}(1/2,+)$};
			
			\draw[blue] (0,2.3) -- (1,1.3);
			\draw[blue] (2,2.3) -- (1,1.3);
			\draw[blue] (1,1.3) -- (2,0.3);
			\draw[blue] (4,2.3) -- (2,0.3);
			\draw[blue] (2,0.3) -- (2.5,-0.2);
			
			\node[below] at (2.5,-0.2) {$\operatorname{spin}(1/2,+)$};
			
			\node[left] at (1.5,0.5) {$\operatorname{spin}(j_1,+)$};
				
			\node[below, draw] at (2.5,-1.2) {$(\bullet\bullet)\bullet$};
		\end{scope}
		
		\begin{scope}[xshift=8cm]
			\node[above] at (0,2.3) {$\operatorname{spin}(1/2,+)$};
			\node[above] at (2,2.3) {$\operatorname{spin}(1/2,+)$};
			\node[above] at (4,2.3) {$\operatorname{spin}(1/2,+)$};
			
			\draw[blue] (2,2.3) -- (3,1.3);
			\draw[blue] (4,2.3) -- (3,1.3);
			\draw[blue] (3,1.3) -- (2,0.3);
			\draw[blue] (0,2.3) -- (2,0.3);
			\draw[blue] (2,0.3) -- (2.5,-0.2);
			
			\node[below] at (2.5,-0.2) {$\operatorname{spin}(1/2,+)$};
			
			\node[right] at (2.5,0.5) {$\operatorname{spin}(j_2,+)$};
			
			\node[below, draw] at (2.5,-1.2) {$\bullet(\bullet\bullet)$};

		\end{scope}
		
	\end{tikzpicture}
	\caption{\centering Two fusion orders of three fundamental representations:
		$(\bullet\bullet)\bullet$ on the left and $\bullet(\bullet\bullet)$ on the right.}
	\label{fig:three_strands_fusion}
\end{figure}

In each basis, the indices $j_1$ and $j_2$ can take the values $0$ or $1$, indicating the spin of the intermediate representation. In terms of the Ising-anyon decomposition \eqref{eq:ising_two_decomp}, these values correspond to the fusion channels $\mathbbm{1}$ and $\psi$, respectively. Since we are interested specifically in the transition matrix between the highest-weight vectors belonging to the space $2\operatorname{spin}(1/2,+)$ in the decomposition \eqref{eq:three_strands_decomposition}, the final representation in both fusion orders is $\operatorname{spin}(1/2,+)$.

To construct the bases explicitly, we use the states $\mid\uparrow\rangle$ and $\mid\downarrow\rangle$ of the representation $\operatorname{spin}(1/2,+)$ introduced in the previous subsection. In the basis $(\bullet\bullet)\bullet$, where the first two strands are fused first, the highest-weight vectors are
\begin{equation}\label{eq:three_strands_left_vectors}
		v_{0\to 1/2}^{(\bullet\bullet)\bullet}
		=
		\mid\downarrow\uparrow\uparrow\rangle
		-
		q\mid\uparrow\downarrow\uparrow\rangle
		,
		\qquad
		v_{1\to 1/2}^{(\bullet\bullet)\bullet}
		=
		q[2]_q\mid\uparrow\uparrow\downarrow\rangle
		-
		q^{-1}\mid\uparrow\downarrow\uparrow\rangle
		-
		\mid\downarrow\uparrow\uparrow\rangle
		.
\end{equation}
In the basis $\bullet(\bullet\bullet)$, where the last two strands are fused first, we obtain
\begin{equation}
	\begin{aligned}
		v_{0\to 1/2}^{\bullet(\bullet\bullet)}
		=
			q\mid\uparrow\uparrow\downarrow\rangle
			-
			\mid\uparrow\downarrow\uparrow\rangle
		,
		\qquad
		v_{1\to 1/2}^{\bullet(\bullet\bullet)}
		=
			[2]_q\mid\downarrow\uparrow\uparrow\rangle
			-
			q^2\mid\uparrow\downarrow\uparrow\rangle
			-
			q\mid\uparrow\uparrow\downarrow\rangle
		.
	\end{aligned}
	\label{eq:three_strands_right_vectors}
\end{equation}
The overall signs of the vectors are chosen so that the transition matrix follows the same phase conventions as the fusion matrix of Ising anyons.

The highest-weight vector of the representation $\operatorname{spin}(3/2,+)$ is the same for both decomposition orders:
\begin{equation}
	v_{1\to 3/2}^{\bullet(\bullet\bullet)}
	=
	v_{1\to 3/2}^{(\bullet\bullet)\bullet}
	=
	\mid\uparrow\uparrow\uparrow\rangle.
\end{equation}
Therefore, the transition between the bases in this summand is trivial.

Comparing the coefficients of the tensor-product basis vectors at $q=e^{\pi i/4}$, we find
\begin{equation}
	\begin{pmatrix}
		v_{0\to 1/2}^{(\bullet\bullet)\bullet}\\[2mm]
		v_{1\to 1/2}^{(\bullet\bullet)\bullet}\\[2mm]
		v_{1\to 3/2}^{(\bullet\bullet)\bullet}
	\end{pmatrix}
	=
	\check{\mathsf{F}}^{\bullet(\bullet\bullet) \to (\bullet\bullet)\bullet}
	\begin{pmatrix}
		v_{0\to 1/2}^{\bullet(\bullet\bullet)}\\[2mm]
		v_{1\to 1/2}^{\bullet(\bullet\bullet)}\\[2mm]
		v_{1\to 3/2}^{\bullet(\bullet\bullet)}
	\end{pmatrix},
	\qquad
	\check{\mathsf{F}}^{\bullet(\bullet\bullet) \to (\bullet\bullet)\bullet}
	=
	\left(
	\begin{array}{cc|c}
		1/\sqrt{2} & 1/\sqrt{2} & 0\\
		1/\sqrt{2} & -1/\sqrt{2} & 0\\
		\hline
		0 & 0 & 1
	\end{array}
	\right),
\end{equation}
Here, 
\begin{equation}
	\left. \check{\mathsf{F}}^{\bullet(\bullet\bullet) \to (\bullet\bullet)\bullet}\right|_{\text{phys}}
	=
	\frac{1}{\sqrt{2}}
	\begin{pmatrix}
		1 & 1 \\
		1 & -1 
	\end{pmatrix}.
	\label{eq:three_strands_racah}
\end{equation}
denotes the physical part of the transition matrix $\check{\mathsf{F}}^{\bullet(\bullet\bullet) \to (\bullet\bullet)\bullet}$, corresponding to the transformation between highest-weight vectors of representations with nonzero quantum dimension.

Since the lower-weight vectors are obtained from the highest-weight vectors by the action of the same operator $\Delta^{(2)}(F)$, the same transition matrix acts on them as well. Although the vectors \eqref{eq:three_strands_left_vectors} and \eqref{eq:three_strands_right_vectors} are not normalized, the matrix \eqref{eq:three_strands_racah} is unitary, which is important for operations in topological quantum computation. The same matrix can also be obtained using normalized vectors.

The normalization is usually performed as follows. The normalization factors are first computed without substituting a specific numerical value of $q$, assuming that $q$ is real and positive. The resulting expressions are then analytically continued to the required value of $q$. Applying this procedure to the vectors \eqref{eq:three_strands_left_vectors} and \eqref{eq:three_strands_right_vectors}, one finds that the transition matrix takes the form
\begin{equation}
	\left.\check{\mathsf{F}}^{\bullet(\bullet\bullet) \to (\bullet\bullet)\bullet}\right|_{\text{phys}}(q)
	=
	\frac{1}{[2]_q}
	\begin{pmatrix}
		1 & \sqrt{[3]_q}\\
		\sqrt{[3]_q} & -1
	\end{pmatrix}
	.
	\label{eq:three_strands_ising_F}
\end{equation}
This matrix coincides with the Racah matrix \cite{KirillovReshetikhin1989, MironovMorozovMorozov2012} up to multiplication of the highest-weight vectors by factors $\pm 1$. After analytic continuation to $q\to\exp(i\pi/4)$, one obtains the matrix \eqref{eq:three_strands_racah}. Thus, the physical part $\check{\mathsf{F}}^{\bullet(\bullet\bullet) \to (\bullet\bullet)\bullet}|_{\text{phys}}$ of the transition matrix $\check{\mathsf{F}}^{\bullet(\bullet\bullet) \to (\bullet\bullet)\bullet}$ reproduces the fusion matrix of Ising anyons~\eqref{eq:ising_fusion_matrix}.

Let us now explain why the transition matrix $\check{\mathsf{F}}^{\bullet(\bullet\bullet) \to (\bullet\bullet)\bullet}$ between the two bases is needed when computing braiding operations. On the full space $\operatorname{spin}(1/2,+)^{\otimes3}$, braiding of the first two strands and braiding of the last two strands are represented by the operators
\begin{equation}
	\check R_1^{\bullet\bullet\bullet}
	=
	\check R^{\bullet\bullet}\otimes\operatorname{id},
	\qquad
	\check R_2^{\bullet\bullet\bullet}
	=
	\operatorname{id}\otimes\check R^{\bullet\bullet}.
\end{equation}
We denote their matrices on the three-dimensional space of highest-weight vectors in the basis
$(v_{0\to 1/2}^{(\bullet\bullet)\bullet},
v_{1\to 1/2}^{(\bullet\bullet)\bullet},
v_{1\to 3/2}^{(\bullet\bullet)\bullet})$
by $\check{R}_1^{(\bullet\bullet)\bullet}$ and
$\check{R}_2^{(\bullet\bullet)\bullet}$, respectively.

In the basis $(\bullet\bullet)\bullet$, the fusion channel of the first two strands is fixed, so the first braiding operation is described by the diagonal matrix found in the previous subsection:
\begin{equation}
	\check{R}_1^{(\bullet\bullet)\bullet}
	=
	\left(
	\begin{array}{cc|c}
		-q^{-3/2} & 0 & 0 \\
		0 & q^{1/2} & 0 \\
		\hline
		0 & 0 & q^{1/2}
	\end{array}
	\right).
\end{equation}
The second braiding operation is diagonal in the basis $\bullet(\bullet\bullet)$, consisting of the highest-weight vectors $(v_{0\to 1/2}^{\bullet(\bullet\bullet)},v_{1\to 1/2}^{\bullet(\bullet\bullet)},v_{1\to 3/2}^{\bullet(\bullet\bullet)})$:
\begin{equation}
	\check{R}_2^{\bullet(\bullet\bullet)}
	=
	\left(
	\begin{array}{cc|c}
		-q^{-3/2} & 0 & 0 \\
		0 & q^{1/2} & 0 \\
		\hline
		0 & 0 & q^{1/2}
	\end{array}
	\right).
\end{equation}
To compute its action in the basis $(\bullet\bullet)\bullet$, one must first transform to the basis $\bullet(\bullet\bullet)$, apply $\check{R}_2^{\bullet(\bullet\bullet)}$, and then transform back:
\begin{equation}
	\check{R}_2^{(\bullet\bullet)\bullet}
	=
	\left(\check{\mathsf{F}}^{\bullet(\bullet\bullet) \to (\bullet\bullet)\bullet}\right)^{-1}\check{R}_2^{\bullet(\bullet\bullet)}\check{\mathsf{F}}^{\bullet(\bullet\bullet) \to (\bullet\bullet)\bullet} 
	= 
	\check{\mathsf{F}}^{\bullet(\bullet\bullet) \to (\bullet\bullet)\bullet}\check{R}_2^{\bullet(\bullet\bullet)}\check{\mathsf{F}}^{\bullet(\bullet\bullet) \to (\bullet\bullet)\bullet}
	.
	\label{eq:three_strands_second_braiding}
\end{equation}
The last equality follows from the fact that $\left(\check{\mathsf{F}}^{\bullet(\bullet\bullet) \to (\bullet\bullet)\bullet}\right)^{-1}=\check{\mathsf{F}}^{\bullet(\bullet\bullet) \to (\bullet\bullet)\bullet}$. For the chosen value of $q$, we obtain
\begin{equation}\label{eq:three_strands_R}
	\check{R}_1^{(\bullet\bullet)\bullet}
	=
	e^{i\pi/8}
	\left(
	\begin{array}{cc|c}
		i & 0 & 0\\
		0 & 1 & 0\\
		\hline
		0 & 0 & 1
	\end{array}
	\right),
	\qquad
	\check{R}_2^{(\bullet\bullet)\bullet}
	=
	\frac{e^{i\pi/8}}{2}
	\left(
	\begin{array}{cc|c}
		1+i & i-1 & 0\\
		i-1 & 1+i & 0\\
		\hline
		0 & 0 & 2
	\end{array}
	\right).
\end{equation}
These matrices satisfy the braid-group relation
\begin{equation}
	\check{R}_2^{(\bullet\bullet)\bullet}
	\check{R}_1^{(\bullet\bullet)\bullet}
	\check{R}_2^{(\bullet\bullet)\bullet}
	=
	\check{R}_1^{(\bullet\bullet)\bullet}
	\check{R}_2^{(\bullet\bullet)\bullet}
	\check{R}_1^{(\bullet\bullet)\bullet}.
\end{equation}
Both braiding matrices in \eqref{eq:three_strands_R} have a block-diagonal form:
\begin{equation}
	\check R_i^{(\bullet\bullet)\bullet}
	=
	\left(
	\begin{array}{c|c}
		\left.\check R_i^{(\bullet\bullet)\bullet}\right|_{\mathrm{phys}} & 0\\
		\hline
		0 & e^{i\pi/8}
	\end{array}
	\right),
	\qquad i=1,2,
\end{equation}
where the upper-left $2\times2$ block acts on the two copies of $\operatorname{spin}(1/2,+)$, while the lower-right $1\times1$ block corresponds to the representation $\operatorname{spin}(3/2,+)$.

When computing the observable for the closure of a three-strand braid, one has to evaluate the quantum trace of the corresponding word in the braiding operators. In the basis of highest-weight vectors, this is equivalent to weighting each block by the quantum dimension of the corresponding final representation. Therefore, the contribution of the lower-right block is always proportional to
\begin{equation}
	\operatorname{qdim}\operatorname{spin}(3/2,+)
	=
	[4]_q
	=
	0.
\end{equation}
Therefore, the representation $\operatorname{spin}(3/2,+)$ does not contribute to the observable, and when computing observables the braiding operators may be restricted to the physical two-dimensional sector $2\operatorname{spin}(1/2,+)$. The restrictions $\check R_1^{(\bullet\bullet)\bullet}|_{\mathrm{phys}}$ and $\check R_2^{(\bullet\bullet)\bullet}|_{\mathrm{phys}}$ of the matrices $\check R_1^{(\bullet\bullet)\bullet}$ and $\check R_2^{(\bullet\bullet)\bullet}$ to the $2\operatorname{spin}(1/2,+)$ sector coincide with the corresponding braiding operators for three Ising anyons, up to reversal of the braiding orientation and an overall phase factor. In particular,
\begin{equation}\label{eq:three_strands_R_comparison}
	\left.
	\check R_i^{(\bullet\bullet)\bullet}
	\right|_{\mathrm{phys}}
	=
	e^{i\pi/2}
	\left(
	\check R_i^{\mathrm{Ising}}
	\right)^{-1},
	\qquad i=1,2,
\end{equation}
under the conventions for the braiding orientation adopted above.

Thus, the additional representation $\operatorname{spin}(3/2,+)$ appearing in the decomposition $\operatorname{spin}(1/2,+)^{\otimes3}$ does not produce any additional contribution to the observables. After restricting to the physical sector, the three-strand braiding operators reproduce the corresponding operators for Ising anyons up to an overall phase and the choice of braiding orientation. Therefore, for three strands, no discrepancy between the two descriptions arises at the level of the observables considered here.

\subsection{Four strands and apparent contradictions}\label{sec:four_strands}
For four strands, we choose the pairwise basis $(\bullet\bullet)(\bullet\bullet)$. In this basis, the first and second strands, as well as the third and fourth strands, are decomposed first (Fig.~\ref{fig:four_strand_pairwise_basis}). This basis is convenient both for the plat closure and for the ordinary braid closure. In the former case, it explicitly singles out the state in which both pairs fuse into the vacuum. In the latter case, it is convenient for computing the braid matrix, since the braiding operations of the first and the last pairs are diagonal, while the quantum trace is independent of the choice of basis.

\begin{figure}[h!]
	\centering
	\begin{tikzpicture}[
		scale=1.05,
		fusion/.style={blue, very thick},
		extlabel/.style={font=\scriptsize},
		intlabel/.style={font=\small}
		]
		
		\draw[fusion] (-3.0,2.2) -- (-2.2,1.4);
		\draw[fusion] (-1.4,2.2) -- (-2.2,1.4);
		
		\draw[fusion] (1.4,2.2) -- (2.2,1.4);
		\draw[fusion] (3.0,2.2) -- (2.2,1.4);
		
		\draw[fusion] (-2.2,1.4) -- (-0.8,0.0);
		\draw[fusion] (2.2,1.4) -- (0.8,0.0);
		
		\draw[fusion] (-0.8,0.0) -- (0,-0.8);
		\draw[fusion] (0.8,0.0) -- (0,-0.8);
		\draw[fusion] (0,-0.8) -- (0,-2.0);
		
		\node[extlabel, above=3pt] at (-3.0,2.2)
		{$\operatorname{spin}(1/2,+)$};
		
		\node[extlabel, above=3pt] at (-1.4,2.2)
		{$\operatorname{spin}(1/2,+)$};
		
		\node[extlabel, above=3pt] at (1.4,2.2)
		{$\operatorname{spin}(1/2,+)$};
		
		\node[extlabel, above=3pt] at (3.0,2.2)
		{$\operatorname{spin}(1/2,+)$};
		
		\node[intlabel] at (-1.45,0.85) {$i$};
		\node[intlabel] at (1.45,0.85) {$j$};
		
		\node[intlabel, right=4pt] at (0,-1.45) {$k$};
		
		\node[draw, font=\small] at (0,-2.6)
		{$(\bullet\bullet)(\bullet\bullet)$};
		
	\end{tikzpicture}
	
	\caption{\centering
		Fusion tree corresponding to the pairwise basis
		$(\bullet\bullet)(\bullet\bullet)$ for four fundamental representations.
		The labels $i$ and $j$ denote the intermediate representations
		obtained by fusing the first and second, and the third and fourth
		strands, respectively, while $k$ denotes the resulting representation.
	}
	\label{fig:four_strand_pairwise_basis}
\end{figure}

For three strands, no contradictions with the description in terms of Ising anyons arose. This observation suggests that a similar picture should persist for four strands.

Indeed, let a knot $\mathcal{K}$ be represented as the ordinary closure of a three-strand braid. By Markov's theorem (Fig.~\ref{fig:markov_stabilization}), stabilization does not change the isotopy class of the closed knot. Therefore, any knot that can be represented as the closure of a three-strand braid can also be represented as the ordinary closure of a suitable four-strand braid.
\begin{figure}[h!]
	\centering
	\begin{tikzpicture}[
		scale=1.0,
		every node/.style={font=\small},
		braid/.style={blue, thick},
		box/.style={draw, thick, minimum width=1.2cm, minimum height=1.0cm, fill=white}
		]
		
		\node[box, align=center] (L) at (0,0) {3-strand\\braid};
		
		\draw[braid] (-0.35,1.35) -- (-0.35,0.5);
		\draw[braid] ( 0.00,1.35) -- ( 0.00,0.5);
		\draw[braid] ( 0.35,1.35) -- ( 0.35,0.5);
		
		\draw[braid] (-0.35,-0.5) -- (-0.35,-1.35);
		\draw[braid] ( 0.00,-0.5) -- ( 0.00,-1.35);
		\draw[braid] ( 0.35,-0.5) -- ( 0.35,-1.35);
		
		
		\draw[->, thick] (1.35,0) -- (3.85,0);
		\node at (2.60,0.35) {Markov stabilization};
		
		\node[box, align=center] (R) at (5.75,0.15) {3-strand\\braid};
		
		\draw[braid] (5.30,1.45) -- (5.30,0.65);
		\draw[braid] (5.65,1.45) -- (5.65,0.65);
		\draw[braid] (6.00,1.45) -- (6.00,0.65);
		
		\draw[braid] (6.75,1.45) -- (6.75,-0.35);
		
		\draw[braid] (5.30,-0.35) -- (5.30,-1.55);
		\draw[braid] (5.65,-0.35) -- (5.65,-1.55);
		
		\draw[braid]
		(6.75,-0.35)
		to[out=-90,in=90]
		(6.00,-0.85)
		to[out=-90,in=90]
		(6.00,-1.55);
		
		\draw[white, line width=4pt]
		(6.00,-0.35)
		to[out=-90,in=90]
		(6.75,-0.85);
		
		\draw[braid]
		(6.00,-0.35)
		to[out=-90,in=90]
		(6.75,-0.85)
		to[out=-90,in=90]
		(6.75,-1.55);
		
		
	\end{tikzpicture}
	
	\caption{\centering
		Markov stabilization of a $3$-strand braid.
		The braid acts on the first three strands, while the fourth
		strand is added trivially and then braided with the third one.
	}
	\label{fig:markov_stabilization}
\end{figure}
Moreover, some knots that admit a representation as the closure of a three-strand braid also admit a four-strand plat representation. An example is the knot $5_2$ considered in Sec.~\ref{subsec:ChSbasics}: it can be represented both as the ordinary closure of a three-strand braid and as the plat closure of a four-strand braid. Since the three-strand description has already been shown to agree with the Ising-anyon picture, it is natural to expect that no new contradictions will arise upon passing to four strands.

However, this argument by itself does not constitute a proof.
First, not every knot that can be represented as the closure of a three-strand braid admits a four-strand plat representation. Therefore, the comparison between the two types of closure that is possible for the knot $5_2$ cannot be extended directly to an arbitrary knot.

Second, although Markov stabilization is a valid topological transformation of a braid, from the viewpoint of quantum-algebra representations the passage from three to four strands is nontrivial. It replaces the space $\operatorname{spin}(1/2,+)^{\otimes 3}$ by $\operatorname{spin}(1/2,+)^{\otimes 4}$, and at the root of unity $q=e^{i\pi/4}$ tensor products of representations are no longer completely reducible in the usual sense. In particular, for three strands we have
\begin{equation}
\operatorname{spin}(1/2,+)^{\otimes 3}
=
2\,\operatorname{spin}(1/2,+)
\oplus
\operatorname{spin}(3/2,+),
\end{equation}
and the additional representation $\operatorname{spin}(3/2,+)$ has zero quantum dimension and remains a separate direct summand. However, after adding one more fundamental representation, the structure of the decomposition changes substantially:
\begin{equation}\label{eq:4strand_real_decomp}
\operatorname{spin}(1/2,+)^{\otimes 4}
=
2\,\operatorname{spin}(0,+)
\oplus
2\,\operatorname{spin}(1,+)
\oplus
\operatorname{ind}(1,+).
\end{equation}
Here, the reducible but indecomposable representation
$\operatorname{ind}(1,+)$ appears for the first time; it also has zero quantum dimension.
Thus, the passage from three to four strands is not reduced to simply adding one more tensor factor: the structure of the zero-quantum-dimension sector changes qualitatively.

Therefore, the fact that the three-strand braiding matrices reproduce the corresponding matrices for Ising anyons does not by itself guarantee that the same correspondence will automatically remain valid for four strands after Markov stabilization. It is necessary to examine separately how the braiding operators act on the new direct summands, in particular on $\operatorname{ind}(1,+)$, and to verify that the resulting zero-quantum-dimension sector does not affect the observables under consideration.

One might also expect, in accordance with the decomposition \eqref{eq:4strand_real_decomp}, that the braiding matrices have a block-diagonal form with three blocks: a $2\times2$ block corresponding to $2\operatorname{spin}(0,+)$, a $2\times2$ block corresponding to $2\operatorname{spin}(1,+)$, and a $2\times2$ block corresponding to $\operatorname{ind}(1,+)$. We will show below that this expectation is incorrect. An interesting and rather unexpected feature, both from the viewpoint of the braiding matrices and of representation theory, is that the braiding operators mix certain vectors belonging to $2\operatorname{spin}(1,+)$ with vectors belonging to $\operatorname{ind}(1,+)$.

Thus, the four-strand case must be investigated directly. We now consider the explicit form of the braiding operators $\check R_1^{\bullet\bullet\bullet\bullet}$, $\check R_2^{\bullet\bullet\bullet\bullet}$, and $\check R_3^{\bullet\bullet\bullet\bullet}$ in the pairwise basis $(\bullet\bullet)(\bullet\bullet)$. To construct them, it is convenient to introduce several different orders of successive decomposition of the tensor product $\operatorname{spin}(1/2,+)^{\otimes4}$ and to compute the transition matrices between the corresponding highest-weight bases.

Before constructing these bases explicitly, let us compare the structure of the fourfold tensor product in three related descriptions. For four Ising anyons, the fusion rules \eqref{eq:ising_fusion_rules} give
\begin{equation}\label{eq:fourIsingAnyons}
\sigma^{\times 4}
=
2\,\mathbbm{1}
+
2\,\psi.
\end{equation}
Thus, the fusion space contains two channels with total charge $\mathbbm{1}$ and two channels with total charge $\psi$. Let us now consider the fourfold tensor product of the fundamental representation $\operatorname{spin}(1/2,+)$ for a generic value of the parameter $q$ that is not a root of unity. In this case, the representation theory is completely reducible, and the usual spin-addition rule gives
\begin{equation}
\operatorname{spin}(1/2,+)^{\otimes4}
=
2\,\operatorname{spin}(0,+)
\oplus
3\,\operatorname{spin}(1,+)
\oplus
\operatorname{spin}(2,+).
\end{equation}
Upon specializing to the root of unity $q=\exp(i\pi/4)$, the structure of the decomposition changes. One of the three summands $\operatorname{spin}(1,+)$ and the summand $\operatorname{spin}(2,+)$, which for generic $q$ are independent irreducible representations, cease to exist as two independent direct summands and instead combine into a single reducible but indecomposable representation:
\begin{equation}
\operatorname{spin}(1,+)
\oplus
\operatorname{spin}(2,+)
\quad\longrightarrow\quad
\operatorname{ind}(1,+).
\end{equation}
In this case,
\begin{equation}
	\dim\operatorname{ind}(1,+)=3+5=8.
\end{equation}

This merging of the two representations is also important for the subsequent construction of the basis. The representation $\operatorname{ind}(1,+)$ contains two linearly independent highest-weight vectors which, for generic $q$, would belong to the distinct summands $\operatorname{spin}(1,+)$ and $\operatorname{spin}(2,+)$. At the root of unity, however, they no longer define two independent direct summands: the corresponding parts of the space are combined into a single indecomposable structure.

One can verify directly that applying the lowering operator once to the highest-weight vector in the channel
$\operatorname{spin}(1/2,+)^{\otimes4}\to\operatorname{spin}(2,+)$
produces, at $q=\exp(i\pi/4)$, a highest-weight vector belonging to the $\operatorname{spin}(1,+)$ sector of the same fourfold tensor product.

On the other hand, the quantum dimension of $\operatorname{ind}(1,+)$ is zero. Therefore, if the zero-quantum-dimension sector is discarded when comparing with Ising anyons, one is left with
\begin{equation}
	\left. \operatorname{spin}(1/2,+)^{\otimes 4} \vphantom{\big|} \right|_{\mathrm{physical}}
	=
	2\,\operatorname{spin}(0,+)\oplus 2\,\operatorname{spin}(1,+).
\end{equation}
Here, the notation
$
\left. \operatorname{spin}(1/2,+)^{\otimes 4} \right|_{\text{physical}}
$
means that we retain only the physical sector, that is, we discard the indecomposable summand $\operatorname{ind}(1,+)$ with zero quantum dimension. In this sector, the decomposition of $\operatorname{spin}(1/2,+)^{\otimes4}$ exactly reproduces the decomposition of the fusion space of four Ising anyons~\eqref{eq:fourIsingAnyons}.

Thus, already at the level of the direct-sum decomposition, one can see that the entire difference from the Ising-anyon picture is concentrated in the additional indecomposable sector $\operatorname{ind}(1,+)$ of zero quantum dimension. We now need to determine how this sector appears in the explicit transition matrices and braiding operators.

To construct the matrices corresponding to the braiding operators
$\check R_1^{\bullet\bullet\bullet\bullet}$,
$\check R_2^{\bullet\bullet\bullet\bullet}$, and
$\check R_3^{\bullet\bullet\bullet\bullet}$
in the pairwise basis $(\bullet\bullet)(\bullet\bullet)$, we will also need two other bases:
$((\bullet\bullet)\bullet)\bullet$ and
$(\bullet(\bullet\bullet))\bullet$. These bases are represented graphically in Fig.~\ref{fig:four_strand_bases}. They correspond to different orders of successive decomposition of the tensor product $\operatorname{spin}(1/2,+)^{\otimes 4}$. The transition matrices between these bases will then allow us to express all braiding operators in the same basis $(\bullet\bullet)(\bullet\bullet)$.

\begin{figure}[h!]
	\centering
	\begin{tikzpicture}[
		fusion/.style={blue, very thick},
		ext/.style={font=\small},
		ch/.style={font=\small}
		]
		
		
		\coordinate (L1) at (-7.4,2.2);
		\coordinate (L2) at (-6.6,1.4);
		\coordinate (L3) at (-5.8,0.6);
		\coordinate (L4) at (-5.0,-0.2);
		
		\coordinate (LV1) at (-8.2,3.0);
		\coordinate (LV2) at (-6.6,3.0);
		\coordinate (LV3) at (-5.0,3.0);
		\coordinate (LV4) at (-3.4,3.0);
		
		\draw[fusion] (LV1) -- (L1);
		\draw[fusion] (LV2) -- (L1);
		
		\draw[fusion] (L1) -- (L2);
		
		\draw[fusion] (LV3) -- (L2);
		
		\draw[fusion] (L2) -- (L3);
		
		\draw[fusion] (LV4) -- (L3);
		
		\draw[fusion] (L3) -- (L4);
		
		\node[ext, above] at (LV1) {$V$};
		\node[ext, above] at (LV2) {$V$};
		\node[ext, above] at (LV3) {$V$};
		\node[ext, above] at (LV4) {$V$};
		
		\node[ch, anchor=north] at (-7.4,2.0) {$i$};
		\node[ch, anchor=north] at (-6.6,1.2) {$j$};
		\node[ch, anchor=north] at (-5.8,0.4) {$k$};
		
		\node[draw, ext] at (-5.8,-0.85)
		{$((\bullet\bullet)\bullet)\bullet$};

		
		\coordinate (P1) at (-1.2,2.1);
		\coordinate (P2) at (1.2,2.1);
		\coordinate (P3) at (0.0,0.9);
		\coordinate (P4) at (1.2,-0.3);
		
		\coordinate (PV1) at (-2.1,3.0);
		\coordinate (PV2) at (-0.3,3.0);
		\coordinate (PV3) at (0.3,3.0);
		\coordinate (PV4) at (2.1,3.0);
		
		\draw[fusion] (PV1) -- (P1);
		\draw[fusion] (PV2) -- (P1);
		
		\draw[fusion] (PV3) -- (P2);
		\draw[fusion] (PV4) -- (P2);
		
		\draw[fusion] (P1) -- (P3);
		\draw[fusion] (P2) -- (P3);
		
		\draw[fusion] (P3) -- (P4);
		
		\node[ext, above] at (PV1) {$V$};
		\node[ext, above] at (PV2) {$V$};
		\node[ext, above] at (PV3) {$V$};
		\node[ext, above] at (PV4) {$V$};
		
		\node[ch, anchor=north] at (-1.2,1.9) {$i$};
		\node[ch, anchor=north] at (1.2,1.9) {$j$};
		\node[ch, anchor=north] at (0.0,0.7) {$k$};
		
		\node[draw, ext] at (0.0,-0.85)
		{$(\bullet\bullet)(\bullet\bullet)$};

		
		\coordinate (R1) at (6.6,2.2);
		\coordinate (R2) at (5.8,1.4);
		\coordinate (R3) at (6.6,0.6);
		\coordinate (R4) at (7.4,-0.2);
		
		\coordinate (RV1) at (4.2,3.0);
		\coordinate (RV2) at (5.8,3.0);
		\coordinate (RV3) at (7.4,3.0);
		\coordinate (RV4) at (9.0,3.0);
		
		\draw[fusion] (RV2) -- (R1);
		\draw[fusion] (RV3) -- (R1);
		
		\draw[fusion] (R1) -- (R2);
		
		\draw[fusion] (RV1) -- (R2);
		
		\draw[fusion] (R2) -- (R3);
		
		\draw[fusion] (RV4) -- (R3);
		
		\draw[fusion] (R3) -- (R4);
		
		\node[ext, above] at (RV1) {$V$};
		\node[ext, above] at (RV2) {$V$};
		\node[ext, above] at (RV3) {$V$};
		\node[ext, above] at (RV4) {$V$};
		
		\node[ch, anchor=north] at (6.6,2.0) {$i$};
		\node[ch, anchor=north] at (5.8,1.2) {$j$};
		\node[ch, anchor=north] at (6.6,0.4) {$k$};
		
		\node[draw, ext] at (6.6,-0.85)
		{$(\bullet(\bullet\bullet))\bullet$};
		
	\end{tikzpicture}
	
	\caption{\centering
		Three decomposition orders of the tensor product $\operatorname{spin}(1/2,+)^{\otimes4}$:
		the left-associated basis $((\bullet\bullet)\bullet)\bullet$,
		the pairwise basis $(\bullet\bullet)(\bullet\bullet)$, and the right-associated
		basis $(\bullet(\bullet\bullet))\bullet$.
		Here, $i$, $j$, and $k$ denote the intermediate and final fusion channels, while $V$ denotes $\operatorname{spin}(1/2,+)$.
	}
	\label{fig:four_strand_bases}
\end{figure}

We now proceed to construct the highest-weight bases corresponding to the three decomposition orders of the tensor product $\operatorname{spin}(1/2,+)^{\otimes 4}$. For each of the three fusion trees, we denote the corresponding highest-weight vectors by $v$. The superscript indicates the type of basis, that is, the fusion order, while the subscripts specify the intermediate and final fusion channels denoted by $i$, $j$, and $k$ in Fig.~\ref{fig:four_strand_bases}. Thus, we use the following notation:
\begin{equation}
v^{((\bullet\bullet)\bullet)\bullet}_{i\to j\to k},
\qquad
v^{(\bullet\bullet)(\bullet\bullet)}_{i\times j\to k},
\qquad
v^{(\bullet(\bullet\bullet))\bullet}_{i\to j\to k}.
\end{equation}
Below, we write out the explicit formulas for the corresponding highest-weight vectors for each of the three types of bases. For the left-associated basis $((\bullet\bullet)\bullet)\bullet$, we obtain the following highest-weight vectors:
\begin{equation}
\begin{aligned}
	v^{((\bullet\bullet)\bullet)\bullet}_{0\to 1/2\to 0}
	&=
	\lvert\downarrow\uparrow\downarrow\uparrow\rangle
	-q\lvert\uparrow\downarrow\downarrow\uparrow\rangle
	-q\lvert\downarrow\uparrow\uparrow\downarrow\rangle
	+q^2\lvert\uparrow\downarrow\uparrow\downarrow\rangle,
	\\[2mm]
	v^{((\bullet\bullet)\bullet)\bullet}_{1\to 1/2\to 0}
	&=
	-[2]_q\lvert\downarrow\downarrow\uparrow\uparrow\rangle
	+q^2\lvert\downarrow\uparrow\downarrow\uparrow\rangle
	+q\lvert\uparrow\downarrow\downarrow\uparrow\rangle
	\\[2mm]
	&\quad
	+q\lvert\downarrow\uparrow\uparrow\downarrow\rangle
	+\lvert\uparrow\downarrow\uparrow\downarrow\rangle
	-q^2[2]_q\lvert\uparrow\uparrow\downarrow\downarrow\rangle,
	\\[2mm]
	v^{((\bullet\bullet)\bullet)\bullet}_{0\to 1/2\to 1}
	&=
	\lvert\downarrow\uparrow\uparrow\uparrow\rangle
	-q\lvert\uparrow\downarrow\uparrow\uparrow\rangle,
	\\[2mm]
	v^{((\bullet\bullet)\bullet)\bullet}_{1\to 1/2\to 1}
	&=
	q[2]_q\lvert\uparrow\uparrow\downarrow\uparrow\rangle
	-q^{-1}\lvert\uparrow\downarrow\uparrow\uparrow\rangle
	-\lvert\downarrow\uparrow\uparrow\uparrow\rangle,
	\\[2mm]
	v^{((\bullet\bullet)\bullet)\bullet}_{1\to 3/2\to 1}
	&=
	\lvert\downarrow\uparrow\uparrow\uparrow\rangle
	+q^{-1}\lvert\uparrow\downarrow\uparrow\uparrow\rangle
	+q^{-2}\lvert\uparrow\uparrow\downarrow\uparrow\rangle
	-q[3]_q\lvert\uparrow\uparrow\uparrow\downarrow\rangle,
	\\[2mm]
	v^{((\bullet\bullet)\bullet)\bullet}_{1\to 3/2\to 2}
	&=
	\lvert\uparrow\uparrow\uparrow\uparrow\rangle.
\end{aligned}
\end{equation}
For the right-associated basis $(\bullet(\bullet\bullet))\bullet$, we have
\begin{equation}
\begin{aligned}
	v^{(\bullet(\bullet\bullet))\bullet}_{0\to 1/2\to 0}
	&=
	-\lvert\downarrow\downarrow\uparrow\uparrow\rangle
	+q\lvert\downarrow\uparrow\downarrow\uparrow\rangle
	+q\lvert\uparrow\downarrow\uparrow\downarrow\rangle
	-q^2\lvert\uparrow\uparrow\downarrow\downarrow\rangle,
	\\[2mm]
	v^{(\bullet(\bullet\bullet))\bullet}_{1\to 1/2\to 0}
	&=
	\lvert\downarrow\downarrow\uparrow\uparrow\rangle
	+q^{-1}\lvert\downarrow\uparrow\downarrow\uparrow\rangle
	-q[2]_q\lvert\uparrow\downarrow\downarrow\uparrow\rangle
	\\[2mm]
	&\quad
	-q[2]_q\lvert\downarrow\uparrow\uparrow\downarrow\rangle
	+q^3\lvert\uparrow\downarrow\uparrow\downarrow\rangle
	+q^2\lvert\uparrow\uparrow\downarrow\downarrow\rangle,
	\\[2mm]
	v^{(\bullet(\bullet\bullet))\bullet}_{0\to 1/2\to 1}
	&=
	q\lvert\uparrow\uparrow\downarrow\uparrow\rangle
	-\lvert\uparrow\downarrow\uparrow\uparrow\rangle,
	\\[2mm]
	v^{(\bullet(\bullet\bullet))\bullet}_{1\to 1/2\to 1}
	&=
	[2]_q\lvert\downarrow\uparrow\uparrow\uparrow\rangle
	-q^2\lvert\uparrow\downarrow\uparrow\uparrow\rangle
	-q\lvert\uparrow\uparrow\downarrow\uparrow\rangle,
	\\[2mm]
	v^{(\bullet(\bullet\bullet))\bullet}_{1\to 3/2\to 1}
	&=
	\lvert\downarrow\uparrow\uparrow\uparrow\rangle
	+q^{-1}\lvert\uparrow\downarrow\uparrow\uparrow\rangle
	+q^{-2}\lvert\uparrow\uparrow\downarrow\uparrow\rangle
	-q[3]_q\lvert\uparrow\uparrow\uparrow\downarrow\rangle,
	\\[2mm]
	v^{(\bullet(\bullet\bullet))\bullet}_{1\to 3/2\to 2}
	&=
	\lvert\uparrow\uparrow\uparrow\uparrow\rangle.
\end{aligned}
\end{equation}
Let us arrange the vectors of the left- and right-associated bases in the following order:
\begin{equation}
	\mathbf v^{((\bullet\bullet)\bullet)\bullet}=
	\left(
	v^{((\bullet\bullet)\bullet)\bullet}_{0\to 1/2\to 0},
	v^{((\bullet\bullet)\bullet)\bullet}_{1\to 1/2\to 0},
	v^{((\bullet\bullet)\bullet)\bullet}_{0\to 1/2\to 1},
	v^{((\bullet\bullet)\bullet)\bullet}_{1\to 1/2\to 1},
	v^{((\bullet\bullet)\bullet)\bullet}_{1\to 3/2\to 1},
	v^{((\bullet\bullet)\bullet)\bullet}_{1\to 3/2\to 2}
	\right)^{\mathsf T},
\end{equation}
\begin{equation}
	\mathbf v^{(\bullet(\bullet\bullet))\bullet}=
	\left(
	v^{(\bullet(\bullet\bullet))\bullet}_{0\to 1/2\to 0},
	v^{(\bullet(\bullet\bullet))\bullet}_{1\to 1/2\to 0},
	v^{(\bullet(\bullet\bullet))\bullet}_{0\to 1/2\to 1},
	v^{(\bullet(\bullet\bullet))\bullet}_{1\to 1/2\to 1},
	v^{(\bullet(\bullet\bullet))\bullet}_{1\to 3/2\to 1},
	v^{(\bullet(\bullet\bullet))\bullet}_{1\to 3/2\to 2}
	\right)^{\mathsf T}.
\end{equation}
Among the six highest-weight vectors constructed above, two belong to the indecomposable representation $\operatorname{ind}(1,+)$. They can be identified conveniently by the intermediate fusion channel. Namely, the vectors
\begin{equation}
	v^{((\bullet\bullet)\bullet)\bullet}_{1\to 3/2\to 1},
	\qquad
	v^{((\bullet\bullet)\bullet)\bullet}_{1\to 3/2\to 2}
\end{equation}
in the left-associated basis and, respectively,
\begin{equation}
	v^{(\bullet(\bullet\bullet))\bullet}_{1\to 3/2\to 1},
	\qquad
	v^{(\bullet(\bullet\bullet))\bullet}_{1\to 3/2\to 2}
\end{equation}
in the right-associated basis pass through the intermediate representation
$\operatorname{spin}(3/2,+)$.

The final index in the notation
\begin{equation}
v^{((\bullet\bullet)\bullet)\bullet}_{1\to 3/2\to 1},
\qquad
v^{((\bullet\bullet)\bullet)\bullet}_{1\to 3/2\to 2}
\end{equation}
has a natural interpretation inherited from the classical case.
For generic $q$, the corresponding highest-weight vectors would belong to the representations $\operatorname{spin}(1,+)$ and $\operatorname{spin}(2,+)$, respectively. Thus, the final index $1$ or $2$ indicates the representation to which the given highest-weight vector would belong in the classical decomposition.

At $q=e^{i\pi/4}$, these two channels no longer correspond to two separate direct summands: both highest-weight vectors belong to the same indecomposable representation $\operatorname{ind}(1,+)$. The remaining four highest-weight vectors $v^{((\bullet\bullet)\bullet)\bullet}_{i\to j\to k}$ (or $v^{(\bullet(\bullet\bullet))\bullet}_{i\to j\to k}$), for which the intermediate channel is $j=1/2$, belong to the physical sector
\begin{equation}
	2\,\operatorname{spin}(0,+)
	\oplus
	2\,\operatorname{spin}(1,+).
\end{equation}

For four Ising anyons, the transition between the bases
$((\bullet\bullet)\bullet)\bullet$ and $(\bullet(\bullet\bullet))\bullet$
acts independently in the two possible final fusion channels,
$\mathbbm{1}$ and $\psi$. Therefore, the corresponding transition matrix has a block-diagonal form
\begin{equation}
	\hat{\mathcal{F}}_{\mathrm{Ising}}^{((\bullet\bullet)\bullet)\bullet \to (\bullet(\bullet\bullet))\bullet}
	=
	\hat{\mathcal{F}}^{\sigma\sigma\sigma}_{\sigma}\oplus \hat{\mathcal{F}}^{\sigma\sigma\sigma}_{\sigma},
	\qquad
	\hat{\mathcal{F}}^{\sigma\sigma\sigma}_{\sigma}=
	\frac{1}{\sqrt{2}}
	\begin{pmatrix}
		1&1\\
		1&-1
	\end{pmatrix}.
\end{equation}
For representations of the quantum algebra $U_q(\mathfrak{sl}_2)$ at
$q=e^{i\pi/4}$, one obtains a six-dimensional transition matrix
\begin{equation}
	\check{\mathsf{F}}^{((\bullet\bullet)\bullet)\bullet\to (\bullet(\bullet\bullet))\bullet}
	=
	\check{\mathsf{F}}^{\bullet(\bullet\bullet) \to (\bullet\bullet)\bullet}|_{\text{phys}}\oplus \check{\mathsf{F}}^{\bullet(\bullet\bullet) \to (\bullet\bullet)\bullet}|_{\text{phys}}\oplus \operatorname{I}_{2}.
\end{equation}
The first two blocks act on the physical sector
\begin{equation}
2\,\operatorname{spin}(0,+)
\oplus
2\,\operatorname{spin}(1,+),
\end{equation}
whereas the last block $I_2$ acts trivially on the two highest-weight vectors belonging to the additional zero-quantum-dimension representation $\operatorname{ind}(1,+)$. Therefore,
\begin{equation}
	\check{\mathsf{F}}^{((\bullet\bullet)\bullet)\bullet\to (\bullet(\bullet\bullet))\bullet}|_{\text{phys}}
	=
	\check{\mathsf{F}}^{\bullet(\bullet\bullet) \to (\bullet\bullet)\bullet}|_{\text{phys}}\oplus \check{\mathsf{F}}^{\bullet(\bullet\bullet) \to (\bullet\bullet)\bullet}|_{\text{phys}}
	=
	\hat{\mathcal{F}}^{\sigma\sigma\sigma}_{\sigma}\oplus \hat{\mathcal{F}}^{\sigma\sigma\sigma}_{\sigma}
	=
	\hat{\mathcal{F}}_{\mathrm{Ising}}^{((\bullet\bullet)\bullet)\bullet \to (\bullet(\bullet\bullet))\bullet}.
\end{equation}
Thus, the physical sector of the transition matrix between the bases
$((\bullet\bullet)\bullet)\bullet$ and $(\bullet(\bullet\bullet))\bullet$
exactly reproduces the corresponding result for four Ising anyons. We also note that, for computational convenience, the highest-weight vectors were chosen so that
\begin{equation}
	\check{\mathsf{F}}^{((\bullet\bullet)\bullet)\bullet\to (\bullet(\bullet\bullet))\bullet}
	=
	\left(
	\check{\mathsf{F}}^{((\bullet\bullet)\bullet)\bullet\to (\bullet(\bullet\bullet))\bullet}
	\right)^{-1}
	=
	\check{\mathsf{F}}^{(\bullet(\bullet\bullet))\bullet\to ((\bullet\bullet)\bullet)\bullet}
\end{equation}

We now construct the pairwise basis $(\bullet\bullet)(\bullet\bullet)$. We obtain
\begin{equation}
\begin{aligned}
	v^{(\bullet\bullet)(\bullet\bullet)}_{0\times 0\to 0}
	&=
	\lvert\downarrow\uparrow\downarrow\uparrow\rangle
	-q\lvert\downarrow\uparrow\uparrow\downarrow\rangle
	-q\lvert\uparrow\downarrow\downarrow\uparrow\rangle
	+q^2\lvert\uparrow\downarrow\uparrow\downarrow\rangle,
	\\[2mm]
	v^{(\bullet\bullet)(\bullet\bullet)}_{1\times 1\to 0}
	&=
	-[2]_q\lvert\downarrow\downarrow\uparrow\uparrow\rangle
	+q^2\lvert\downarrow\uparrow\downarrow\uparrow\rangle
	+q\lvert\downarrow\uparrow\uparrow\downarrow\rangle
	\\[2mm]
	&\quad
	+q\lvert\uparrow\downarrow\downarrow\uparrow\rangle
	+\lvert\uparrow\downarrow\uparrow\downarrow\rangle
	-q^2[2]_q\lvert\uparrow\uparrow\downarrow\downarrow\rangle,
	\\[2mm]
	v^{(\bullet\bullet)(\bullet\bullet)}_{0\times 1\to 1}
	&=
	\lvert\downarrow\uparrow\uparrow\uparrow\rangle
	-q\lvert\uparrow\downarrow\uparrow\uparrow\rangle,
	\\[2mm]
	v^{(\bullet\bullet)(\bullet\bullet)}_{1\times 0\to 1}
	&=
	\lvert\uparrow\uparrow\downarrow\uparrow\rangle
	-q\lvert\uparrow\uparrow\uparrow\downarrow\rangle,
	\\[2mm]
	v^{(\bullet\bullet)(\bullet\bullet)}_{1\times 1\to 1}
	&=
	-\lvert\downarrow\uparrow\uparrow\uparrow\rangle
	-q^{-1}\lvert\uparrow\downarrow\uparrow\uparrow\rangle
	+q^2\lvert\uparrow\uparrow\downarrow\uparrow\rangle
	+q\lvert\uparrow\uparrow\uparrow\downarrow\rangle,
	\\[2mm]
	v^{(\bullet\bullet)(\bullet\bullet)}_{1\times 1\to 2}
	&=
	\lvert\uparrow\uparrow\uparrow\uparrow\rangle.
\end{aligned}
\end{equation}
Let us arrange the vectors of the pairwise basis in the following order:
\begin{equation}
	\mathbf v^{(\bullet\bullet)(\bullet\bullet)}=
	\left(
	v^{(\bullet\bullet)(\bullet\bullet)}_{0\times 0\to 0},
	v^{(\bullet\bullet)(\bullet\bullet)}_{1\times 1\to 0},
	v^{(\bullet\bullet)(\bullet\bullet)}_{0\times 1\to 1},
	v^{(\bullet\bullet)(\bullet\bullet)}_{1\times 0\to 1},
	v^{(\bullet\bullet)(\bullet\bullet)}_{1\times 1\to 1},
	v^{(\bullet\bullet)(\bullet\bullet)}_{1\times 1\to 2}
	\right)^{\mathsf T}.
\end{equation}
Among the vectors of the pairwise basis, two belong to the indecomposable representation $\operatorname{ind}(1,+)$:
\begin{equation}
	v^{(\bullet\bullet)(\bullet\bullet)}_{1\times 1\to 1},
	\qquad
	v^{(\bullet\bullet)(\bullet\bullet)}_{1\times 1\to 2}.
\end{equation}
This can be seen directly from the correspondence with the fusion rules of Ising anyons. Recall that
\begin{equation}
\operatorname{spin}(0,+)\longleftrightarrow \mathbbm{1},
\qquad
\operatorname{spin}(1,+)\longleftrightarrow \psi.
\end{equation}
Therefore, the physical part of the product of the two intermediate channels
$i=j=1$ must reproduce the fusion rule
\begin{equation}
	\psi\times\psi=\mathbbm{1},
\end{equation}
that is, in the notation of quantum-algebra representations
\begin{equation}
	1\times1\longrightarrow0.
\end{equation}
Indeed, the highest-weight vector
\begin{equation}
v^{(\bullet\bullet)(\bullet\bullet)}_{1\times1 \to 0}
\end{equation}
belongs to the physical representation $\operatorname{spin}(0,+)$.
The remaining highest-weight vectors arising from the fusion of the same intermediate channels $1\times1$, namely
\begin{equation}
v^{(\bullet\bullet)(\bullet\bullet)}_{1\times 1\to 1},
\qquad
v^{(\bullet\bullet)(\bullet\bullet)}_{1\times 1\to 2},
\end{equation}
do not have corresponding fusion channels in the Ising-anyon theory. A direct calculation shows that these highest-weight vectors are highest-weight vectors of the representation $\operatorname{ind}(1,+)$. Thus, the channel
\begin{equation}
1\times1\longrightarrow1
\end{equation}
should not be interpreted as an additional physical summand
$\operatorname{spin}(1,+)$.

Let us now consider the transition between the pairwise basis
$(\bullet\bullet)(\bullet\bullet)$ and the left-associated basis
$((\bullet\bullet)\bullet)\bullet$. For four Ising anyons, the corresponding transition matrix can be obtained from the pentagon equation for the $F$-matrices. With the conventions for the $F$-symbols adopted above, the result takes a particularly simple form:
\begin{equation}
	\hat{\mathcal{F}}_{\mathrm{Ising}}^{\,(\bullet\bullet)(\bullet\bullet)
		\rightarrow ((\bullet\bullet)\bullet)\bullet}
	=
	I_4.
\end{equation}
Thus, in the chosen bases, the transition between
$(\bullet\bullet)(\bullet\bullet)$ and $((\bullet\bullet)\bullet)\bullet$
for four Ising anyons is the identity transformation.

At first sight, one might expect the corresponding transition matrix
for representations of the quantum algebra to be the identity as well,
now of size $6\times6$. This expectation is based on two observations.

First, the physical sectors of the two bases should reproduce the fusion
space of four Ising anyons and therefore, up to possible phase factors
multiplying individual highest-weight vectors, should be related by the
same transition matrix. Since for Ising anyons this matrix is $I_4$, it is
natural to expect the transition between the corresponding physical
highest-weight vectors to be the identity as well.

Second, one might assume that the highest-weight vectors of the physical representations do not mix, under a change of basis, with the highest-weight vectors belonging to $\operatorname{ind}(1,+)$. Under this assumption, the transition matrix would decompose into independent physical and nonphysical blocks and, with the chosen normalization of the basis vectors, would take the form $I_6$.

However, a direct calculation shows that the transition matrix between the bases $(\bullet\bullet)(\bullet\bullet)$ and $((\bullet\bullet)\bullet)\bullet$ has a more complicated structure: changing the decomposition order leads to mixing between the physical sector and the $\operatorname{ind}(1,+)$ sector. For the orderings of the basis vectors introduced above, direct comparison gives
\begin{equation}
	\mathbf{v}^{((\bullet\bullet)\bullet)\bullet}
	=
	\check{\mathsf{F}}^{(\bullet\bullet)(\bullet\bullet) \to ((\bullet\bullet)\bullet)\bullet}\; \mathbf{v}^{(\bullet\bullet)(\bullet\bullet)},
\end{equation}
where
\begin{equation}
	\check{\mathsf{F}}^{(\bullet\bullet)(\bullet\bullet) \to ((\bullet\bullet)\bullet)\bullet}
	=
	\left(
	\begin{array}{cccc|cc}
		1&0&0&0&0&0\\
		0&1&0&0&0&0\\
		0&0&1&0&0&0\\
		0&0&0&1&1&0\\
		\hline
		0&0&0&0&-1&0\\
		0&0&0&0&0&1
	\end{array}
	\right).
	\label{eq:LP_transition}
\end{equation}
In particular, the first three physical basis vectors coincide:
\begin{equation}
	v^{((\bullet\bullet)\bullet)\bullet}_{0\to 1/2\to 0}
	=
	v^{(\bullet\bullet)(\bullet\bullet)}_{0\times 0\to 0},
	\qquad
	v^{((\bullet\bullet)\bullet)\bullet}_{1\to 1/2\to 0}
	=
	v^{(\bullet\bullet)(\bullet\bullet)}_{1\times 1\to 0},
\end{equation}
\begin{equation}
	v^{((\bullet\bullet)\bullet)\bullet}_{0\to 1/2\to 1}
	=
	v^{(\bullet\bullet)(\bullet\bullet)}_{0\times 1\to 1},
\end{equation}
whereas for the fourth physical vector we obtain
\begin{equation}
	v^{((\bullet\bullet)\bullet)\bullet}_{1\to 1/2\to 1}
	=
	v^{(\bullet\bullet)(\bullet\bullet)}_{1\times 0\to 1}
	+
	v^{(\bullet\bullet)(\bullet\bullet)}_{1\times 1\to 1}.
	\label{eq:physical_ind_mixing}
\end{equation}
The second vector on the right-hand side of
\eqref{eq:physical_ind_mixing} belongs to
$\operatorname{ind}(1,+)$. The physical $4\times4$ block of the matrix
\eqref{eq:LP_transition} has the form
\begin{equation}
	\left.\check{\mathsf{F}}^{(\bullet\bullet)(\bullet\bullet) \to ((\bullet\bullet)\bullet)\bullet} \right|_{\mathrm{phys}}
	=
	I_4,
\end{equation}
and fully reproduces the result for four Ising anyons.
However, the full six-dimensional transition matrix is not the identity:
one of the physical highest-weight vectors in the two bases differs by a highest-weight vector belonging to $\operatorname{ind}(1,+)$.
For this reason, the zero-quantum-dimension sector must be retained when constructing the full transition matrix. This is an essential difference from the result obtained in the previous section~\ref{sec:three_strands}. It shows that the nonphysical sector can no longer simply be discarded as it was before.

As above, the highest-weight vectors were chosen, without loss of generality, so that the following relation holds
\begin{equation}
	\check{\mathsf{F}}^{(\bullet\bullet)(\bullet\bullet) \to ((\bullet\bullet)\bullet)\bullet}
	=
	\left(
	\check{\mathsf{F}}^{(\bullet\bullet)(\bullet\bullet) \to ((\bullet\bullet)\bullet)\bullet}
	\right)^{-1}
	=
	\check{\mathsf{F}}^{((\bullet\bullet)\bullet)\bullet \to (\bullet\bullet)(\bullet\bullet)}
\end{equation}
for computational convenience.

Equation~\eqref{eq:LP_transition} gives rise to an apparent contradiction. The fact that, under a simple change in the decomposition order, a highest-weight vector of a physical representation acquires a component belonging to $\operatorname{ind}(1,+)$ is somewhat unexpected. Indeed, if the physical sector is to correspond to the state space of Ising anyons, one would naturally expect it to be closed under all physically meaningful transformations. Mixing with the zero-quantum-dimension sector, on the contrary, appears at first sight to cast doubt on this correspondence.

However, we will show below that this contradiction is only apparent. Although, at the level of the full six-dimensional matrices, the braiding operators do mix the physical sector with $\operatorname{ind}(1,+)$, after performing the corresponding braid closure the additional sector does not alter the physical results. In particular, we will show that the probability distributions of the observables coincide with the corresponding distributions for Ising anyons. Thus, the agreement with the Ising-anyon theory should not be sought at the level of each intermediate basis transformation separately, but rather at the level of the quantities obtained after the full action of the braiding operators followed by the braid closure or the plat closure.

We now express the braiding operators
$\check R_1^{\bullet\bullet\bullet\bullet}$, $\check R_2^{\bullet\bullet\bullet\bullet}$, and $\check R_3^{\bullet\bullet\bullet\bullet}$ in the pairwise basis
$(\bullet\bullet)(\bullet\bullet)$. The operators $\check R_1^{(\bullet\bullet)(\bullet\bullet)}$ and $\check R_3^{(\bullet\bullet)(\bullet\bullet)}$ act on the first and second strands, and on the third and fourth strands, respectively. Therefore, in the pairwise basis, where both of these pairs are fused directly, their matrices are diagonal. Recall that the braiding matrix acts on the fusion vectors
$\operatorname{spin}(1/2,+)^{\otimes 2} \to \operatorname{spin}(0,+)$
and
$\operatorname{spin}(1/2,+)^{\otimes 2} \to \operatorname{spin}(1,+)$
with eigenvalues $-q^{-3/2}$ and $q^{1/2}$, respectively. For the chosen ordering of the basis vectors, we obtain
\begin{equation}
	\check R_1^{(\bullet\bullet)(\bullet\bullet)}
	=
	\operatorname{diag}
	\left(
	-q^{-3/2},q^{1/2},-q^{-3/2},q^{1/2},q^{1/2},q^{1/2}
	\right),
	\label{eq:R1_pair_basis}
\end{equation}
since the first index in the pairwise basis specifies the fusion channel of the first and second strands. Similarly,
\begin{equation}
	\check R_3^{(\bullet\bullet)(\bullet\bullet)}
	=
	\operatorname{diag}
	\left(
	-q^{-3/2},q^{1/2},q^{1/2},-q^{-3/2},q^{1/2},q^{1/2}
	\right),
	\label{eq:R3_pair_basis}
\end{equation}
where the eigenvalue is determined by the second index, that is, by the fusion channel of the third and fourth strands. To construct $\check R_2^{(\bullet\bullet)(\bullet\bullet)}$, one must pass to the right-associated basis $(\bullet(\bullet\bullet))\bullet$, in which the second and third strands are fused directly. In this basis,
\begin{equation}
	\check R_2^{(\bullet(\bullet\bullet))\bullet}
	=
	\operatorname{diag}
	\left(
	-q^{-3/2},q^{1/2},-q^{-3/2},q^{1/2},q^{1/2},q^{1/2}
	\right).
	\label{eq:R2_right_basis}
\end{equation}
The transition matrix from the right-associated basis $(\bullet(\bullet\bullet))\bullet$ to the pairwise basis $(\bullet\bullet)(\bullet\bullet)$ is
\begin{equation}
	\check{\mathsf F}^{(\bullet(\bullet\bullet))\bullet\to (\bullet\bullet)(\bullet\bullet)}
	=
	\check{\mathsf F}^{((\bullet\bullet)\bullet)\bullet \to (\bullet\bullet)(\bullet\bullet)} \; \check{\mathsf F}^{(\bullet(\bullet\bullet))\bullet \to ((\bullet\bullet)\bullet)\bullet}.
	\label{eq:FPR}
\end{equation}

Thus, the matrix of the operator $\check R_2^{(\bullet\bullet)(\bullet\bullet)}$ in the pairwise basis $(\bullet\bullet)(\bullet\bullet)$ takes the form
\begin{equation}
		\check{R}_2^{(\bullet\bullet)(\bullet\bullet)}
		=
		\frac{e^{i\pi/8}}{2}
		\begin{pmatrix}
			1+i & i-1 & 0   & 0   & 0   & 0 \\
			i-1 & 1+i & 0   & 0   & 0   & 0 \\
			0   & 0   & 1+i & i-1 & i-1 & 0 \\
			0   & 0   & i-1 & 1+i & i-1 & 0 \\
			0   & 0   & 0   & 0   & 2   & 0 \\
			0   & 0   & 0   & 0   & 0   & 2
		\end{pmatrix}
	.
	\label{eq:R2_pairwise}
\end{equation}
Among the three braiding matrices, only $\check R^{(\bullet\bullet)(\bullet\bullet)}_2$ is non-diagonal. For the chosen ordering of the basis vectors, its structure is
\begin{equation}
	\check R^{(\bullet\bullet)(\bullet\bullet)}_2
	=
	\left(
	\begin{array}{c|c}
		\begin{matrix}
			\text{physical part}\\[-1mm]
			4\times4
		\end{matrix}
		&
		\begin{matrix}
			\text{mixing with}\\[-1mm]
			\operatorname{ind}(1,+)\\[-1mm]
			4\times2
		\end{matrix}
		\\ \hline
		\begin{matrix}
			\text{zeros}\\[-1mm]
			2\times4
		\end{matrix}
		&
		\begin{matrix}
			\text{nonphysical part}\\[-1mm]
			2\times2
		\end{matrix}
	\end{array}
	\right).
	\label{eq:R2_block_structure}
\end{equation}
Moreover, as before, its physical part differs from the corresponding braiding matrix for Ising anyons only by replacing the matrix with its inverse and multiplying by an overall phase. Thus, it remains only to show that only the physical part contributes to the observables. This will be demonstrated below for both the plat closure and the braid closure.

\subsubsection{Four-strand plat closure}\label{sec:four_strands_plat}

We now consider the four-strand plat closure. In this case, it is particularly important that the only non-diagonal braiding operator, $\check R^{(\bullet\bullet)(\bullet\bullet)}_2$, does not mix the vectors of the representations $2\operatorname{spin}(0,+)$ with the remaining vectors in the decomposition \eqref{eq:4strand_real_decomp}. Therefore, after an appropriate grouping of the basis vectors, its matrix decomposes into two independent blocks:
\begin{equation}
	\check R^{(\bullet\bullet)(\bullet\bullet)}_2
	=
	\begin{pmatrix}
		\left. \check R_2^{(\bullet\bullet)(\bullet\bullet)} \right|_{2\operatorname{spin}(0,+)} & 0\\
		0 & \left. \check R_2^{(\bullet\bullet)(\bullet\bullet)} \right|_{\text{not } 2\operatorname{spin}(0,+)}
	\end{pmatrix}.
	\label{eq:R2_plat_blocks}
\end{equation}

The first block acts on the two physical highest-weight vectors with total charge $\operatorname{spin}(0,+)$. The second block contains two physical highest-weight vectors with total charge $\operatorname{spin}(1,+)$ and two highest-weight vectors belonging to $\operatorname{ind}(1,+)$. It is precisely in this second block that the mixing between the physical and nonphysical sectors described above occurs. For the plat closure, the initial state is the state in which the first and second strands, as well as the third and fourth strands, are created from the vacuum:
\begin{equation}
	\lvert\varnothing\otimes\varnothing\rangle
	=
	v^{(\bullet\bullet)(\bullet\bullet)}_{0\times 0\to 0}.
	\label{eq:plat_initial_state}
\end{equation}
For the chosen ordering of the basis vectors, this state is represented by the vector
\begin{equation}
	\lvert\varnothing\otimes\varnothing\rangle
	=
	\begin{pmatrix}
		1&0&0&0&0&0
	\end{pmatrix}^{\mathsf T}.
\end{equation}
Thus, the initial state lies entirely within the two-dimensional block with total charge $\operatorname{spin}(0,+)$. The operators $\check R^{(\bullet\bullet)(\bullet\bullet)}_1$, $\check R^{(\bullet\bullet)(\bullet\bullet)}_2$, and $\check R^{(\bullet\bullet)(\bullet\bullet)}_3$ preserve the total charge $\operatorname{spin}(0,+)$. Therefore, for an arbitrary braid word $\check{B}^{(\bullet\bullet)(\bullet\bullet)}_\text{pl}$ the state
\begin{equation}
	\check{B}^{(\bullet\bullet)(\bullet\bullet)}_\text{pl}\lvert\varnothing\otimes\varnothing\rangle
\end{equation}
remains in the two-dimensional physical sector with total charge $\operatorname{spin}(0,+)$. In particular, it never acquires a component belonging to $\operatorname{ind}(1,+)$.

The braiding matrices $\check R_{i}^{(\bullet\bullet)(\bullet\bullet)}$ coincide with the corresponding braiding matrices for Ising anyons up to the choice of braiding orientation and an overall phase factor. Therefore, the probability density
\begin{equation}\label{eq:plat_amplitude_four}
	\langle\varnothing\otimes\varnothing\rvert \check{B}^{(\bullet\bullet)(\bullet\bullet)}_\text{pl}\lvert\varnothing\otimes\varnothing\rangle
\end{equation}
will, up to a phase factor, give the same probability density as the corresponding knot with opposite crossings in the Ising-anyon theory. Therefore, for the four-strand plat closure, the mixing of physical highest-weight vectors with $\operatorname{ind}(1,+)$ found above does not manifest itself at all: the initial state lies in a different invariant block. The probability densities for all measurements defined in this two-dimensional state space coincide with the corresponding probability densities for four Ising anyons up to an overall phase.

\subsubsection{Four-strand braid closure}\label{sec:four_strands_braid}

We now consider the ordinary closure of a four-strand braid. In the chosen pairwise basis, the braiding operators have the following common block structure
\begin{equation}
	\check R^{(\bullet\bullet)(\bullet\bullet)}_i=
	\begin{pmatrix}
		\left.\check{R}^{(\bullet\bullet)(\bullet\bullet)}_i \right|_{\text{phys}} & \left.\check{R}^{(\bullet\bullet)(\bullet\bullet)}_i \right|_{\text{mix}}\\
		0 & \left.\check{R}^{(\bullet\bullet)(\bullet\bullet)}_i \right|_{\text{nonphys}}
	\end{pmatrix},
	\qquad i=1,2,3,
	\label{eq:R_block_structure_braid_closure}
\end{equation}
where $\check{R}^{(\bullet\bullet)(\bullet\bullet)}_i |_{\text{phys}}$ is the physical $4\times4$ block, $\check{R}^{(\bullet\bullet)(\bullet\bullet)}_i |_{\text{mix}}$ is the $4\times2$ block describing the mixing with $\operatorname{ind}(1,+)$, and $\check{R}^{(\bullet\bullet)(\bullet\bullet)}_i |_{\text{nonphys}}$ is the nonphysical $2\times2$ block of $\operatorname{ind}(1,+)$. For $\check R^{(\bullet\bullet)(\bullet\bullet)}_1$ and $\check R^{(\bullet\bullet)(\bullet\bullet)}_3$, the block $\check{R}^{(\bullet\bullet)(\bullet\bullet)}_i |_{\text{mix}}$ vanishes, whereas for $\check R^{(\bullet\bullet)(\bullet\bullet)}_2$ it is nonzero. In all three cases, the nonphysical block is proportional to the identity matrix:
\begin{equation}
	\left.\check{R}^{(\bullet\bullet)(\bullet\bullet)}_i \right|_{\text{nonphys}}\sim I_2.
	\label{eq:unphysical_scalar_block}
\end{equation}
This structure is preserved under matrix multiplication. Therefore, an arbitrary braid word $\check{B}_\text{br}$ is represented by a matrix of the same form:
\begin{equation}
	\check{B}^{(\bullet\bullet)(\bullet\bullet)}_\text{br}=
	\begin{pmatrix}
		\left.\check{B}^{(\bullet\bullet)(\bullet\bullet)}_\text{br}\right|_{\text{phys}} & \left.\check{B}^{(\bullet\bullet)(\bullet\bullet)}_\text{br}\right|_{\text{mix}} \\[2mm]
		0 & \left.\check{B}^{(\bullet\bullet)(\bullet\bullet)}_\text{br}\right|_{\text{nonphys}}
	\end{pmatrix}.
	\label{eq:braid_word_block_structure}
\end{equation}
where the part $\check{B}^{(\bullet\bullet)(\bullet\bullet)}_{\mathrm{br}}|_{\mathrm{phys}}$ of the matrix $\check{B}^{(\bullet\bullet)(\bullet\bullet)}_{\mathrm{br}}$ consists only of products of the physical parts $\check{R}^{(\bullet\bullet)(\bullet\bullet)}_i|_{\mathrm{phys}}$ of the braiding operators and $\check{B}^{(\bullet\bullet)(\bullet\bullet)}_{\mathrm{br}}|_{\mathrm{nonphys}}$ consists only of products of $\check{R}_i^{(\bullet\bullet)(\bullet\bullet)}|_{\mathrm{nonphys}}$. The observable corresponding to the braid closure is defined by the quantum trace of the matrix $\check{B}^{(\bullet\bullet)(\bullet\bullet)}_\text{br}$
\begin{equation}
	\operatorname{Tr}_q \check{B}^{(\bullet\bullet)(\bullet\bullet)}_\text{br}
	=
	\operatorname{Tr}
	\left(
	\mathfrak{D}^{(\bullet\bullet)(\bullet\bullet)}_q \check{B}^{(\bullet\bullet)(\bullet\bullet)}_\text{br}
	\right),
	\label{eq:quantum_trace_four_strands}
\end{equation}
where the diagonal matrix $\mathfrak{D}^{(\bullet\bullet)(\bullet\bullet)}_q$ contains the quantum dimensions of the corresponding final channels:
\begin{equation}
	\mathfrak{D}^{(\bullet\bullet)(\bullet\bullet)}_q=
	\operatorname{diag}
	\left(
	[1]_q,[1]_q,[3]_q,[3]_q,[3]_q,[5]_q
	\right).
\end{equation}
At $q=e^{i\pi/4}$,
\begin{equation}
	[1]_q=[3]_q=1,
	\qquad
	[5]_q=-1,
\end{equation}
therefore,
\begin{equation}
	\mathfrak{D}^{(\bullet\bullet)(\bullet\bullet)}_q=
	\operatorname{diag}
	\left(
	1,1,1,1,1,-1
	\right)
	=
	\begin{pmatrix}
		I_4&0\\
		0&\left.\mathfrak{D}^{(\bullet\bullet)(\bullet\bullet)}\right|_{\text{nonphys}}
	\end{pmatrix},
	\qquad
	\left.\mathfrak{D}^{(\bullet\bullet)(\bullet\bullet)}\right|_{\text{nonphys}}
	=
	\begin{pmatrix}
		1&0\\
		0&-1
	\end{pmatrix}.
	\label{eq:Dq_four_strands}
\end{equation}
The last two coefficients correspond to the two highest-weight vectors of
$\operatorname{ind}(1,+)$ which, for generic $q$, would belong to the channels
$\operatorname{spin}(1,+)$ and $\operatorname{spin}(2,+)$. Their quantum dimensions satisfy
\begin{equation}
	[3]_q+[5]_q=1-1=0,
\end{equation}
which reproduces the vanishing quantum dimension of
$\operatorname{ind}(1,+)$. Therefore, the nonphysical block $\check{B}^{(\bullet\bullet)(\bullet\bullet)}_\text{br}|_{\text{nonphys}}$ does not contribute to the trace at all:
\begin{equation}
	\operatorname{Tr}_q \check{B}^{(\bullet\bullet)(\bullet\bullet)}_\text{br}
	=
	\operatorname{Tr}\left(\check{B}^{(\bullet\bullet)(\bullet\bullet)}_{\mathrm{br}}|_{\mathrm{phys}}\right)
	+
	\operatorname{Tr}\left(\mathfrak{D}^{(\bullet\bullet)(\bullet\bullet)}|_{\text{nonphys}} \check{B}^{(\bullet\bullet)(\bullet\bullet)}_{\mathrm{br}}|_{\mathrm{nonphys}}\right)
	=
	\operatorname{Tr}\left( \check{B}^{(\bullet\bullet)(\bullet\bullet)}_{\mathrm{br}}|_{\mathrm{phys}}\right),
	\label{eq:ind_vanishes_quantum_trace}
\end{equation}
where the last equality holds because the matrix
$\check{B}^{(\bullet\bullet)(\bullet\bullet)}_{\mathrm{br}}|_{\mathrm{nonphys}}$ is proportional to the identity, while
$\mathfrak{D}^{(\bullet\bullet)(\bullet\bullet)}|_{\text{nonphys}}=\operatorname{diag}(1,-1)$.
Thus, although at the level of the braiding-operator matrices the physical highest-weight vectors may mix with the highest-weight vectors of $\operatorname{ind}(1,+)$, this mixing does not affect the quantum trace. The upper-triangular structure of the matrices prevents the mixing block from contributing to the diagonal, while the contribution of the nonphysical block itself vanishes because of the zero quantum dimension of $\operatorname{ind}(1,+)$.

Therefore, for the ordinary closure of a four-strand braid, the observable obtained from representations of $U_q(\mathfrak{sl}_2)$ at $q=e^{i\pi/4}$ also coincides with the corresponding observable for Ising anyons up to an overall phase factor and the choice of braiding orientation
\begin{equation}\label{eq:trace_vs_ising_four}
	\operatorname{Tr}_q \check{B}^{(\bullet\bullet)(\bullet\bullet)}_\text{br}
	=
	e^{i\Phi}
	\operatorname{Tr}
	\mathcal B^{(\bullet\bullet)(\bullet\bullet)}_{\mathrm{Ising}}.
\end{equation}

\subsection{Arbitrary number of strands and the zero-quantum-dimension sector}
\label{sec:arbitrary-strands}

The work of Etingof and Ostrik~\cite{EtingofOstrik2022} presents, in
categorical language, a general approach to discarding negligible
homomorphisms and representations. A homomorphism
$f:\mathfrak{W}_1\to\mathfrak{W}_2$ is called negligible if
\begin{equation}
	\label{eq:s34:negligible}
	\operatorname{Tr}_{q, \mathfrak{W}_1}(g\circ f)=0
	\qquad\text{for every homomorphism  }g:\mathfrak{W}_2\to \mathfrak{W}_1.
\end{equation}
Here the quantum trace is the one adopted in the preceding subsections. A
representation is called negligible if its identity map is negligible.
In other words, the quantum trace of every endomorphism of such a
representation vanishes. For the finite-dimensional indecomposable
representations considered here, this condition is equivalent to the
vanishing of the quantum dimension (Lemma~\ref{lem:indecomposable-endomorphism} below). A finite direct sum of negligible
representations is again negligible.

Applied to our problem, these results explain why, as the number of strands
grows, the $\operatorname{ind}$-type representations and their mixing with
the physical representations do not contribute to the quantities computed
by the braid closure and by the plat closure. We formulate the required
properties in the language of representations of the quantum algebra:
\begin{enumerate}
	\item A linear combination of negligible homomorphisms with the same
	domain and the same target is negligible.
	\item The composition of a negligible homomorphism with arbitrary
	homomorphisms on the left and on the right remains negligible, whenever
	the composition is defined.
	\item The tensor product of a negligible homomorphism with an arbitrary
	homomorphism, in either order, remains negligible.
	\item Any homomorphism that factors through a negligible representation
	is negligible. By such a factorization we mean the successive application
	of a homomorphism into this representation and a homomorphism out of it.
\end{enumerate}
The last property follows from the second one and from the definition of a
negligible representation. It also covers the homomorphisms connecting a
negligible representation with the physical ones.

In the case of three strands, the representation $\operatorname{spin}(3/2,+)$
has zero quantum dimension and is therefore negligible. Consequently, the
part of the braiding operator that acted as an endomorphism of
$\operatorname{spin}(3/2,+)$ was a negligible homomorphism. Its inclusion
into the full operator on the three-strand space preserves negligibility by
property~2.

Adding one more strand corresponds to taking the tensor product with the
identity map on the space of that strand. By property~3, negligibility is
preserved under this operation. By property~2, it is also preserved under
all subsequent braidings and changes of decomposition. Therefore, upon
passing to four strands, all contributions that passed through the channel
$\operatorname{spin}(3/2,+)$ remain negligible, including those that have
become mixed with the physical representations. This is precisely how the
negligible part associated with the $\operatorname{ind}$-type
representation arises.

As further strands are added, the nonphysical part of the decomposition is
built from summands that are already negligible and from new
$\operatorname{spin}(3/2,+)$ channels. The same properties apply to them.
This allows one to extend the elimination of the nonphysical part, together
with all mixing homomorphisms, to an arbitrary number of strands. For the
braid closure, as well as for the plat closure with an even number of
strands, such contributions vanish.

Relying on the general framework of Etingof and Ostrik, we now give a direct
proof of the statements needed in the present work, without resorting to
the categorical formalism. To this end, we use properties of endomorphisms
of indecomposable representations, embeddings, projections, and the quantum
trace.

\subsubsection{Braid closure for an arbitrary number of strands}
\label{sec:braid-closure-general}
We first formulate a lemma on endomorphisms of an indecomposable
representation. 
\begin{lemma}\label{lem:indecomposable-endomorphism}
Let $Z$ be a finite-dimensional indecomposable
representation of the quantum algebra over $\mathbb C$, and let
$\xi\in\operatorname{End}_{U_q}(Z)$. Then there exist $\lambda\in\mathbb C$
and a nilpotent endomorphism $\nu$ such that
\begin{equation}
	\label{eq:s34:indecomposable-form}
	\xi=\lambda\operatorname{id}_Z+\nu,
	\qquad \nu^s=0
\end{equation}
for some integer $s\geq1$. For the quantum trace used in this work, one has
\begin{equation}
	\label{eq:s34:indecomposable-trace}
	\operatorname{Tr}_{q,Z}(\xi)
	=\operatorname{Tr}_Z(K_Z\xi)
	=\lambda\operatorname{qdim}Z.
\end{equation}
In particular, if the quantum dimension of $Z$ vanishes, then the quantum
trace of every endomorphism of $Z$ vanishes as well.
\end{lemma}
\begin{proof}
Since the endomorphism $\xi$ is a
homomorphism of representations, it commutes with the action $\rho(x)$ of
every element $x$ of the quantum algebra:
\begin{equation}
	\xi\,\rho(x)=\rho(x)\,\xi .
\end{equation}
Let us show that $\xi$ has a single eigenvalue. Let us prove it by contradiction. Assume that $\xi$ has several distinct eigenvalues. To each eigenvalue $\lambda$ one
associates the generalized eigenspace
\begin{equation}
	Z_\lambda = \ker (\xi-\lambda I)^d, \qquad \text{where } d=\dim(Z).
\end{equation}
This generalized eigenspace $Z_\lambda$ is a subrepresentation. Indeed, take
a vector $v\in Z_\lambda$. Then for the action $\rho(x)$ of any element $x$
on the space $Z$,
\begin{equation}
	(\xi-\lambda I)^d\,\rho(x)v
	=\rho(x)\,(\xi-\lambda I)^d v=0 \qquad\Rightarrow\qquad \rho(x)v\in Z_\lambda.
\end{equation}
If there are several distinct eigenvalues, then $Z$ has several distinct
generalized eigenspaces, and $Z$ decomposes into their direct sum. This
contradicts the indecomposability of $Z$. Hence the endomorphism $\xi$ has
only one eigenvalue on $Z$. Moreover, by the Cayley--Hamilton theorem,
\begin{equation}
	(\xi-\lambda I)^d=0.
\end{equation}
Thus $\nu=\xi-\lambda\operatorname{id}_Z$ is nilpotent, and since $\nu$
commutes with $K_Z$, the operator $K_Z\nu$ is nilpotent as well, so that its
ordinary trace vanishes. Hence
\begin{equation}
	\label{eq:s34:trace-proof}
		\operatorname{Tr}_{q,Z}(\xi)
		=\operatorname{Tr}_Z\bigl(K_Z
		(\lambda\operatorname{id}_Z+\nu)\bigr)
		=\lambda\operatorname{Tr}_Z(K_Z)
		+\operatorname{Tr}_Z(K_Z\nu)
		=\lambda\operatorname{qdim}Z.
\end{equation}
\end{proof}

We now turn to tensor products of a larger number of fundamental
representations. Denote the full space of $n$ strands by
\begin{equation}
	\label{eq:s34:tensor-power}
	V_n=\operatorname{spin}(1/2,+)^{\otimes n}.
\end{equation}
To obtain the decompositions of the fifth and sixth tensor powers, we use
Tables~\ref{tab:spin_tensor_products} and~\ref{tab:mult_table_spin_ind}. In
particular, we shall need the rules
\begin{equation}
	\label{eq:s34:ind-times-fundamental}
	\begin{aligned}
		\operatorname{ind}(1,+)\otimes\operatorname{spin}(1/2,+)
		&\simeq\operatorname{ind}(1/2,+)
		\oplus2\,\operatorname{spin}(3/2,+),\\
		\operatorname{ind}(1/2,+)\otimes\operatorname{spin}(1/2,+)
		&\simeq\operatorname{ind}(0,+)\oplus\operatorname{ind}(1,+).
	\end{aligned}
\end{equation}
Multiplying the known four-strand decomposition~\eqref{eq:4strand_real_decomp}
by one more fundamental representation, we obtain
\begin{equation}
	\label{eq:s34:V5}
	V_5\simeq
	4\,\operatorname{spin}(1/2,+)
	\oplus4\,\operatorname{spin}(3/2,+)
	\oplus\operatorname{ind}(1/2,+).
\end{equation}
Repeating this operation, we find
\begin{equation}
	\label{eq:s34:V6}
		V_6\simeq{}
		4\,\operatorname{spin}(0,+)\oplus4\,\operatorname{spin}(1,+)\oplus5\,\operatorname{ind}(1,+)
		\oplus\operatorname{ind}(0,+).
\end{equation}

In the decomposition~\eqref{eq:s34:V5}, the physical part consists of four
copies of $\operatorname{spin}(1/2,+)$. The remaining summands, namely four
copies of $\operatorname{spin}(3/2,+)$ and one representation
$\operatorname{ind}(1/2,+)$, have zero quantum dimension. In the
decomposition~\eqref{eq:s34:V6}, the physical part consists of four copies
of $\operatorname{spin}(0,+)$ and four copies of $\operatorname{spin}(1,+)$.
The zero-quantum-dimension sector consists of five copies of
$\operatorname{ind}(1,+)$ and one representation $\operatorname{ind}(0,+)$.
All multiplicities here refer to individual direct summands, each of which
is an indecomposable representation.

For the ordinary braid closure, one computes the quantum trace of the full
braiding operator. However, the presence of zero-quantum-dimension summands
in the decompositions~\eqref{eq:s34:V5} and~\eqref{eq:s34:V6} does not by
itself allow one to apply the Lemma~\ref{lem:indecomposable-endomorphism} directly to the full
operator: the lemma concerns an endomorphism of a single indecomposable
representation.

Each operator $\check R_i$ is an endomorphism of the whole space $V_n$, but
in general it does not preserve each chosen direct summand separately. In
particular, after the action of $\check R_i$, a vector from an
$\operatorname{ind}$-type representation may acquire a component in a
physical direct summand, and a physical vector may acquire a component in
an $\operatorname{ind}$-type representation. This is precisely the mixing
we have already encountered in the four-strand case
(Sec.~\ref{sec:four_strands}). Therefore, the restriction of the full
operator to a chosen $\operatorname{ind}$-type representation cannot
automatically be regarded as an endomorphism of that representation: its
image may leave the chosen summand.

Moreover, the braiding operators may connect different isomorphic copies of
the same representation. For example, five copies of
$\operatorname{ind}(1,+)$ appear in~\eqref{eq:s34:V6}. Their direct sum is a
decomposable representation, so the Lemma~\ref{lem:indecomposable-endomorphism} cannot be applied to it as to a
single indecomposable representation either.

Thus, in order to apply Lemma~\ref{lem:indecomposable-endomorphism}, one has to extract from the full operator
an endomorphism of each individual indecomposable summand, taking into
account the homomorphisms between different summands. To this end, we now
introduce the embedding and projection homomorphisms associated with the
chosen decomposition of $V_n$.

Let us fix a decomposition of the full space $V_n$ into a direct sum of
indecomposable representations:
\begin{equation}
	\label{eq:s34:decomposition}
	V_n=\bigoplus_{r}Z_r.
\end{equation}
The index $r$ labels the individual direct summands, including the
different copies of isomorphic representations. For each summand we
introduce the embedding and the projection
\begin{equation}
	\label{eq:s34:inclusions-projections}
	\tilde{\mathfrak{p}}_r:Z_r\longrightarrow V_n,
	\qquad
	\mathfrak{p}_r:V_n\longrightarrow Z_r.
\end{equation}
The embedding $\tilde{\mathfrak{p}}_r$ regards a vector of $Z_r$ as a vector
of the full space, while the projection $\mathfrak{p}_r$ extracts its
component in the chosen summand. These maps satisfy the relations
\begin{equation}
	\label{eq:s34:projection-inclusion-relations}
	{\mathfrak{p}}_r\tilde{\mathfrak{p}}_r=\operatorname{id}_{Z_r},
	\qquad
	{\mathfrak{p}}_r\tilde{\mathfrak{p}}_s=0\quad(r\ne s).
\end{equation}
The corresponding projector onto the $r$-th summand, now acting on the whole
space $V_n$, is denoted by $Q_r$:
\begin{equation}
	\label{eq:s34:projector-relations}
	\begin{gathered}
		Q_r=\tilde{\mathfrak{p}}_r {\mathfrak{p}}_r,\qquad Q_r^2=Q_r,\qquad
		Q_rQ_s=0\quad(r\ne s),\\
		\sum_{r}Q_r=\operatorname{id}_{V_n}.
	\end{gathered}
\end{equation}
The operator $Q_r$ keeps the component in the $r$-th summand and annihilates
all the others. Since the decomposition~\eqref{eq:s34:decomposition} is a
direct sum of representations of the quantum algebra, the embeddings
$\tilde{\mathfrak{p}}_r$, the projections $\mathfrak{p}_r$, and the projectors
$Q_r$ are homomorphisms of representations: they are compatible with the
action of all generators, including the action of $K$.

We now formulate one more lemma, which allows us to take into account the
mixing of different direct summands. 
\begin{lemma}\label{lem:zero-channel-trace}
Let $Z_r$ be an indecomposable direct
summand of $V_n$ with zero quantum dimension, and let
$Q_r=\tilde{\mathfrak{p}}_r\,\mathfrak{p}_r$ be the corresponding projector.
Then for any $\mathbf{G},\mathbf{H}\in\operatorname{End}_{U_q}(V_n)$
\begin{equation}
	\label{eq:s34:zero-channel-trace}
	\operatorname{Tr}_{q,V_n}(\mathbf{G}Q_r \mathbf{H})=0.
\end{equation}
\end{lemma}

\begin{proof}
Since $\mathfrak{p}_r$ and $\mathbf{H}$ are homomorphisms of
representations, $\mathfrak{p}_r\mathbf{H}K_{V_n}=K_{Z_r}\mathfrak{p}_r\mathbf{H}$.
Therefore
\begin{equation}
	\label{eq:s34:zero-channel-proof}
	\begin{aligned}
		\operatorname{Tr}_{q,V_n}(\mathbf{G}Q_r \mathbf{H})
		&=\operatorname{Tr}_{V_n}(K_{V_n}\mathbf{G} \tilde{\mathfrak{p}}_r {\mathfrak{p}}_r \mathbf{H})\\
		&=\operatorname{Tr}_{Z_r}({\mathfrak{p}}_r \mathbf{H} K_{V_n}\mathbf{G} \tilde{\mathfrak{p}}_r)\\
		&=\operatorname{Tr}_{Z_r}(K_{Z_r} {\mathfrak{p}}_r \mathbf{H} \mathbf{G} \tilde{\mathfrak{p}}_r)\\
		&=\operatorname{Tr}_{q,Z_r}({\mathfrak{p}}_r \mathbf{H} \mathbf{G} \tilde{\mathfrak{p}}_r)=0.
	\end{aligned}
\end{equation}
In the second line we have used the cyclicity of the ordinary trace for
rectangular matrices, with the corresponding change of the space over which
the trace is taken. The composition
$\mathfrak{p}_r\mathbf{H}\mathbf{G}\tilde{\mathfrak{p}}_r$ is an endomorphism
of the entire indecomposable representation $Z_r$. Hence the last equality
follows from Lemma~\ref{lem:indecomposable-endomorphism} and the vanishing of the quantum
dimension of $Z_r$. Note that $\mathbf{G}$ and $\mathbf{H}$ may mix different
direct summands: after composition with $\tilde{\mathfrak{p}}_r$ and
$\mathfrak{p}_r$, the source and the target are nevertheless the same
space $Z_r$.
\end{proof}

Let us introduce the projectors onto the physical and the nonphysical
parts:
\begin{equation}
	\label{eq:s34:physical-projector-sums}
	Q_{\mathrm{phys}}=
	\sum_{r\in \text{phys}}Q_r,
	\qquad
	Q_{\mathrm{nonphys}}=
	\sum_{r\in \text{nonphys}}Q_r,
	\qquad
	Q_{\mathrm{phys}}+Q_{\mathrm{nonphys}}=\operatorname{id}_{V_n}.
\end{equation}
By linearity of the trace, the lemma just proved gives
\begin{equation}
	\label{eq:s34:nf-projector-trace}
	\operatorname{Tr}_{q,V_n}(\mathbf{G}Q_{\mathrm{nonphys}} \mathbf{H})
	=
	\sum_{r\in\text{nonphys}}
	\operatorname{Tr}_{q,V_n}(\mathbf{G}Q_r \mathbf{H})=0.
\end{equation}
Thus, for the entire nonphysical part the result is obtained by summing over
the individual indecomposable summands. 
\begin{corollary}\label{cor:braid-physical-reduction}
For any
$\mathbf{G},\mathbf{H}\in\operatorname{End}_{U_q}(V_n)$ one has
\begin{equation}\label{eq:s34:trace-physical-reduction}
	\operatorname{Tr}_{q,V_n}(\mathbf{G}\mathbf{H})
	=
	\operatorname{Tr}_{q,V_n}(\mathbf{G}Q_{\mathrm{phys}}\mathbf{H})
	=
	\operatorname{Tr}_{q,V_n}(Q_{\mathrm{phys}}\mathbf{G}Q_{\mathrm{phys}}\mathbf{H} Q_{\mathrm{phys}})
	=
	\operatorname{Tr}_{q,V_n}(\left.\mathbf{G}\right|_{\mathrm{physical}}\left.\mathbf{H}\right|_{\mathrm{physical}})
\end{equation}
\end{corollary}
It follows that, for any product of braiding matrices under the quantum
trace, each braiding matrix can be reduced to its physical part, while the
mixing part and the nonphysical part can be discarded. This does not affect
the quantum trace. We have thus shown that, for an arbitrary number of
strands, the braid closure leads to no contradiction between Ising anyons
and the representation theory of the quantum algebra $U_q(\mathfrak{sl}_2)$
at $q=\exp(i\pi/4)$. The reduced braiding matrices coincide, up to the
choice of braiding orientation and an overall phase, with the braiding
matrices of Ising anyons.

\subsubsection{Plat closure for an even number of strands}
\label{sec:plat-closure-general}

Consider the plat closure of $2n$ strands. The initial state is obtained by
creating $n$ pairs from the vacuum, each pair carrying total vacuum charge.
We denote this state by $|\varnothing^{2n}\rangle$, where the superscript
indicates the number of strands. The creation of the pairs and their
subsequent fusion into the vacuum are given by the homomorphisms of
representations
\begin{equation}
	\label{eq:s34:plat-vacuum-maps}
	|\varnothing^{2n}\rangle:\mathbf1\longrightarrow V_{2n},
	\qquad
	\langle\varnothing^{2n}|:V_{2n}\longrightarrow\mathbf1,
\end{equation}
where $\mathbf1=\operatorname{spin}(0,+)$ is the trivial one-dimensional
representation. Here the bra-vector denotes the invariant fusion functional,
compatible with the action of the quantum algebra.

Each braiding matrix is an endomorphism of $V_{2n}$ and therefore preserves
the invariance of the initial state. Consequently, for the plat closure it
suffices to consider the copies of $\operatorname{spin}(0,+)$ and the
invariant subspaces inside the summands $\operatorname{ind}(0,+)$. The
latter may mix with the physical vacuum states, so the preservation of
invariance alone does not yet allow one to discard the nonphysical part.

Let an arbitrary braid word have the form
\begin{equation}
	\label{eq:s34:plat-word-amplitude}
	\begin{aligned}
		\check B_{\mathrm{pl}}&=T_\ell\cdots T_1,
		\qquad T_a=\check R_{i_a}^{\varepsilon_a},
		\qquad \varepsilon_a=\pm1
		.
	\end{aligned}
\end{equation}
We prove that the amplitude
$\langle\varnothing^{2n}|\check B_{\mathrm{pl}}|\varnothing^{2n}\rangle$ is
determined by the physical parts of the braiding operators only.

The following statement is the analogue of Lemma~\ref{lem:zero-channel-trace}.
\begin{lemma}\label{lem:plat-zero-channel}
Let $Z_r$ be an indecomposable direct summand of $V_{2n}$ with zero quantum
dimension, and let $Q_r=\tilde{\mathfrak{p}}_r\,\mathfrak{p}_r$ be the
corresponding projector. Then for any
$\mathbf{G},\mathbf{H}\in\operatorname{End}_{U_q}(V_{2n})$
\begin{equation}
	\label{eq:s34:plat-zero-channel}
	\langle\varnothing^{2n}|\mathbf{G}Q_r \mathbf{H}|\varnothing^{2n}\rangle=0.
\end{equation}
\end{lemma}
\begin{proof}
To prove this, we introduce the homomorphisms
\begin{equation}
	\label{eq:s34:plat-channel-maps}
	\begin{aligned}
		\mathfrak{a}&={\mathfrak{p}}_r \mathbf{H}|\varnothing^{2n}\rangle:\mathbf1\longrightarrow Z_r,\\
		\mathfrak{b}&=\langle\varnothing^{2n}|\mathbf{G}\tilde{\mathfrak{p}}_r:Z_r\longrightarrow\mathbf1.
	\end{aligned}
\end{equation}
Their composition $\mathfrak{b}\mathfrak{a}$ is a scalar operator on
$\mathbf1$. Identifying this operator with the corresponding number and
taking into account that $\operatorname{qdim}\mathbf1=1$, we obtain
\begin{equation}
	\label{eq:s34:plat-zero-channel-proof}
	\begin{aligned}
		\langle\varnothing^{2n}|\mathbf{G}Q_r \mathbf{H}|\varnothing^{2n}\rangle
		&=\mathfrak{b}\mathfrak{a}=\operatorname{Tr}_{q,\mathbf1}(\mathfrak{b}\mathfrak{a})\\
		&=\operatorname{Tr}_{q,Z_r}(\mathfrak{a}\mathfrak{b})=0.
	\end{aligned}
\end{equation}
Here we have used the cyclicity of the quantum trace for homomorphisms of
representations, with the corresponding change of the space over which the
trace is taken. The composition $\mathfrak{a}\mathfrak{b}$ is an endomorphism
of the entire indecomposable representation $Z_r$. Hence the last equality
follows from Lemma~\ref{lem:indecomposable-endomorphism}. Note that $\mathbf{G}$ and
$\mathbf{H}$ may mix physical and nonphysical summands. 
\end{proof}

For the projector $Q_{\mathrm{nonphys}}$ onto the sum of all nonphysical
summands, the statement of Lemma~\ref{lem:plat-zero-channel} is obtained by summing over the individual
$Z_r$. Let us now insert the decomposition of the identity operator
$Q_{\mathrm{phys}}+Q_{\mathrm{nonphys}}$ between neighbouring factors of the
braid word, as well as between this word and the vacuum bra- and
ket-vectors. After expanding the products, every term containing at least
one projector $Q_{\mathrm{nonphys}}$ vanishes by Lemma~\ref{lem:plat-zero-channel}: the
remaining operators to the left and to the right of the chosen projector
are absorbed into $\mathbf{G}$ and $\mathbf{H}$, respectively. Thus,
\begin{equation}
	\label{eq:s34:plat-physical-reduction}
		\langle\varnothing^{2n}|\check B_{\mathrm{pl}}
		|\varnothing^{2n}\rangle=
		\langle\varnothing^{2n}|
		Q_{\mathrm{phys}}T_\ell Q_{\mathrm{phys}}
		\cdots Q_{\mathrm{phys}}T_1Q_{\mathrm{phys}}
		|\varnothing^{2n}\rangle.
\end{equation}

Consequently, each braiding matrix can be replaced by its physical block,
that is, by the operator
$T_a|_\text{phys} = Q_{\mathrm{phys}}T_aQ_{\mathrm{phys}}$:
\begin{equation}
	\langle\varnothing^{2n}|\check B_{\mathrm{pl}}
	|\varnothing^{2n}\rangle=
	\langle\varnothing^{2n}\big|
	\;
	T_\ell |_\text{phys}
	\cdots T_1 |_\text{phys}
	\;
	\big|\varnothing^{2n}\rangle.
\end{equation}
All contributions passing through the nonphysical summands, including the
transitions from the physical part to the nonphysical one and back, vanish.
The remaining expression involves only physical states with total vacuum
charge.

Taking into account the established correspondence between the physical
braiding matrices and the braiding matrices of Ising anyons, and with a
consistent normalization of pair creation and fusion, the plat-closure
amplitudes coincide up to an overall phase and the choice of braiding
orientation. The corresponding normalized probabilities coincide as well.

\section{Conclusion}
\label{sec:conclusion}

In this work we have shown how the representations of the quantum algebra
$U_q(\mathfrak{sl}_2)$ at $q=e^{i\pi/4}$ reproduce the fusion rules, the
transformations between fusion bases, and the braiding operations of Ising
anyons. We have considered the correspondence between the three physical
representations and the topological charges $\mathbbm{1}$, $\sigma$, and
$\psi$. Their quantum dimensions coincide, and the decomposition rules for
tensor products turn into the fusion rules of the anyons once the direct
summands of zero quantum dimension are discarded.

For systems of two, three, and four strands, we have explicitly constructed
the decompositions of tensor products of the fundamental representation and
the braiding matrices. For three and four strands, the transitions between
different fusion bases have also been computed. After the normalizations are
matched, the physical parts of these transitions reproduce the corresponding
matrices for Ising anyons. The braiding matrices agree with the Ising ones
up to an overall phase and the choice of braiding orientation.

Particular attention has been paid to four strands, where an indecomposable
representation of $\operatorname{ind}$ type appears in the decomposition.
We have shown that a change of the decomposition order and the action of
the braiding operators may mix the physical states with components of this
representation. Therefore, the physical results cannot be justified merely
by the coincidence of quantum dimensions or by simply discarding the extra
rows and columns of the full matrices. One has to take into account the
action of these matrices in the closed compositions that define the
observables under consideration.

For an arbitrary number of strands, we have given a direct proof that the
representations of zero quantum dimension do not contribute, including
their mixing with the physical summands. The proof relies on the properties
of negligible homomorphisms, of the quantum trace, and of endomorphisms of
indecomposable representations. Embeddings and projections make it possible
to apply these properties to each individual direct summand, even when the
braiding operators do not preserve it. For the ordinary braid closure of an
arbitrary number of strands, all terms passing through the nonphysical
summands have vanishing quantum trace. For the plat closure of any even
number of strands, the corresponding vacuum matrix elements vanish in the
same way. Thus, in both cases, each braiding matrix can be replaced by its
physical block without changing the quantity being computed.

The results obtained justify the use of the physical blocks of the
quantum-algebraic matrices in computing the considered observables of
$SU(2)_2$ Chern--Simons theory and in comparing them with the description
in terms of Ising anyons. With a consistent normalization of the closures,
the corresponding amplitudes agree up to an overall phase and the choice of
braiding orientation, and the normalized probabilities coincide. Thus, the
presence of additional representations in the full tensor products, and of
the mixing with them, does not alter the results of the anyonic operations
under consideration.

\section{Outlook}

A natural continuation of the present work is the study of Fibonacci anyons
from the viewpoint of representations of the quantum algebra. In the
conventions adopted here, one may consider $U_q(\mathfrak{sl}_2)$ at
$q=e^{i\pi/5}$ and tensor products of the representation
$\operatorname{spin}(1,+)$. After the negligible summands are discarded, the
representations $\operatorname{spin}(0,+)$ and $\operatorname{spin}(1,+)$
reproduce the two topological charges of this model and the fusion rule
$\tau\times\tau=\mathbbm{1}+\tau$. It would be of interest to carry out an
analysis of the full tensor products similar to the one performed in this
work, to construct the braiding matrices and the transitions between fusion
bases, and to trace the role of the zero-quantum-dimension representations
and of their mixing with the physical states in the computation of
observables.

Another direction is to obtain a general decomposition formula for tensor
products involving representations of $\operatorname{ind}$ type in the
adopted notation. In particular, it would be of interest to write down
uniform rules for products of the types
$\operatorname{spin}\otimes\operatorname{ind}$ and
$\operatorname{ind}\otimes\operatorname{ind}$, which determine the
indecomposable direct summands and their multiplicities in terms of the
parameters of the representations and the order of the root of unity. Such
formulas would make it possible to pass from individual tables to a
systematic description of tensor products and to obtain decompositions for
a larger number of factors.

\section{Acknowledgments}
The research was carried out within the state assignment of Ministry of Science and Higher Education of the Russian Federation for IITP RAS. Work of A.Belov was also supported by the Foundation for the Advancement of Theoretical Physics and Mathematics “BASIS” (grant no. 23-1-1-48-2). We are grateful for very useful discussions with L. Bishler.

\newpage
\bibliographystyle{unsrt}
\bibliography{references}

\end{document}

%% file: pictures/5_1_knot.tex
\begin{tikzpicture}[scale=1.1]
	\begin{knot}[
		consider self intersections=true,
		every strand/.style={
			draw=blue,
			line width=1pt,
			line cap=round,
			line join=round
		},
		clip radius=4pt,
		end tolerance=1pt,
		flip crossing/.list={2,4}
		]
		
		\strand
		(-0.1989,0.7290)
		.. controls (-1.6835,2.1428) and (0.5165,3.3341)
		.. (0.6598,1.7783)
		-- (0.6405,1.4473)
		.. controls (0.4884,0.9345) and (-0.9381,0.1381)
		.. (-0.1340,-0.3300)
		-- (0.2890,-0.5978)
		.. controls (0.6765,-1.2099) and (-0.1764,-1.9663)
		.. (-1.2550,-1.6648)
		.. controls (-3.3050,-1.0832) and (-1.9479,1.5797)
		.. (-0.9264,1.5972)
		-- (-0.4235,1.6790)
		.. controls (3.8325,1.9768) and (1.8635,-2.6742)
		.. (0.0134,-1.8251)
		-- (-0.3420,-1.4738)
		.. controls (-1.1599,-0.7331) and (0.7238,-1.1745)
		.. (0.1066,0.4721)
		-- cycle;
		
	\end{knot}
\end{tikzpicture}

%% file: pictures/4_1_braid.tex
\begin{tikzpicture}[scale=0.75]
	\rcross(0,0);
	\rcross(0,1);
	\rcross(0,2);
	\rcross(1,3);
	\lcross(0,4);
	\rcross(1,5);
	\draw[blue, thick] (2,0)--(2,3);
	\draw[blue, thick] (0,3)--(0,4);
	\draw[blue, thick] (2,4)--(2,5);
	\draw[blue, thick] (0,5)--(0,6);
	
	\draw[blue, thick] (4.5,0)--(4.5,6);
	\draw[blue, thick] (3.5,0)--(3.5,6);
	\draw[blue, thick] (4,0)--(4,6);	
	
	\draw[draw = white, double = blue, ultra thick] (2, 6) .. controls (2, 6.5) and (3.5, 6.5) .. (3.5, 6);
	\draw[draw = white, double = blue, ultra thick] (1, 6) .. controls (1, 6.75) and (4, 6.75) .. (4, 6);
	\draw[draw = white, double = blue, ultra thick] (0, 6) .. controls (0, 7) and (4.5, 7) .. (4.5, 6);
	
	\draw[draw = white, double = blue, ultra thick] (2, 0) .. controls (2, -0.25) and (3.5, -0.25) .. (3.5, 0);
	\draw[draw = white, double = blue, ultra thick] (1, 0) .. controls (1, -0.5) and (4, -0.5) .. (4, 0);
	\draw[draw = white, double = blue, ultra thick] (0, 0) .. controls (0, -0.75) and (4.5, -0.75) .. (4.5, 0);
	
	\draw[dashed, black, thick, fill=none] (-0.5,0) rectangle (2.5,6);
	
	\node[rotate=90] at (-1,3) {$\check{B}_\text{br} = \check{R}_1\check{R}_1\check{R}_1\check{R}_2\check{R}_1^{-1}\check{R}_2 $};
\end{tikzpicture}

%% file: pictures/4_1_morse_link.tex
\begin{tikzpicture}[scale=0.75]
	\lcross(1,0);
	\rcross(2,1);
	\rcross(2,2);
	\lcross(1,3);
	\lcross(1,4);
	
	\draw[blue, thick] (0,0)--(0,5);
	\draw[blue, thick] (1,1)--(1,3);
	\draw[blue, thick] (3,0)--(3,1);
	\draw[blue, thick] (3,3)--(3,5);
	
	\draw[blue, thick] (0, 5) .. controls (0, 5.5) and (1, 5.5) .. (1, 5);
	\draw[blue, thick] (2, 5) .. controls (2, 5.5) and (3, 5.5) .. (3, 5);
	
	\draw[blue, thick] (0, 0) .. controls (0, -0.5) and (1, -0.5) .. (1, 0);
	\draw[blue, thick] (2, 0) .. controls (2, -0.5) and (3, -0.5) .. (3, 0);
	
	\draw[dashed,black, thick, fill=none] (-0.5,0) rectangle (3.5,5);
	
	\node[rotate=90] at (-1,2.5) {$\check{B}_\text{pl} = \check{R}_2^{-1}\check{R}_3\check{R}_3\check{R}_2^{-1}\check{R}_2^{-1}$};
	
	\node[rotate=0] at (1.5,6) {$|\varnothing\otimes\varnothing\rangle$};
	
	\node[rotate=0] at (1.5,-1) {$\langle\varnothing\otimes\varnothing|$};
	
\end{tikzpicture}

%% file: pictures/spin_fund.tex
\input{pictures/decomposition_pic_styles}
\input{pictures/ydiagram_style}
\node[vector] (L0) {$ \lvert\uparrow\rangle_{1/2} $};

\node[vector, below=of L0] (L1) {$ \lvert\downarrow\rangle_{1/2} $};
\draw[edge] (L0) to[bend right = 30] (L1);
\draw[edge] (L1) to[bend right = 30] (L0);

\node[frame, fit=(L0)(L1), label=north:{Spin(1/2,+)}] {};

%% file: pictures/spin_sym.tex
\input{pictures/decomposition_pic_styles}
\input{pictures/ydiagram_style}
\node[vector] (L0) {$ \lvert\uparrow\rangle_1 $};

\node[vector, below=of L0] (L1) {$ \lvert0\rangle_1$};
\draw[edge] (L0) to[bend right = 30] (L1);
\draw[edge] (L1) to[bend right = 30] (L0);

\node[vector, below=of L1] (L2) {$ \lvert\downarrow\rangle_1 $};
\draw[edge] (L1) to[bend right = 30] (L2);
\draw[edge] (L2) to[bend right = 30] (L1);

\node[frame, fit=(L0)(L1)(L2), label=north:{Spin(1,+)}] {};

%% file: pictures/spin_zero.tex
\input{pictures/decomposition_pic_styles}
\input{pictures/ydiagram_style}
\node[vector] (L0) {$ \lvert\Uparrow\rangle_{3/2} $};

\node[vector, below=of L0] (L1) {$ \lvert\uparrow\rangle_{3/2} $};
\draw[edge] (L0) to[bend right = 30] (L1);
\draw[edge] (L1) to[bend right = 30] (L0);

\node[vector, below=of L1] (L2) {$ \lvert\downarrow\rangle_{3/2} $};
\draw[edge] (L1) to[bend right = 30] (L2);
\draw[edge] (L2) to[bend right = 30] (L1);

\node[vector, below=of L2] (L3) {$ \lvert\Downarrow\rangle_{3/2} $};
\draw[edge] (L2) to[bend right = 30] (L3);
\draw[edge] (L3) to[bend right = 30] (L2);

\node[frame, fit=(L0)(L1)(L2)(L3), label=north:{Spin(3/2,+)}] {};

%% file: pictures/ind_fund.tex
\input{pictures/decomposition_pic_styles}
\input{pictures/ydiagram_style}
\node[vector] (L0) {$ \lvert+1;+3/2\rangle $};

\node[vector, below=of L0] (L1) {$ \left( -i \right) \lvert +1;+1/2 \rangle + \lvert0;+3/2\rangle $};
\draw[edge] (L0) to[bend right = 30] (L1);
\draw[edge] (L1) to[bend right = 30] (L0);

\node[vector, below=of L1] (L2) {$ \left( -1 \right) \lvert +1;-1/2 \rangle + \left( -i + 1 \right) \lvert 0;+1/2 \rangle + \lvert-1;+3/2\rangle $};
\node[vector, right=of L2] (R2) {$ \left( -i - \frac{1}{2} \right) \lvert +1;-1/2 \rangle + \lvert0;+1/2\rangle $};
\draw[edge] (L1) to[bend right = 30] (L2);
\draw[edge] (R2) to[bend right = 15] (L1.east);

\node[vector, below=of L2] (L3) {$ \left( i \right) \lvert +1;-3/2 \rangle + \left( -i \right) \lvert 0;-1/2 \rangle + \lvert-1;+1/2\rangle $};
\node[vector, right=of L3] (R3) {$ \left( \frac{1}{2} i - 1 \right) \lvert +1;-3/2 \rangle + \left( -i + \frac{1}{2} \right) \lvert 0;-1/2 \rangle + \lvert-1;+1/2\rangle $};
\draw[edge] (L2) to[bend right = 30] (L3);
\draw[edge] (L3) to[bend right = 30] (L2);
\begin{scope}
 \coordinate (fork) at ($ (L2)!0.5!(R2) + (0.5cm,-0.8cm) $);
\draw[-, semithick] (R3.north west) to [out=155, in=-90] (fork);
\draw[edge] (fork) to [out=90, in=-25] (L2.south east);
\draw[edge] (fork) to [out=90, in=-135] (R2.south west);
\end{scope}
\draw[edge] (R2) to[bend left = 30] (R3);

\node[vector, below=of L3] (L4) {$ \left( -\frac{1}{2} i - \frac{1}{2} \right) \lvert 0;-3/2 \rangle + \left( \frac{1}{2} \right) \lvert -1;-1/2 \rangle $};
\draw[edge] (R3) to[bend left = 15] (L4.east);
\draw[edge] (L4) to[bend right = 30] (L3);

\node[vector, below=of L4] (L5) {$ \left( -\frac{1}{2} \right) \lvert -1;-3/2 \rangle $};
\draw[edge] (L4) to[bend right = 30] (L5);
\draw[edge] (L5) to[bend right = 30] (L4);

\node[frame, fit=(L0)(L1)(L2)(R2)(L3)(R3)(L4)(L5), label=north:{spin(1,+) $\otimes$ spin(3/2,+) $\to$ ind(1/2,+)}] {};